\documentclass[aps,prx,twocolumn,superscriptaddress,final,amsmath,amssymb]{revtex4-2}

\usepackage{graphicx}% Include figure files
\usepackage{dcolumn}% Align table columns on decimal point
\usepackage{bm}% bold math
\usepackage{hyperref}% add hypertext capabilities
\usepackage[usenames,dvipsnames]{xcolor}

\definecolor{vibrant}{HTML}{CC6600}
\definecolor{muted}{HTML}{003AB8}

\hypersetup{
    colorlinks,
    linkcolor = vibrant,
    citecolor = muted,
    urlcolor = muted
}

\usepackage{braket}
\usepackage{enumitem}
\usepackage[english]{babel}

\newcommand{\tr}{\mathrm{Tr}}

\newcommand{\pD}[2]{\frac{\partial #1}{\partial #2}}
\newcommand{\al}[1]{\begin{align}#1\end{align}}
\newcommand{\eq}[1]{\begin{equation}#1\end{equation}}
\newcommand{\conj}[1]{\overline{#1}}

\newcommand{\Z}{\mathbb{Z}}

\newcommand{\mzero}[2]{\frac{1}{\sqrt{2}}\left(\ket{\uparrow_{#1}\downarrow_{#2}} \pm \ket{\downarrow_{#1}\uparrow_{#2}}\right)}
\newcommand{\singlet}[2]{\frac{1}{\sqrt{2}}\left(\ket{\uparrow_{#1}\downarrow_{#2}} - \ket{\downarrow_{#1}\uparrow_{#2}}\right)}
\newcommand{\triplet}[2]{\frac{1}{\sqrt{2}}\left(\ket{\uparrow_{#1}\downarrow_{#2}} + \ket{\downarrow_{#1}\uparrow_{#2}}\right)}
\newcommand{\upaligned}[2]{\ket{\uparrow_{#1}\uparrow_{#2}}}
\newcommand{\downaligned}[2]{\ket{\downarrow_{#1}\downarrow_{#2}}}
\newcommand{\XX}[2]{X_{#1}X_{#2} + Y_{#1}Y_{#2}}

\newcommand{\Heff}{H_\mathrm{eff}}

\newcommand{\spinUp}[3]{\Uparrow^{#1}_{\mathbf{#2}, \mathbf{#3}}}

\newcommand{\spinups}[1]{\uparrow^{\otimes\mathbf{#1}}}
\newcommand{\spinDown}[3]{\Downarrow^{#1}_{\mathbf{#2}, \mathbf{#3}}}

\newcommand{\spindowns}[1]{\downarrow^{\otimes\mathbf{#1}}}

\begin{document}

% \preprint{APS/xxx}

\title{Superspin Renormalization and Slow Relaxation in Random Spin Systems}


\author{Yi J. Zhao*}
\affiliation{Department of Physics, University of California, Berkeley, California 94720, USA}
\email{yi\_zhao@berkeley.edu}
\author{Samuel J. Garratt}
\affiliation{Department of Physics, University of California, Berkeley, California 94720, USA}
\author{Joel E. Moore}
\affiliation{Department of Physics, University of California, Berkeley, California 94720, USA}

\affiliation{Materials Sciences Division, Lawrence Berkeley National Laboratory, Berkeley, California 94720, USA}

\date{\today}

\begin{abstract}
We develop an excited-state real-space renormalization group (RSRG-X) formalism to describe the dynamics of conserved densities in randomly interacting spin-$\frac{1}{2}$ systems. Our formalism is suitable for systems with $\textrm{U}(1)$ and $\mathbb{Z}_2$ symmetries, and we apply it to chains of randomly positioned spins with dipolar $XX+YY$ interactions, as arise in Rydberg quantum simulators and other platforms. The formalism generates a sequence of effective Hamiltonians which provide approximate descriptions for dynamics on successively smaller energy scales. These effective Hamiltonians involve ``superspins'': two-level collective degrees of freedom constructed from (anti)aligned microscopic spins. Conserved densities can then be understood as relaxing via coherent collective spin flips. For the well-studied simpler case of randomly interacting nearest-neighbor $XX+YY$ chains, the superspins reduce to single spins. Our formalism also leads to a numerical method capable of simulating the dynamics up to an otherwise inaccessible combination of large system size and late time. Focusing on disorder-averaged infinite-temperature autocorrelation functions, in particular the local spin survival probability $\overline{S_p}(t)$, we demonstrate quantitative agreement in results between our algorithm and exact diagonalization (ED) at low but nonzero frequencies. Such agreement holds for chains with nearest-neighbor, next-nearest-neighbor, and long-range dipolar interactions. Our results indicate decay of $\overline{S_p}(t)$ slower than any power law and feature no significant deviation from the $\sim 1/ \log^2(t)$ asymptote expected from the infinite-randomness fixed-point of the nearest-neighbor model. We also apply the RSRG-X formalism to two-dimensional long-range systems of moderate size and find slow late-time decay of $\overline{S_p}(t)$.
\end{abstract}

\maketitle

%\tableofcontents
\section{Introduction}
\label{sec:introduction}
Recent years have witnessed the development of experimental platforms enabling high-fidelity local control over and readout from many-body spin systems. Many of these systems, such as nitrogen-doped diamond~\cite{Choi17, Kucsko18, Zu21, Davis23} as well as arrays of Rydberg atoms~\cite{Browaeys_review}, ultracold magnetic atoms~\cite{Su23, bloch2023trapping, gruen2024optical}, and polar molecules~\cite{cornish2024quantum}, feature long-range, often dipolar interactions between spins that are randomly or arbitrarily positioned. These developments call for progress in theories that can yield predictions for the dynamics of systems with irregular geometries and long-range interactions.

In this work, we study the late-time local relaxation of conserved densities in such systems. For concreteness, we examine the relaxation of the conserved spin density in systems with random $XX+YY$ interactions. We discover a dramatic impact of the randomness on the infinite-temperature autocorrelation function of the $Z$ component of the spin, whose average over realizations of the randomness (or, in a large system, over space) features heavy late-time tails decaying slower than any power law. In experiments, this autocorrelation function can be measured after the preparation of randomly initialized spins and has served as a probe of their ensuing dynamics~\cite{Davis23, Zu21, Peng23, Titus_experiment_24}. We put an emphasis on chains with dipolar $XX+YY$ interactions and positional randomness, as are particularly relevant to experiments on Rydberg atoms~\cite{Browaeys_review}, including for example early work emulating a three-site spin chain~\cite{Browaeys_chain}. 

A longer version of such an array of Rydberg atoms, with some fraction of sites left empty as typically occurs in experiments, will turn out to realize a kind of random quantum Hamiltonian that has been of great theoretical interest for decades. The analysis of disordered Hamiltonians has been generally advanced by real-space, strong-randomness renormalization group (RSRG) approaches~\cite{IM05, RA13, IM18, Huse23}. Particularly, the ground state and low-energy excited states of one-dimensional spin chains under random nearest-neighbor interactions have been intensively studied with RSRG and shown to possess universal properties in many cases~\cite{MDH79, DM80, BL82, Fisher92, Fisher94, Westerberg95, Fisher95, Westerberg97, DMH00, MDH01, RKF02, DH02, RM04, Mohdeb20, Roberts21}. The above-mentioned experiments, however, can hardly access low temperatures due to the lack of clear cooling protocols.

On the theory side, several works have generalized the ground-state RSRG to describe, under specific setups, generic excited states (RSRG-X) or non-equilibrium dynamics~\cite{VA13, Pek14, VA14, HM14, VPP15, YQX16, VFPP16, Monthus18, Protopopov20, MVK22, Titus_theory_22, Mohdeb23, BVG24, Aramthottil24}. An RSRG-X formalism centers on an iterative scheme to update an effective Hamiltonian $H_\mathrm{eff}$ to lower energy scales. In a random system, these updates involve identifying a strong coupling in $H_\mathrm{eff}$ and approximating all others as perturbations. One then updates $H_\mathrm{eff}$ to describe dynamics within each eigenspace of the strong coupling. Physically, such treatment can be viewed as ``integrating out'' the fast degrees of freedom. Typically, the randomness of the system improves along RSRG-X, leading to more trustworthy predictions at lower energy scales and hence later times. We propose an RSRG-X formalism to analyze infinite-temperature spin relaxation under the experimentally relevant setups.

A common challenge for applying RSRG-X, either numerically or analytically, is the complexity of $\Heff$.  Generally, the iterative steps of RSRG-X generate new classes of interactions which are not present in the original model. In this work, we identify a situation where the symmetries of the system, here global $\textrm{U}(1)$ and $\Z_2$, severely restrict the forms of interactions allowed in $\Heff$. This enables us to formulate an RSRG-X formalism applicable to a wide range of randomly interacting systems, including those featuring long-range $XX+YY$ and $ZZ$ interactions. In our formalism, $\Heff$ describes dynamics as coherent flipping of spin pairs and nontrivial \textit{superspins}, two-level collective degrees of freedom consisting of multiple (anti)aligned spins. These relaxation processes are examples of many-body resonances, which are known to control slow dynamics in many-body localized systems \cite{Gopalakrishnan2015,Crowley2022,Garratt2021,Garratt2022,Long2023}. As we will comment on, there is no contribution to dynamics from nontrivial superspins in systems that describe non-interacting fermions after a Jordan-Wigner transformation, such as nearest-neighbor $XX+YY$ models. Besides the above physical picture, our RSRG-X generates a numerical method for simulating late-time dynamics at system sizes beyond the reach of exact diagonalization (ED).

Our work extends previous efforts to study the dynamics of randomly interacting spin chains with RSRG approaches. Early works have focused on low-temperature dynamics \cite{Fisher95, DMH00, MDH01, DH02}, where it has been argued that nontrivial superspins do not play a significant role~\cite{MDH01}. The non-equilibrium quench dynamics of interacting spins chains respecting $\text{U}(1)$ and $\Z_2$ symmetries have been later studied using RSRG-X in Ref.~\cite{VA13}; there, the initial states are chosen to suppress contributions from nontrivial superspins. RSRG-X techniques have since been developed for Hamiltonians that are sums of Pauli strings (i.e., tensor products of Pauli matrices $X, Y, Z$) that have statistically independent strengths \cite{Pek14, VA14, YQX16, Monthus18}, as well as for nearest-neighbor Heisenberg models \cite{Protopopov20}. Unfortunately, these RSRG-X developments do not apply to the general (including long-range) random $XX+YY$ models of our interest. For nearest-neighbor $XX+YY$ models, Ref.~\cite{HM14} has established an RSRG-X scheme for all of the excited states and calculated their entanglement entropies, while in models which do not map to non-interacting fermions Refs.~\cite{Mohdeb20, MVK22, Titus_theory_22, Mohdeb23, BVG24, Aramthottil24} have formulated  simplified RSRG-X schemes which neglect contributions from nontrivial superspins.

This work is organized as follows.
In Sec.~\ref{sec:overview}, we introduce the models, define the key quantities of interest, and summarize their behaviors. We also include a brief description of our formalism for those familiar with the principles of RSRG-X.
In Sec.~\ref{sec:review}, we review RSRG-X and, as an example, its well-established application to the random nearest-neighbor $XX+YY$ models \cite{HM14}.
In Sec.~\ref{sec:NN}, we apply the formalism to arrive at a two-body spin-flip picture of dynamics in the nearest-neighbor models, detailing how this picture leads to both analytical and numerical predictions, which we verify with ED.
In Sec.~\ref{sec:LR}, we propose our RSRG-X formalism with superspins for models with interactions beyond the nearest neighbors, including the dipolar $XX+YY$ model. We describe a many-body spin-flip picture of dynamics given by RSRG-X and the picture's resulting numerical method and predictions, which we also benchmark with ED.
In Sec.~\ref{sec:2d}, we present results for the long-range model generalized to two spatial dimensions. We summarize our results and put open questions in Sec.~\ref{sec:conclusion}.

\section{Overview}
\label{sec:overview}
Here, we collect definitions of the models and physical quantities studied in this work. We also outline their behaviors. Following this, we provide a brief description of our RSRG-X formalism for models with interactions beyond the nearest neighbors. This description is intended for readers familiar with the general framework, anticipating a complete explanation in Secs.~\ref{sec:review} and \ref{sec:LR:RSRGX}. 

\subsection{Models}\label{sec:overview:models}

We study many-body dynamics generated by Hamiltonians of the form
\eq{
    H = \sum_{(j, k)} J_{jk}\big( X_j X_k + Y_j Y_k), \label{equ:overview:model}
}
where the index $j=1,\ldots,N$ labels the sites, each hosting a spin-$\frac{1}{2}$ degrees of freedom. The operators $X_j$, $Y_j$, and $Z_j$ are Pauli matrices on site $j$. Our focus is on systems in one spatial dimension with positional randomness and dipolar interactions: the positions $x_j$ of sites are random on the interval $[0,L)$ up to a hard-core constraint $|x_j-x_k|>\delta=1/40$, and we set $J_{jk}=|x_j - x_k|^{-3}$. The thermodynamic limit reads $L = N \rightarrow \infty$, fixing the typical energy scale $(L/N)^{-3} = 1$. This \textit{long-range model} is motivated by experiments on resonantly interacting nitrogen vacancy centers~\cite{Zu21} (which in reality also feature $ZZ$ anisotropies) and Rydberg atom arrays~\cite{Browaeys_review}. In practice, our calculations use periodic boundary conditions (PBC), with the interaction $J_{jk}$ given by the larger of $\left|x_k-x_j\right|^{-3}$ and $|L-\left|x_j-x_k\right||^{-3}$. We will also investigate the model generalized to two spatial dimensions, with spins randomly positioned on a $L \times L$ torus with PBC --- we defer a detailed definition to Sec.~\ref{sec:2d}.

To guide our investigation of the dipolar chain, we first study its \textit{nearest-neighbor} counterparts, i.e., models with $J_{jk}=0$ for $|j-k|\ge 2$. Here, we label the sites so that $0\leq x_1<x_2<\cdots<x_{N}<L$, the indices being defined modulo $N$ in light of PBC. Such models admit a simplified RSRG-X formalism leading to analytically controlled analyses, a simplification also expected from the free-fermion nature. There are two types of randomness that we consider, (i) $J_{j,j+1}$ drawn from the ``fixed point'' distribution, whose functional form is invariant under the analytic RSRG-X [see Equ.~\eqref{equ:RG_fpdistribution} and Sec.~\ref{sec:NN:analytics}]; and (ii) $J_{j,j+1}$ as in the dipolar model. Additionally, we study the \textit{next-nearest-neighbor model} where interactions $J_{jk}$ with $|j-k|\ge 3$ are truncated from the dipolar chain.

Finally, we mention two important global symmetries of all the systems we consider: a $\textrm{U}(1)$ symmetry associated with conservation of total spin, and a $\Z_2$ symmetry associated with invariance of $H$ under a global spin flip. The generators of the $\textrm{U}(1)$ and $\Z_2$ symmetries are, respectively,
\eq{
    M = \frac{1}{2}\sum_j Z_j, \;\;\; P=\prod_{j} X_j. \label{equ:overview:symmetries}
}
Note that the commutators $\left[M, H\right] = \left[P, H\right] = 0$. As we will see, the presence of both two symmetries tremendously reduces the possible forms of interactions appearing the effective Hamiltonians along RSRG-X, enabling a formalism that is self-contained (Sec.~\ref{sec:LR:RSRGX}).

\subsection{Autocorrelation Functions}\label{sec:overview:autocorrelations}

The central quantity that we are interested in is the infinite-temperature spin survival probability. This quantity is defined as
\eq{
    S_p(t) = \frac{1}{2} + \frac{1}{2}C(t)  \label{equ:Spt_Czz},
}
where $C(t)$ is the infinite-temperature autocorrelation function of the local spin,
\eq{
    C(t) = 2^{-N} \text{Tr}\left[Z_j(t)Z_j(0)\right],\label{equ:Czz_def}
}
where on the left the site dependence is implicit. Here, $Z_j(t)=e^{iHt}Z_j e^{-iHt}$ is a Heisenberg operator, and $Z_j$ acts on the probed site $j$. It will often be convenient to work in the frequency domain
\eq{
    S_p(t) = \int_{-\infty}^{+\infty} d\omega\, \mathcal{S}_p(\omega) e^{-i\omega t}, \label{equ:fourier}
}
where we have defined $\mathcal{S}_p(\omega)$ by Fourier transform. For convenience, we refer to both $S_p(t)$ and its Fourier transform $\mathcal{S}_p(\omega)$ as the infinite-temperature \textit{spin survival probability}; the notation clarifies which domain we concern.  In practice, to wash out fluctuations, we generally restrict ourselves to the disorder-averaged functions $\overline{S_p}(t)$, $\overline{\mathcal{S}_p}(\omega)$. 

For the nearest-neighbor chains, we show that the standard RSRG-X formalism, which we review in Sec.~\ref{sec:review}, leads to a late-time decay of the spin survival probability of the form
\begin{align}
    \overline{S_p}(t) \simeq B + \frac{A}{\log^2(t/t_0)},\label{equ:asymptote}
\end{align}
in the thermodynamic limit and at $t \gg 1$. As we show in Appendix~\ref{appendix:asymptote}, this decay corresponds to
\begin{align}
    \overline{\mathcal{S}_p}(\omega) \simeq B \delta(\omega) + \frac{A}{|\omega| \log^3|\omega_0/\omega|}\label{equ:overview:Sp_freq},
\end{align}
at $|\omega| \ll 1$. The behavior in Eqs.~(\ref{equ:asymptote}, \ref{equ:overview:Sp_freq}) has been previously identified at low temperatures \cite{MDH01,IJR00}. We refer to the nonuniversal constants $A$ and $B$ as the \textit{amplitude} and \textit{baseline} of the decay, respectively. 

Throughout this work, we present the low-frequency $\overline{\mathcal{S}_p}(\omega)$ results on the log-log scales of $\omega \overline{\mathcal{S}_p} (\omega)$ versus $\omega$. On such scales, the $\log|\omega|$ corrections~\eqref{equ:overview:Sp_freq}
%and equivalently the late-time $\log t$ decays
would appear as a curved line, whereas the more common power-law decays of the form $\overline{S_p}(t) \sim t^{-c}$ with $c>0$, converting to $\omega \overline{\mathcal{S}_p} (\omega) \sim \omega^{c}$, would correspond to straight lines. The scales we use therefore serve as a sensitive probe of the possibly slower-than-power-law dynamics.

We calculate $\overline{\mathcal{S}_p}(\omega)$ using RSRG-X for nearest-neighbor chains in Secs.~\ref{sec:NN:analytics},~\ref{sec:NN:numerics}, and for next-nearest-neighbor and long-range chains in Sec.~\ref{sec:LR:RSRGX}. In all cases, we observe low-frequency behaviors that are consistent with the one in Equ.~\eqref{equ:overview:Sp_freq}. Our RSRG-X formalism is not expected to capture infinite-time behaviors, and hence we should not expect an accurate estimate for the baseline $B$. However, we find remarkable quantitative agreement between the RSRG-X and ED results at late times up to a baseline shift, or equivalently in the regime of low but nonzero frequencies. This agreement holds even when there are significant finite-size effects in both results. In Sec.~\ref{sec:2d}, we extend the analysis to two-dimensional long-range models:  we find such agreement only for rapidly decaying interactions, and our results at small system sizes suggest subdiffusive spin relaxation.

\subsection{Renormalization Group Formalism}\label{sec:overviewRSRG}

Here, we provide a brief description of our RSRG-X formalism for those familiar with the general technique. Given a Hamiltonian, RSRG-X involves repeatedly identifying a strong local term (e.g., the strongest field or interaction) and defining a new effective Hamiltonian describing dynamics within each eigenspace of this term. In practice, the effective Hamiltonian can be calculated by treating all the terms overlapping with the strong term as perturbations. 

Nearest-neighbor $XX+YY$ models in one spatial dimension are simple because the effective Hamiltonians generated along RSRG-X remain nearest-neighbor. This property does not apply once interactions beyond the nearest neighbors are present. Nevertheless, we are able to describe the effective Hamiltonian compactly in terms of \textit{superspins}, two-level systems whose two basis states correspond to flipped configurations of physical spins (anti)aligned along the $Z$ direction. A superspin-$m$ is a two-level system whose states differ in their total $M$ magnetization (\ref{equ:overview:symmetries}) by $2m$, so a microscopic spin-$\frac{1}{2}$ is a trivial example of a superspin-$\frac{1}{2}$. One example of a nontrivial superspin-$\frac{1}{2}$ is a two-level system whose two basis states are $\ket{\uparrow\uparrow\downarrow}$ and $\ket{\downarrow\downarrow\uparrow}$.

Our RSRG-X formalism generates effective Hamiltonians describing interactions between superspins, and we design our formalism so that these interactions obey the $\textrm{U}(1)$ and $\Z_2$ symmetries (\ref{equ:overview:symmetries}) inherited along RSRG-X. In general, these interactions can be quite complex, involving large numbers of superspins. This kind of effective Hamiltonian nevertheless represents an important simplification in that it only involves two-level systems, i.e., superspins, so it can be represented in terms of standard Pauli operators associated with these systems.

The rule that we use to generate the effective Hamiltonians is as follows. At each RSRG-X step, we (a) find the strongest \textit{one- or two-}superspin coupling, (b) define subspaces of the overall Hilbert space corresponding to the eigenspaces of this coupling, and (c) generate effective Hamiltonians for each of these subspaces. Importantly, this rule is not the same as discarding couplings involving more than two superspins, as the RSRG-X will often convert, e.g., three- and four-superspin couplings into one- and two-superspin couplings, which can then play an important role in the subsequent steps. 

For the nearest-neighbor $XX+YY$ chains, these rules reduce to the standard RSRG-X formalism applicable in that setting~\cite{HM14}. This reduction arises from the special property that, although superspin-$1$'s are generated, in nearest-neighbor chains these have no dynamics: the effective Hamiltonians, which map to free-fermion models, only contain interactions involving the original spin-$\frac{1}{2}$'s. When $XX+YY$ interactions beyond the nearest neighbors are introduced, superspins formed from multiple spin-$\frac{1}{2}$'s acquire nontrivial dynamics. The fact that our formalism preserves the interacting superspin structure is a consequence of the $\mathrm{U}(1)$ and $\Z_2$ symmetries~\eqref{equ:overview:symmetries}.

\section{The RSRG-X Formalism: A Review}
\label{sec:review}
We begin by reviewing the general RSRG-X approach and its application established in Ref.~\cite{HM14} to nearest-neighbor $XX+YY$ chains. We will follow this approach to calculate dynamical quantities in the nearest-neighbor models of our concern in Sec.~\ref{sec:NN}, and in Sec.~\ref{sec:LR} we will extend it to models with interactions beyond the nearest neighbors (\ref{equ:overview:model}). In Sec.~\ref{sec:review:generalities}, we set up important general notations and terminologies of RSRG-X, which we adopt in the rest of this work. In Sec.~\ref{sec:review:NN}, we review the application to nearest-neighbor chains \cite{HM14}.

\subsection{The General Procedure}
\label{sec:review:generalities}
The general RSRG approach for systems with random interactions proceeds by iteratively updating an effective Hamiltonian $\Heff$ to involve fewer degrees of freedom; initially, $\Heff$ is set as the Hamiltonian of the entire system. Every RG step starts by identifying a strong (usually, the strongest) interaction $H_0$ in $\Heff$ and is based on a premise that $H_0$ is much stronger than all other interactions $H_1$ that spatially overlap with $H_0$. Typically, this \textit{RG premise} of strong randomness is satisfied to better extent as the RSRG proceeds. Formally, one divides $\Heff$ into
\eq{
    \Heff = H_0 + H_1 + H_\perp, \label{equ:H_separation}
}
where $H_\perp$ collects interactions spatially away from $H_0$; it follows that $[H_0, H_\perp] = 0$. The RG premise then reads $\left\Vert H_0\right\Vert \gg \left\Vert H_1\right\Vert$. Here and throughout this paper, $\left\Vert \cdots \right\Vert$ refers to the operator (infinity) norm. Under this premise, one can treat $H_1$ perturbatively over the strong interaction $H_0$. The interactions $H_\perp$, on the other hand, are unaffected by this treatment. 

It is convenient to introduce the RSRG-X formalism as a generalization of the early RSRG approach for approximating ground states \cite{DM80, Fisher92, Fisher94}. In the ground-state RSRG, at every RG step one identifies $H_0$, $H_1$, and $H_\perp$ as above (\ref{equ:H_separation}). One approximates, based on the above RG premise, the ground state of $H_\mathrm{eff}$ to be within the ground-state subspace $V$ of $H_0$. One then derives a new effective Hamiltonian $H_\mathrm{eff}^{(V)}$ by constraining the other interactions $H_1$ and $H_\perp$ to $V$, by perturbation theory. The next RG step then proceeds with the updated effective Hamiltonian, $H_\mathrm{eff}^{(V)}$. The iterative RG steps terminate when $V$ is a one-dimensional subspace, at which point the approximate ground state is fully determined.

In addition to approximating the ground states, the RSRG-X formalism is capable of yielding approximate excited eigenstates, which we call \textit{RSRG-X states}. At every RG step in RSRG-X, one makes a branching choice by picking one of the eigenspaces $V$ of the strong interaction $H_0$. One then derives an $\Heff^{(V)}$ by constraining $H_1$ and $H_\perp$ to the subspace $V$. The RG steps terminate when $V$ is spanned by only one state, namely an RSRG-X state. Note that RSRG-X reduces to ground-state RSRG if one branches into the ground-state subspace at every RG step.

For concreteness, we outline the degenerate perturbation theory used to update $\Heff$ at every RG step~\cite{VA13, Pek14, VA14, HM14, VPP15, VFPP16, Monthus18, Protopopov20, Mohdeb20, MVK22, Mohdeb23}; we will follow this approach throughout this work. With $H_0$, $H_1$, $H_\perp$ specified (\ref{equ:H_separation}) together with the branching choice $V$, one can find $\Heff^{(V)}$ by~\cite{Sachdev11}
\al{
    \left\langle i\left|\Heff^{(V)}\right| j \right\rangle &=  E_i \delta_{i j}+\left\langle i\left|\left(H_1+H_\perp\right)\right| j\right\rangle\nonumber \\
    +\frac{1}{2}\sum_{\xi} &\left\langle i \left|H_1\right| \xi \right\rangle\left\langle \xi \left|H_1\right| j \right\rangle \left(\frac{1}{E_i-E_\xi}+\frac{1}{E_j-E_\xi}\right) \label{equ:degen_perturb}.
}
Here $i, j$ index a basis of $V$, while $\xi$ are eigenstates of $H_0$ that are not in $V$. For all of these indices, $E_a = \bra{a} H_0 \ket{a}$.

In summary, for every sequence of branching choices, the RG steps lead to an approximate eigenstate, which we refer to as an RSRG-X state. One may iterate over the branching choices to obtain enough RSRG-X states to describe the dynamics of interest.

\subsection{Application: Nearest-Neighbor $XX+YY$ Chains}
\label{sec:review:NN}
Under the general framework we have set up in Sec.~\ref{sec:review:generalities}, we now review the application established in Ref.~\cite{HM14} of RSRG-X to the nearest-neighbor models

\eq{
    H = \sum_{j}J_j \left(X_j X_{j+1} + Y_j Y_{j+1}\right), \label{equ:H_NN}
}
under PBC, with generic random interactions $J_j$. This model~\eqref{equ:H_NN} is non-interacting and maps to a free fermion model. For example, in sectors where $\frac{1}{2}\sum_{j}\left(1-Z_j\right)$ is odd, per Jordan-Wigner transformation,
\eq{
    H = \sum_j J_j \left(c_j^\dagger c_{j+1} + h.c. \right). \label{equ:H_NN_fermion}
}
In Sec.~\ref{sec:review:RG_steps}, we detail the RG steps and demonstrate that the non-interacting nature of the model is, as expected, preserved along RSRG-X. We then describe the RSRG-X states in Sec.~\ref{sec:review:RSRGX_states} and their distributions in Sec.~\ref{sec:review:analytics}, paving the way for our computation of dynamical quantities in Sec.~\ref{sec:NN}.

\subsubsection{The RG Steps}
\label{sec:review:RG_steps}
Following the general procedure described in Sec.~\ref{sec:review:generalities}, the first RG step begins with $\Heff$ equal to the original system Hamiltonian~\eqref{equ:H_NN}. We identify the interaction corresponding to the largest $|J_j|$ as $H_0$. Without loss of generality, denote $H_0 = \Omega(X_1 X_2 + Y_1 Y_2)$, where $|\Omega| = \max_i\left\{|J_j|\right\}$. We label the site left (right) of the sites 1, 2 by subscript $L$ ($R$). As in Equ.~\eqref{equ:H_separation}, $\Heff$ divides into the strong interaction $H_0$, its neighboring interactions $H_1 = J_L\left(X_L X_1 + Y_L Y_1\right) + J_R \left(X_2 X_R + Y_2 Y_R\right)$, and other interactions denoted by $H_\perp$, which commute with $H_0$. Under this setup, the RG premise (Sec.~\ref{sec:review:generalities}) reads $|\Omega| \gg |J_L|, |J_R|$, which will be shown to be self-consistent at later RG steps (Sec.~\ref{sec:review:analytics}).

To update $\Heff$ following the perturbation theory~\eqref{equ:degen_perturb}, consider the spectrum of $H_0$, which alone has two non-degenerate eigenstates, $\mzero{1}{2}$ of energies $\pm 2\Omega$, and two degenerate ones, $\upaligned{1}{2}, \downaligned{1}{2}$ of energy $0$. Alternatively in terms of the whole Hilbert space, $H_0$ features three eigenspaces at corresponding energies $E$ gapped by the order of $|\Omega|$, namely
\begin{alignat}{3}
    &V_+ &&=  \mathcal{H}_{\conj{12}} \otimes \left\{\triplet{1}{2}\right\}, \;\;\; & E &= 2\Omega \nonumber \\
    &V_0 &&=  \mathcal{H}_{\conj{12}} \otimes \mathrm{span}\left\{\upaligned{1}{2}, \downaligned{1}{2}\right\}, \;\;\;  & E &= 0 \nonumber \\
    &V_- &&= \mathcal{H}_{\conj{12}} \otimes \left\{\singlet{1}{2}\right\}, \;\;\; & E &= -2\Omega,  \label{equ:review:RG_subspace}
\end{alignat}
where $\mathcal{H}_{\conj{12}}$ denotes the Hilbert space on sites other than $1$, $2$. The updated $\Heff$ can be derived~\eqref{equ:degen_perturb} once one specifies a branching choice (Sec.~\ref{sec:review:generalities}). The results are
\al{
    \Heff^{(V_+)} &= 2\Omega + J_{LR}\left(\XX{L}{R}\right) + H_\perp \nonumber \\
    \Heff^{(V_0)} &= -J_{LR}\left(\XX{L}{R}\right) + H_\perp  \nonumber \\
    \Heff^{(V_-)} &= -2\Omega + J_{LR}\left(\XX{L}{R}\right) + H_\perp. \label{equ:review:RG_step}
}
where $J_{LR} = J_L J_R / \Omega$. Note that, for all cases, the updated $\Heff$ does not hybridize $\mathcal{H}_{\conj{12}}\otimes \upaligned{1}{2}$ with $\mathcal{H}_{\conj{12}}\otimes \downaligned{1}{2}$ and still takes the form of a nearest-neighbor $XX+YY$ chain. Consequently, we can view the updated $\Heff$ as consisting of two fewer sites --- the sites $1$, $2$ are cast out, and we fix the degrees of freedom on these two sites to be one of the four eigenstates of $H_0$. [Particularly for branching into $V_0$, we choose between $\upaligned{i}{j}$, $\downaligned{i}{j}$ with definite magnetization~\eqref{equ:overview:symmetries}, although in principle a linear combination would also work as they do not hybridize~\eqref{equ:review:RG_step}.] A new interaction of strength $|J_{LR}|$ is generated between the sites $L$ and $R$, which now become nearest neighbors. One can then start another RG step with the updated $\Heff$ and repeat the procedures above.

Since $\Heff$ remains in the same form, the above description of the RG step is \textit{self-contained} in the sense that all the branching choices and consequences have been specified. This allows one to proceed iteratively along the RG steps until there is no site left in $\Heff$, at which point one obtains an RSRG-X state, which we will detail in Sec.~\ref{sec:review:RSRGX_states}.

That the form of $\Heff$ is unchanging along the RG steps naturally follows from the non-interacting nature of the nearest-neighbor model~\eqref{equ:H_NN}. Any interaction hybridizing the two subspaces $\mathcal{H}_{\conj{12}} \otimes \upaligned{1}{2}$, $\mathcal{H}_{\conj{12}} \otimes \downaligned{1}{2}$ while preserving the total number of particles would be interacting in the corresponding fermion language. Concretely, we can condense the update rules (\ref{equ:review:RG_step}) by noting that $H_0 = 0, \pm 2\Omega$ and $Z_1Z_2 = -1, 1$ are fixed in the $V_0, V_\pm$ subspaces (\ref{equ:review:RG_subspace}), respectively, so the updated effective Hamiltonian reads
\eq{
    \Heff^* = H_0 - J_{LR}\left(X_L Z_1Z_2 X_R + Y_L Z_1Z_2 Y_R\right) + H_{\perp}. \label{equ:NN_RG_SB}
}
In the fermion language \footnote{One can also formulate the RSRG-X and derive the same update rule by directly working with the single particle
Hamiltonian in the fermion language.},
\eq{
    \Heff^* = 2\Omega\left(c_1^\dagger c_2 + h.c.\right) - 2 J_{LR}\left(c_L^\dagger c_R + h.c.\right) + H_\perp, \label{equ:NN_RG_fermion}
}
so it is transparent that the updated $\Heff$ stays non-interacting throughout the RG steps.

\subsubsection{The RSRG-X States}
\label{sec:review:RSRGX_states}
The RG steps terminate when there is no site left in $\Heff$. At this stage, an RSRG-X state has been determined: each sequence of branching choices leads to a distinct state. Particularly, every RG step (Sec.~\ref{sec:review:RG_steps}) operates with the strong interaction,
\eq{
    H_0 = \Omega_\alpha\left(X_iX_j+Y_iY_j\right), \label{equ:NN_H0}
}
supported on some pair of sites $\alpha = (i, j)$ and fixes the degrees of freedom on the pair to be one of the four local eigenstates of $H_0$. For every sequence of branching choices, the corresponding RSRG-X state then reads
\begin{widetext}
    \eq{
        \ket{\psi} = \left(\bigotimes_{(i, j)\in D_-} \singlet{i}{j}\right)\otimes \left(\bigotimes_{(i, j)\in D_+} \triplet{i}{j}\right)\otimes \left(\bigotimes_{(i, j)\in D_\uparrow} \upaligned{i}{j} \right)\otimes \left( \bigotimes_{(i, j)\in D_\downarrow} \downaligned{i}{j}\right), \label{equ:RSRGX_state_NN}
    }
\end{widetext}
in terms of $D_\zeta$, the pairs hosting different types (according to the branching choice) of local eigenstates, $\zeta \in \left\{+, -, \uparrow, \downarrow\right\}$. These pairs are disjoint, and the disjoint union $D = \bigsqcup_{\zeta} D_\zeta$ covers the entire original chain (\ref{equ:H_NN}). For future reference, we call the first two tensor product components (\ref{equ:RSRGX_state_NN}) the \textit{resonant parts} of the RSRG-X state, while we call the last two the \textit{frozen part}.

We mention a special property of the structure of the RSRG-X states~\eqref{equ:RSRGX_state_NN} for the nearest-neighbor models~\eqref{equ:H_NN}, namely that the pairing $D$ of sites is fixed given the initial sample of random interactions (\ref{equ:H_NN}) and is therefore shared by all the RSRG-X states for the sample. Indeed, the interaction strengths $|J_j|$ in an $\Heff$ determine which two sites host $H_0$ and thus pair up, as well as the strengths $|J_j|$ in the updated $\Heff$, regardless of branching choice~\eqref{equ:review:RG_step}. As we discuss below, this property will allow for analytical progress in calculations of dynamical quantities~\eqref{equ:H_NN}, as well as a straightforward physical picture (Sec.~\ref{sec:NN:analytics}). 

\subsubsection{An Analytical Description}
\label{sec:review:analytics}
Citing previously established results \cite{Fisher94, RM04, HM14}, we describe how the distribution $P(|J|)$ of the nearest-neighbor interaction strengths $|J_j|$ in $\Heff$ flows along RSRG-X. The results will be useful in our computation of the ensemble-averaged spin relaxation dynamics in Sec.~\ref{sec:NN:analytics}.

The $P(|J|)$ results are developed with the thermodynamic limit in mind: the RG steps are coarse-grained by defining at every step an \textit{RG time} $\Gamma = \log\left(\left|\Omega_0/\Omega_\alpha\right|\right)$. Here, $\Omega_\alpha$ is the strength of the strongest interaction $H_0$ at a step~\eqref{equ:NN_H0}, and $\Omega_0$ is the energy scale of the original chain~\eqref{equ:H_NN}. For simplicity, we drop the $\alpha$ labeling in $\Omega_\alpha$. Note that at every RG step, as one decimates the strongest interaction, one effectively moves forward to smaller $\Omega$, i.e., larger $\Gamma$. We then parametrize the flow of the distribution $P(|J|)$ along RSRG-X by the RG time $\Gamma$, denoting the distribution by $P_\Gamma(|J|)$. 

The flow of $P_\Gamma(|J|)$ at late RG steps has been first derived in the ground-state RSRG for the antiferromagnetic chain [i.e., Equ.~\eqref{equ:H_NN} with $J_j > 0$]~\cite{Fisher94}: at every RG step, branching into the ground-state subspace $V_-$ always introduces a weaker antiferromagnetic interaction $J_{LR} = J_L J_R / \Omega$ between the neighboring sites (\ref{equ:review:RG_step}). It is convenient to introduce the logarithmic interaction strengths $\beta_j=\log(|\Omega/J_j|)$ so that the update rule reads $\beta_{LR} = \beta_{L} + \beta_{R}$ and that at every step the strongest interaction $\Omega$ corresponds to $\beta=0$. We can then write down an integro-differential equation describing the flow of $P_\Gamma(\beta)$,
\al{
    \pD{P_\Gamma(\beta)}{\Gamma} &= \pD{P_\Gamma(\beta)}{\beta} + P_\Gamma(0) \nonumber \\ 
    & \times\int_{\beta_{1, 2} > 0} d\beta_1 d\beta_2\, P_\Gamma(\beta_1)P_\Gamma(\beta_2) \delta(\beta_1 + \beta_2 - \beta). \label{equ:integro_differential}
}
Here on the right-hand side, the first term results from a global shift in $\beta$ due to the decimation of strongest interaction at $\beta=0$, and the second term accounts for the formation of weaker interactions $\beta_{LR}$. Equ.~\eqref{equ:integro_differential} has been shown to feature a fixed point
\eq{
    P^c_\Gamma(\beta) = \frac{1}{\Gamma} e^{-\beta/\Gamma} \label{equ:RG_fpdistribution}
}
that attracts, at late RG times, generic initial randomness distributions \cite{Fisher94}. This result immediately extends to the RSRG-X of the general nearest-neighbor chains we consider~\eqref{equ:H_NN}: the new interaction generated at every RG step has strength $|J_{LR}| = |J_L J_R / \Omega|$ regardless of branching choice~\eqref{equ:review:RG_step} and signs of couplings, so the same fixed point~\eqref{equ:RG_fpdistribution} applies~\cite{HM14}.

With these analytical results, it is straightforward to justify the RG premise of strong randomness at late RG steps (Sec.~\ref{sec:review:generalities}). Indeed for generic initial randomness, there is a certain RG time where $P_\Gamma(\beta)$ is close to $P^c_\Gamma(\beta)$, the fixed-point distribution~\eqref{equ:RG_fpdistribution}; from this step on as the RSRG-X proceeds to larger $\Gamma$, the distribution $P_\Gamma(\beta)\approx P^c_\Gamma(\beta)$ becomes broader, i.e., the effective Hamiltonian features stronger randomness. In this sense, the fixed point~\eqref{equ:RG_fpdistribution} is dubbed an \textit{infinite-randomness fixed point}, and the perturbative treatment in the RSRG-X formalism is \textit{asymptotically exact} at small energy scales.

\section{The Nearest-Neighbor Chain}
\label{sec:NN}

The RSRG-X formalism yields a whole set of approximate eigenstates, from which we can estimate many aspects of the spin dynamics. In this section, we investigate the late-time spin relaxation dynamics at infinite temperature in the nearest-neighbor models~\eqref{equ:H_NN} with RSRG-X. In Sec.~\ref{sec:NN:analytics}, we exploit the structure of RSRG-X states (Sec.~\ref{sec:review:RSRGX_states}) to depict a simple two-body spin-flip picture of the spin dynamics. This picture, together with the analytical results at the infinite-randomness fixed point~\eqref{equ:RG_fpdistribution}, makes it straightforward to derive in the thermodynamic limit the low-frequency (equivalently, late-time) behavior~\eqref{equ:asymptote} of $\mathcal{S}_p(\omega)$, the infinite-temperature spin survival probability~\eqref{equ:Spt_Czz}. In Sec.~\ref{sec:NN:numerics}, we turn to numerically simulating the RSRG-X steps and analyze the resulting disorder-averaged $\overline{\mathcal{S}_p}(\omega)$. We verify the analytical derivations and, strikingly, observe agreement with ED even at system sizes featuring strong finite-size effects.

Previously, a ground-state RSRG formalism has been used to calculate low-temperature dynamical autocorrelation functions~\cite{MDH01}. There, one finds a late-time decay $\sim 1/|\log(t)|^2$ slower than any power law, and this analytical prediction has been verified with ED~\cite{IJR00}. As we will show below, this behavior is also expected for the nearest-neighbor model at infinite temperature.

\subsection{Analytical Predictions}
\label{sec:NN:analytics}

\begin{figure*}
\includegraphics[width=0.95\textwidth]{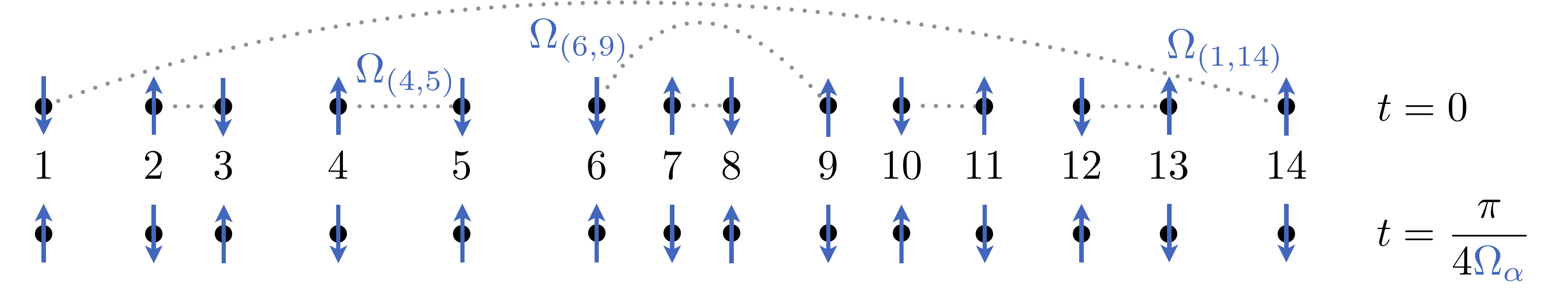}
\caption{The two-body spin-flip picture (Sec.~\ref{sec:NN:analytics}) of the dynamics in a nearest-neighbor chain~\eqref{equ:H_NN} with a single sample of random interactions. The dashed lines show pairing of sites $\alpha = (i, j)$ that has been formed along the RSRG-X. The dynamics are then viewed as decomposed into resonant two-body spin flips within the pairs, each at frequency $4\Omega_\alpha$.} \label{fig:twobody_picture}
\end{figure*}

Here, we illustrate how the RSRG-X formalism provides a two-body spin-flip picture as well as quantitative predictions of the spin dynamics by considering $\overline{\mathcal{S}_p}(\omega)$. This function in the frequency domain~\eqref{equ:fourier} can be decomposed in terms of the full spectrum $\left(\ket{n}, E_n\right)$,
\eq{
    \mathcal{S}_p(\omega) =  \frac{1}{2}\delta(\omega)+\frac{1}{2 D_\mathcal{H}}\sum_{m, n} \left|\braket{m|Z|n}\right|^2 \delta(E_n - E_m - \omega),  \label{equ:Spt_spectral}
}
where $D_\mathcal{H} = 2^N$ is the full Hilbert space dimension. From now on, we approximate the spectrum by the RSRG-X states~\eqref{equ:RSRGX_state_NN}.

It is helpful to comment on the physical picture these RSRG-X states~\eqref{equ:RSRGX_state_NN} offer us before delving into detailed calculations. We have mentioned in Sec.~\ref{sec:review:RSRGX_states} that, for each single sample of initial randomness, all the RSRG-X states share a common spatial structure characterized by the pairing $D$ of sites formed along the RG steps. This enables us to view the spin dynamics as disentangled among the pairs in $D$. To elaborate, consider a pair $\alpha = (i, j) \in D$ that has hosted the strong interaction $H_0$ of energy $\Omega_\alpha$ at some RG step (\ref{equ:NN_H0}). An initial state of the form $\upaligned{i}{j} \otimes \ket{\phi}$ or $\downaligned{i}{j} \otimes \ket{\phi}$, where $\ket{\phi}\in \mathcal{H}_{\conj{ij}}$ encodes some RSRG-X state component elsewhere, does not evolve with time. On the other hand, an initial state of the form $\ket{\uparrow_i\downarrow_j}\otimes \ket{\phi}$ or $\ket{\downarrow_i\uparrow_j}\otimes \ket{\phi}$ features oscillation between the two states at frequency $4\Omega_\alpha$, the energy splitting between the $V_+$ and $V_-$ subspaces~\eqref{equ:review:RG_subspace} of $H_0$. Schematically, we observe here a two-body spin-flip picture where the spin quantum number oscillates within the pair. We illustrate the picture in Fig.~\ref{fig:twobody_picture}.

Mathematically, this picture is corroborated by the fact that, given an RSRG-X state $\ket{n}$ appearing in the spectral sum (\ref{equ:Spt_spectral}), among all the RSRG-X states~\eqref{equ:RSRGX_state_NN} there is only one $\ket{m}$ such that $\braket{m|Z|n}\neq 0$. In fact, this RSRG-X state is $\ket{m}=Z\ket{n}$. If $Z$ resides on the frozen part~\eqref{equ:RSRGX_state_NN} of $\ket{n}$, then $\ket{n} = \ket{m}$, and we have only a zero-frequency contribution. In the case where $Z$ resides on a resonant part, suppose that $\alpha=(i, j)$ is the pair where $Z$ acts (with e.g. $Z$ acting on $i$). Then, the branching sequences leading to $\ket{m}=Z\ket{n}$ and $\ket{n}$ only differ at the RG step where $i$ and $j$ are paired up. Considering this step, we find $|E_m - E_n| = 4|\Omega_\alpha|$, which is the energy splitting between the two eigenspaces $V_{\pm}$ (\ref{equ:review:RG_subspace}). The sum (\ref{equ:Spt_spectral}) over the RSRG-X states then becomes
\eq{
    \mathcal{S}_p(\omega) = \frac{3}{4}\delta(\omega) + \frac{1}{8}\left[\delta\left(\omega - 4\Omega_\alpha\right) + \delta\left( \omega + 4\Omega_\alpha\right)\right]. \label{equ:S_p_freq_D}
}
One may get the same result, more straightforwardly, by using the above-mentioned two-body spin-flip picture and considering only the dynamics within the pair $\alpha$.

The next piece of information we need to evaluate the ensemble averaged $\overline{\mathcal{S}_p}(\omega)$ is the distribution of the energies $\Omega_\alpha$. This distribution at low energies can be inferred from the infinite-randomness fixed point~\eqref{equ:RG_fpdistribution}, as follows. Suppose that, at some RG time $\Gamma = \Gamma_c$, the remaining system has a distribution of interaction strengths $P_{\Gamma_c}$ that is ``close to'' a fixed-point distribution $P_{\Gamma_c}^c$~\eqref{equ:RG_fpdistribution}. Let $\Omega_c = \max\{\left|J_j\right|\}$ be the energy scale, and $\zeta N$ be the system size, at this RG time $\Gamma_c$. Note $0<\zeta\leq1$. From then on, the system remains close to the fixed point as $\Gamma$ increases (from $\Gamma_c$ to infinity) along the RSRG-X. To track the rate at which the pairs $\alpha \in D$ are formed, let $N(\Gamma)$ be the number of sites left in $\Heff$ for $\Gamma > \Gamma_c$, with $N(\Gamma_c) = \zeta N$. The sites are decimated at rates
\eq{
    \frac{dN(\Gamma)}{d\Gamma} = -2N(\Gamma) P_\Gamma^c(\beta = 0) = -\frac{2N(\Gamma)}{\Gamma}.
}
The factor of 2 accounts for the fact that RG steps involve decimation of pairs of sites, while $N(\Gamma)P^c_{\Gamma}(\beta=0) d\Gamma$ is the number of such pairs, whose interactions have been at the upper energy cutoff corresponding to $\beta=0$. Solving this differential equation, we find $N(\Gamma) = \zeta N \Gamma_c^2/\Gamma^2$.

The distribution of pairing energies $\Omega_\alpha$ follows from this result. During the RG time interval $[\Gamma, \Gamma + d\Gamma]$, the number of pairs formed is $(1/2) |dN(\Gamma) / d\Gamma| = \zeta N \Gamma_c^2 / \Gamma^3$, and they all have energy $\Omega = \Omega_c e^{-(\Gamma - \Gamma_c)}$ or equivalently $l = \Gamma$ in terms of a logarithmic energy $l$ which we define to be $l(\Omega) = \Gamma_c + \log\left(\Omega_c / \Omega\right)$. Dividing by the total number of pairs $N/2$, we find the desired energy distribution
\eq{
    g(l) = \frac{2\zeta \Gamma_c^2}{ l^3},
}
for $l \ge \Gamma_c$. The distribution $g(l)$ at $l < \Gamma_c$, i.e., energies $\Omega > \Omega_c$, is determined by the earlier RG steps and is thus non-universal; it is nevertheless normalized as $\int d l g(l)=1$, with $\int_{\Gamma_c}^{\infty} d l g(l) = \zeta$. In the following, we will only be interested in the universal low-energy (large $l$) part of the distribution $g(l)$.

Now we are ready to average over the ensemble of randomness using $\overline{\mathcal{S}_p}(\omega) = \int dl\, g(l) \mathcal{S}_p(\omega)$, with $\mathcal{S}_p(\omega)$ given by Equ.~\eqref{equ:S_p_freq_D} and the interaction energy $\Omega_{\alpha}=\Omega_c e^{-(l-\Gamma_c)}$. Evaluating this integral, we obtain the RSRG-X prediction in the frequency domain~\eqref{equ:overview:Sp_freq} for $|\omega| < 4\Omega_c$,
\eq{
    \overline{\mathcal{S}_p}(\omega) = B \delta(\omega) + \frac{A}{\left|\omega\right|\left(\log\left|\omega_0 / \omega\right|\right)^3}, \label{equ:analytics:Sp_freq}
}
where
\eq{
    A = \frac{\zeta}{4} \Gamma_c^2, \;\;\; B = \frac{3}{4}, \;\;\; \omega_0 = 4 \Omega_c e^{\Gamma_c}. \label{equ:NN_A_t0}
}
We show in Appendix~\ref{appendix:asymptote} that this form of low-frequency spectral function (\ref{equ:analytics:Sp_freq}) transforms in the time domain to the extremely slow decay
\eq{
    \conj{S_p}(t)= B + \frac{A}{\left|\log(t/t_0)\right|^2} \label{equ:analytics:Spt},
}
up to subleading corrections, such that $A$, $B$ are precisely the decay amplitude and baseline~\eqref{equ:asymptote}, respectively. Here, $t_0 = \pi / (2\omega_0)$.

In summary, for the nearest-neighbor $XX+YY$ models with generic randomness~\eqref{equ:H_NN} the RSRG-X formalism analytically predicts a universal form~\eqref{equ:analytics:Spt} for the late-time spin relaxation at infinite temperature. We have derived this form quantitatively in terms of model-dependent parameters in both the low-frequency~\eqref{equ:analytics:Sp_freq} and late-time~\eqref{equ:analytics:Spt} domains. The formalism has also allowed us to arrive at a simple two-body spin-flip picture of the spin dynamics. A generalization of this picture will appear in the RSRG-X formalism we propose for the generic beyond-nearest-neighbor models~\eqref{equ:overview:model} in Sec.~\ref{sec:LR:RG:picture}.

\subsection{Numerical Implementation}
\label{sec:NN:numerics}
We numerically implement the RSRG-X formalism to test the validity of its predictions above (Sec.~\ref{sec:NN:analytics}). Two particular kinds of distributions of randomness for the initial couplings (\ref{equ:H_NN}) are considered. One distribution draws directly from the infinite-randomness fixed point~\eqref{equ:RG_fpdistribution}, allowing us to confirm the analytical calculations (Sec.~\ref{sec:NN:analytics}). The other distribution (Sec.~\ref{sec:NN_lr}) steps closer to the situation where the randomness in the interaction strengths results from positional randomness and the dipolar interactions. For both cases, using the numerical RSRG-X we compute the disorder-averaged infinite-temperature spin survival probability $\overline{\mathcal{S}_p}(\omega)$ in the frequency domain and compare with ED results --- the methods for both approaches are detailed in Appendix~\ref{appendix:numerics}. The ED and numerical RSRG-X show good agreement in the low-frequency (i.e., late-time) regime and, notably, even at small system sizes with strong finite-size effects. At large system sizes, numerical methods verify the analytical prediction in Equ.~\eqref{equ:analytics:Sp_freq}.

\subsubsection{The Fixed-Point Distribution}
\label{sec:NN_fp}
As a first step, we draw initial random interactions~\eqref{equ:H_NN} as positive numbers $J_j > 0$ from the fixed-point distribution~\eqref{equ:RG_fpdistribution} at some $\Gamma = \Gamma_0$ and fixed maximal interaction strength $\Omega_0 = 1$. This is equivalent to setting $J_j = (1 - h_j)^{\Gamma_0}$ for some random numbers $h_j$ sampled uniformly from $[0, 1)$. We expect systems sampled at this distribution to exhibit the behavior predicted in Sec.~\ref{sec:NN:analytics} with the parameters $\zeta=1$ and $\Gamma_c = \Gamma_0$.

\begin{figure}
    \includegraphics[width = 0.48\textwidth]{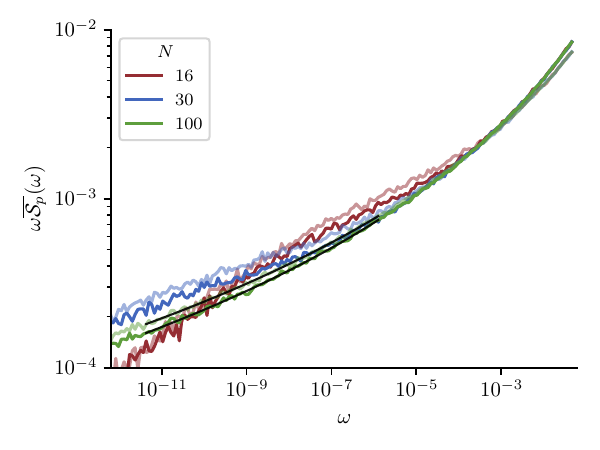}
    \caption{ED (dark) and numerical RSRG-X (light) results for $\overline{\mathcal{S}_p}(\omega)$, the infinite-temperature spin survival probability~\eqref{equ:Spt_Czz}, in the nearest-neighbor chain~\eqref{equ:H_NN} at various system sizes $N$, with initial random interactions drawn from the fixed-point distribution~\eqref{equ:RG_fpdistribution} at fixed $\Gamma = 5$. Solid black lines show fits to the expected universal form~\eqref{equ:overview:Sp_freq} with fitting parameters $A$, $\omega_0$. Each dataset reflects average over more than $5\times 10^4$ random samples. Note that, in contrast, diffusion would result in $\omega \overline{\mathcal{S}_p}(\omega) \sim \omega^{c}$ with $c=\frac{1}{2}$, corresponding to a straight line with a large slope on the plotted scales.}
    \label{fig:NN_g5_SDscale_freq_ED_RG}
\end{figure}

In Fig.~\ref{fig:NN_g5_SDscale_freq_ED_RG}, we present $\overline{\mathcal{S}_p}(\omega)$
results at $\Gamma_0 = 5$ by both ED and numerical RSRG-X for various system sizes. We observe strong finite-size effects at system sizes $N \lesssim 30$ for the range of $\omega$ investigated. Notably, the numerical RSRG-X quantitatively captures these effects. Moreover, for $N\gtrsim 30$ we find good agreement with the low-frequency scaling form~\eqref{equ:analytics:Sp_freq} predicted by the analytical RSRG-X calculation (Sec.~\ref{sec:NN:analytics}). 

\begin{figure}
    \includegraphics[width = 0.48\textwidth]{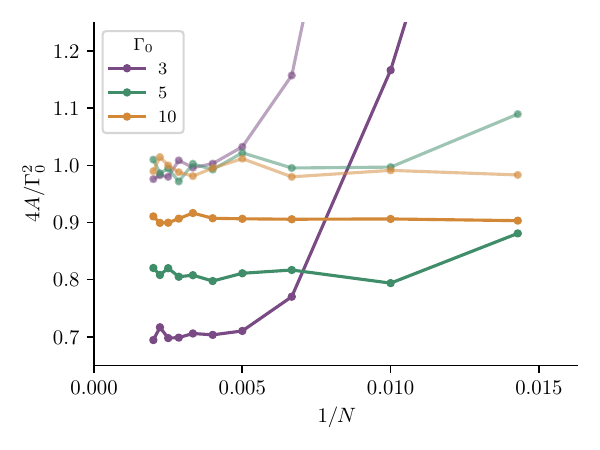}
    \caption{Finite-size behavior of the decay amplitude $A$ extracted from ED (dark) and numerical RSRG-X (light) results for $\overline{\mathcal{S}_p}(\omega)$ by fits as exemplified in Fig.~\ref{fig:NN_g5_SDscale_freq_ED_RG}. Initial random nearest-neighbor interactions are drawn from fixed-point distributions (\ref{equ:RG_fpdistribution}) at various $\Gamma = \Gamma_0$.}
    \label{NN_A_ED_A_RG_varg_varN}
\end{figure}

From fits to our numerical results illustrated in Fig.~\ref{fig:NN_g5_SDscale_freq_ED_RG}, we extract the decay amplitudes $A$. These amplitudes, which in general can differ between RSRG-X and ED, are shown for various $\Gamma_0$ as a function of $N$ in Fig.~\ref{NN_A_ED_A_RG_varg_varN}. For all data, we see clear convergence of $A$ with increasing $N$: the RSRG-X estimates for $A$ approach the analytical prediction~\eqref{equ:NN_A_t0} of $A = \Gamma_0^2/4$, while the ED results converge to values that are moderately away from the analytical predictions. Indeed, RSRG-X is only expected to be quantitatively accurate in the limit of large $\Gamma_0$, i.e., of infinite randomness. We check whether this is the origin of the discrepancy by varying $\Gamma_0$: the larger is $\Gamma_0$, the closer the ED results are to the RSRG-X prediction $A = \Gamma_0^2/4$ (Fig.~\ref{NN_A_ED_A_RG_varg_varN}).

\subsubsection{Truncated Dipolar Interactions}
\label{sec:NN_lr}

\begin{figure}
    \includegraphics[width = 0.48\textwidth]{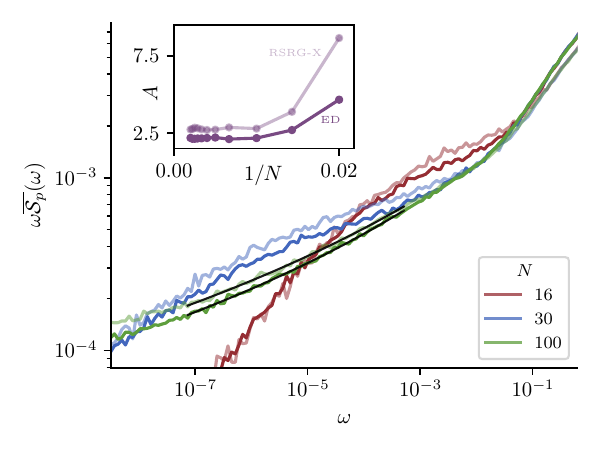}
    \caption{ED (dark) and numerical RSRG-X (light) results for $\overline{\mathcal{S}_p}(\omega)$, the spin survival probability~\eqref{equ:Spt_Czz}, in the nearest-neighbor chain with truncated dipolar interactions. Solid black lines show fits to the universal scaling form~\eqref{equ:overview:Sp_freq}. Each dataset represents average over more than $5\times 10^4$ random samples. (Inset) Decay amplitudes $A$ extracted from the fits.}
    \label{fig:NN_dpl_SDscale_ED_RG}
\end{figure}

Having verified RSRG-X predictions of $\overline{\mathcal{S}_p}(\omega)$ at the strong-randomness fixed point (Sec.~\ref{sec:NN_fp}), we now consider the nearest-neighbor model with truncated dipolar interactions~\eqref{equ:overview:model}. In Fig.~\ref{fig:NN_dpl_SDscale_ED_RG}, we calculate $\overline{\mathcal{S}_p}(\omega)$ and find that, at large $N$, both ED and RSRG-X results agree with the scaling form~\eqref{equ:analytics:Sp_freq} predicted by analytical RSRG-X arguments. As is for the model with interactions drawn from the fixed-point distribution (Sec.~\ref{sec:NN_fp}), there is good agreement between ED and numerical RSRG-X at low frequencies at all system sizes studied, even when finite-size effects are strong (Fig.~\ref{fig:NN_dpl_SDscale_ED_RG}).

Whenever the numerical RSRG-X and ED results manifest the expected scaling form~\eqref{equ:analytics:Sp_freq}, we extract the decay amplitude $A$ by fitting. Although the decay amplitudes $A$ extracted from the two results get closer at larger system sizes, we still find moderate departure in the thermodynamic limit (Fig.~\ref{fig:NN_dpl_SDscale_ED_RG}, inset). Since such departure is seen even in the fixed-point model especially at smaller $\Gamma_0$ (Sec.~\ref{sec:NN_fp}), we attribute it to the fact that the original systems are not exactly featuring infinite randomness. We stress that, however, our results indicate that the dynamics of the truncated dipolar model still appear to be controlled by the infinite-randomness fixed point, as evidenced by the fitted scaling in Fig.~\ref{fig:NN_dpl_SDscale_ED_RG}.

To highlight, for the nearest-neighbor $XX+YY$ chain~\eqref{equ:H_NN} with the two types of randomness realizations (Secs.~\ref{sec:NN_fp},~\ref{sec:NN_lr}), we observe quantitative agreement (Figs.~\ref{fig:NN_g5_SDscale_freq_ED_RG}, \ref{fig:NN_dpl_SDscale_ED_RG}) between numerical RSRG-X and ED results for the low-frequency spin relaxation dynamics at all system sizes. The agreement holds even at small $N$, indicating that RSRG-X has the potential as an efficient numerical approach for predicting late-time dynamics. We will see such power of RSRG-X again in the interacting models (Sec.~\ref{sec:LR}), where ED calculations are limited to very small systems.

\section{Chains with Interactions Beyond the Nearest Neighbors}
\label{sec:LR}
In this section, we move on to consider the experimentally motivated chain with dipolar $XX+YY$ interactions between every pair of randomly positioned sites~\eqref{equ:overview:model}. Because there are interactions beyond the nearest neighbors, the model cannot be transformed into a system of free fermions, and the simple RSRG-X formalism used for the nearest-neighbor model (Sec.~\ref{sec:review:NN}) no longer applies.  Here, we develop a new RSRG-X formalism for more general random spin systems with $\text{U}(1)$ and $\mathbb{Z}_2$ symmetries. The formalism provides a transparent physical picture of the late-time dynamics. This picture then leads to a numerical approach that, as we will verify, successfully simulates the infinite-temperature spin relaxation in the long-range and next-nearest-neighbor models~\eqref{equ:overview:model} we consider. For these chains, the numerical approach has the potential to scale up to systems sizes inaccessible via ED. We describe our RSRG-X formalism in Sec.~\ref{sec:LR:RSRGX} and present numerical results in Sec.~\ref{sec:LR:numerics}.

\begin{figure}
    \includegraphics[width = 0.48\textwidth]{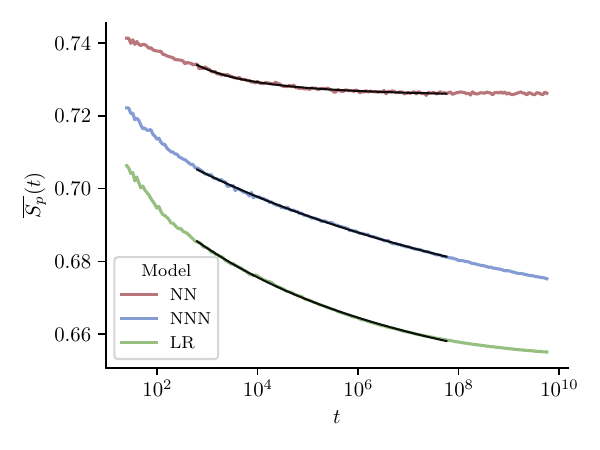}
    \caption{Disorder-averaged infinite-temperature spin survival probability $\overline{S_p}(t)$ at late times, calculated by ED for the nearest-neighbor (NN), next-nearest-neighbor (NNN), and long-range (LR) chains of $N=16$ randomly positioned sites with dipolar interactions~\eqref{equ:overview:model}. The results represent averages over $1\times 10^4$, $1\times 10^3$, and $1\times 10^3$ samples, respectively. Solid black lines show fits of the form $\overline{S_p}(t) = B + A|\log(t/t_0)|^{-2}$.}
    \label{fig:LR_ED}
\end{figure}

Before detailing our RSRG-X formalism, we present ED results at accessible system sizes in Fig.~\ref{fig:LR_ED} to qualitatively illustrate the effect of beyond-nearest-neighbor interactions on the spin survival probability $\overline{S_p}(t)$. At late times, $\overline{S_p}(t)$ is significantly reduced from $3/4$, the asymptotic value predicted by the RSRG-X formalism of the nearest-neighbor model, indicating that the spin spreads beyond the resonant pairs that would be constructed in that case. The beyond-nearest-neighbor interactions result in a breakdown of the two-body spin-flip picture (Sec.~\ref{sec:NN:analytics}) of the non-interacting model. 

Strikingly, we still observe persistent decay of $\conj{S_p}(t)$ to its infinite-time value at surprisingly large time scales, e.g. $t\gtrsim 10^7$, in the beyond-nearest-neighbor models (Fig.~\ref{fig:LR_ED}). The decay fits to the form $\sim B + A/|\log(t/t_0)|^2$ that has been found for the nearest-neighbor models (Sec.~\ref{sec:NN:analytics}), suggesting that the two models might contain low-energy resonances of similar statistics. That such statistics in  the nearest-neighbor model are described by an RSRG-X formalism (Sec.~\ref{sec:review:analytics}) motivates applying RSRG-X to the beyond-nearest-neighbor models~\eqref{equ:overview:model}. Nevertheless, the breakdown of the two-body spin-flip picture held for the nearest-neighbor model demands fundamental modification to the prior RSRG-X formalism.

\subsection{The RSRG-X Formalism}
\label{sec:LR:RSRGX}

We now propose an RSRG-X formalism, under the general framework we have set up in Sec.~\ref{sec:review:generalities}, for models with random interactions beyond the nearest neighbors~\eqref{equ:overview:model}. A major difficulty arises from the complexity of the effective Hamiltonian $\Heff$ updated along the RG steps: not only are there a larger number of interactions to begin with, but new forms of interactions will also appear. For instance, as we will see already in the first RG step, the updated $\Heff$ may contain interactions of $ZZ$ form or even those coupling more than two sites. To give a self-contained RSRG-X formalism, one in principle has to describe the RG steps where the strongest interaction in $\Heff$ takes any of the new forms.

We will show that a self-contained RSRG-X formalism arises if we choose at every RG step the strong interaction $H_0$ to be the strongest one- or two-site interaction in $\Heff$, where each ``site'' now hosts two-level degrees of freedom dubbed superspin, to be defined below. Notably, $\Heff$ bears a structure as a sum of interactions involving these superspins, and our choice of $H_0$ ensures that this structure is preserved. Within our formalism, as we will see, $H_0$ is restricted to only three possibilities; this is a consequence of the $\textrm{U}(1)$ and $\Z_2$ symmetries~\eqref{equ:overview:symmetries}, which we demand be inherited by $\Heff$ along the RG steps. 

In Sec.~\ref{sec:lr:RG:first_step}, we detail the first RG step to showcase the complexity of the updated $\Heff$ and the introduction of superspins. In Sec.~\ref{sec:LR:RG:superspin}, we provide the general superspin definition and notations we adopt in this work. In Sec.~\ref{sec:lr:RG:subsequent_steps}, we move on to list all the possible subsequent RG steps, thus describing our self-contained RSRG-X formalism. We show the RSRG-X states in Sec.~\ref{sec:LR:RG:RSRGX_states} and the ensuing many-body spin-flip picture of the late-time dynamics in Sec.~\ref{sec:LR:RG:picture}.

\subsubsection{The First RG Step: Motivating Superspin}
\label{sec:lr:RG:first_step}
As usual (Sec.~\ref{sec:review:generalities}), the RSRG-X formalism starts by setting $\Heff$ as the original Hamiltonian~\eqref{equ:overview:model} and identifying a strong interaction $H_0$. Without loss of generality, let $H_0 = \Omega\left(\XX{1}{2}\right)$, taking $|\Omega| = \max{|J_{jk}|}$ to be the strongest coupling. To showcase the effect of beyond-nearest-neighbor interactions~\eqref{equ:overview:model}, we consider a four-site system with two other sites indexed by L, R each coupled to both sites $1$, $2$. Separating $\Heff = H_0 + H_1 + H_\perp$ as in Equ.~\eqref{equ:H_separation}, where $H_1$ denotes the interactions overlapping with $H_0$, we have for the four-site system
\al{
    H_1 &= J_L\left(\XX{L}{1}\right) + J_R\left(\XX{R}{2}\right) \nonumber \\
    &+ \epsilon_1\left(\XX{1}{R}\right) + \epsilon_2\left(\XX{2}{L}\right), \label{equ:LR:4spin}
}
and $H_\perp = \epsilon_3 \left(\XX{L}{R}\right)$ collecting the remaining, non-overlapping terms. The RG premise (Sec.~\ref{sec:review:generalities}) allowing for a perturbative treatment of $H_1$ over $H_0$ reads, in this context, $|\Omega| \gg |J_L|, |J_R|, |\epsilon_1|, |\epsilon_2|$.

The interaction $H_0$ in isolation has three eigenstate subspaces, $V_+$, $V_0$, and $V_-$, as defined previously in~\eqref{equ:review:RG_subspace}, gapped by the order of $|\Omega|$. We then apply the degenerate perturbation theory (\ref{equ:degen_perturb}) to update $\Heff$ depending on a choice to branch into one of the three subspaces. Branching into $V_+$ or $V_-$ is straightforward; the subspace has the degrees of freedom fixed on sites $1$, $2$, so the updated $\Heff$ involves a chain of two fewer sites,
\al{
    H_\mathrm{eff}^{(V_+)} &= H_\perp + \frac{(J_L+\epsilon_2)(J_R+\epsilon_1)}{\Omega}\left(\XX{L}{R}\right) \\
    H_\mathrm{eff}^{(V_-)} &= H_\perp + \frac{(J_L-\epsilon_2)(J_R-\epsilon_1)}{\Omega}\left(\XX{L}{R}\right), \nonumber
}
up to some constant-energy terms that may be easily tracked but are irrelevant to this work. Note that setting $\epsilon_1 = \epsilon_2 = 0$ restores to the branching results of the nearest-neighbor model (Sec.~\ref{sec:review}). These expressions agree with the results of an earlier simplified RSRG-X where only $V_\pm$ are branched into~\cite{Mohdeb20, MVK22}.

Branching into $V_0$ is much more complicated. Unlike the case for the nearest-neighbor model (Sec.~\ref{sec:review:NN}), now the updated Hamiltonian $\Heff^{(V_0)}$ may contain interactions that non-trivially couple the two orthogonal subspaces $\mathcal{H}_{\conj{12}}\otimes\left\{\upaligned{1}{2}\right\}$ and $\mathcal{H}_{\conj{12}}\otimes\left\{\downaligned{1}{2}\right\}$ of $V_0$. Physically, such dynamics drive the aligned spins from the strongly interacting pair $(1, 2)$ to the other sites hosting the Hilbert space $\mathcal{H}_{\conj{12}}$. To express $\Heff^{(V_0)}$ compactly, we view $V_0 = \mathcal{H}_{\conj{12}} \otimes \mathcal{H}_\mathrm{new}$ as consisting of a new two-level Hilbert space $\mathcal{H}_\mathrm{new}$ spanned by the basis $\left\{\upaligned{1}{2}, \downaligned{1}{2}\right\}$, under which we introduce the Pauli matrices $
X^{(1)}, Y^{(1)}, Z^{(1)}$, e.g., $Y^{(1)} = -i\upaligned{1}{2}\bra{\downarrow_1\downarrow_2} + h.c.$ Conveniently, we can view $\mathcal{H}_\mathrm{new}$ as the Hilbert space of a new site hosting two always aligned spin-$\frac{1}{2}$'s and in this way call the new site to be superspin-1. Note that the new superspin-1 encodes two-level degrees of freedom and is not to be confused with what is conventionally called a spin-$1$, which is three-level degrees of freedom. Branching into $V_0$ then yields, again up to some constant term,
\begin{widetext}
\al{
    \Heff^{(V_0)} = & H_\perp - \frac{J_L J_R + \epsilon_1 \epsilon_2}{\Omega}\left(\XX{L}{R}\right) + \frac{2\epsilon_2 J_L}{\Omega} Z^{(1)} Z_L + \frac{2\epsilon_1 J_R}{\Omega}Z^{(1)} Z_R \nonumber \\
    & - \frac{\epsilon_1J_L + \epsilon_2 J_R}{\Omega}\left[X^{(1)} \left(X_LX_R - Y_L Y_R\right)+ Y^{(1)} \left(X_LY_R + Y_L X_R\right)\right]. \label{equ:LR:RG_Heff}
}
\end{widetext}
As a check, if we set $\epsilon_1 = \epsilon_2 = 0$, then there is no dynamics involving the two levels of the newly introduced superspin-1, as in the nearest-neighbor model (Sec.~\ref{sec:review}). 

The terms other than the first two in Equ.~\eqref{equ:LR:RG_Heff} are of new forms. In fact, these forms are the only ones allowed by the global $\textrm{U}(1)$ and $\Z_2$ symmetries inherited by the updated $\Heff$. Particularly, in Equ.~\eqref{equ:LR:RG_Heff} the last term is allowed as it keeps track of the collective spin-flip between $\ket{\uparrow_1 \uparrow_2 \downarrow_L \downarrow_R}$ and $\ket{\downarrow_1 \downarrow_2 \uparrow_L \uparrow_R}$, a spin-conserving process. Also, the $ZZ$ interaction between the L (R) site and the new superspin-1 site is consistent with both symmetries. In contrast, spin-flip interactions between these two sites are not allowed because one spin-$\frac{1}{2}$ cannot hop to or from a superspin-$1$ site hosting two such spins, otherwise violating total spin conservation. Furthermore, single-site $Z$ fields are not allowed as they conflict with the $\Z_2$ global spin-flip symmetry.

In summary, we have showcased the first RG step for the interacting $XX+YY$ model~\eqref{equ:overview:model} by examining all the possible branching choices $V_+$, $V_0$, and $V_-$ in a four-site system with beyond-nearest-neighbor interactions. Due to these interactions, the updated $\Heff$ by branching into $V_0$ needs to be described in terms of new two-level degrees of freedom called superspin. Besides, in the updated $\Heff$ all the interactions allowed by the inherited $\textrm{U}(1)$ and $\Z_2$ symmetries appear. Some of these interactions are of forms other than $XX+YY$ and require other treatments should they dominate in the subsequent RG steps.

\subsubsection{Superspin Definition and Notations}
\label{sec:LR:RG:superspin}
We gather some important superspin definition and notations to be adopted throughout the rest of this work. While we have motivated the introduction of superspin in Sec.~\ref{sec:lr:RG:first_step}, we generalize the notation to cover also the more complicated superspins that will appear in the subsequent RG steps (Sec.~\ref{sec:lr:RG:subsequent_steps}).

Generally, a superspin-$m$ encodes two-level degrees of freedom consisting of spin states that are oppositely polarized and have respective eigenvalues $\pm m$ under $\frac{1}{2}\sum_j Z_j$. Note that $m$ takes non-negative integer or half-integer values, as the original systems in our work consist of spin-$\frac{1}{2}$'s. A superspin-$\frac{1}{2}$, for example, may trivially represent a
% \textcolor{red}{[YZ: For easy reference, we could add a qualifier ``primitive'' or ``trivial''?]}
spin-$\frac{1}{2}$ at site $i$ in the original chain, in this case encoding the two levels $\left\{\ket{\uparrow_i}, \ket{\downarrow_i}\right\}$. It may also represent a group of spins totaling $\pm \frac{1}{2}$, encoding such as $\left\{\ket{\uparrow_{i_1} \uparrow_{i_2} \downarrow_{j_1}}, \ket{\downarrow_{i_1} \downarrow_{i_2} \uparrow_{j_1}}\right\}$, $\left\{\ket{\uparrow_{i_1} \uparrow_{i_2} \uparrow_{i_3} \downarrow_{j_1}\downarrow_{j_2}}, \ket{\downarrow_{i_1} \downarrow_{i_2}\downarrow_{i_3} \uparrow_{j_1}\uparrow_{j_2}}\right\}$, etc. For all cases, we denote the two encoded levels by $\left\{\ket{\Uparrow^{m}_{\mathbf{I}, \mathbf{J}}}, \ket{\Downarrow^{m}_{\mathbf{I}, \mathbf{J}}}\right\}$, where $\mathbf{I}$, $\mathbf{J}$ are disjoint sets of indices labeling the sites in the original chain, such that
\eq{
    \ket{\spinUp{m}{I}{J}} = \ket{\spinups{I}\spindowns{J}}, \;\;\; 
    \ket{\spinDown{m}{I}{J}} = \ket{\spindowns{I}\spinups{J}}. \label{equ:superspin_notation}
}
Here, $\spinups{I}$ ($\spindowns{I}$) denotes $\uparrow_{i_1}\cdots \uparrow_{i_{|\mathbf{I}|}}$ ($\downarrow_{i_1}\cdots \downarrow_{i_{|\mathbf{I}|}}$), where $|\mathbf{I}|$ is the number of elements for an index set $\mathbf{I}$. By definition, we require $m = \left(|\mathbf{I}|-|\mathbf{J}|\right)/2$. Under this notation, the superspin-$1$ site introduced in the first RG step (Sec.~\ref{sec:lr:RG:first_step}) corresponds to $m=1$, $\mathbf{I} = \left\{1, 2\right\}$, and $\mathbf{J} = \emptyset$ the empty set. A spin-$\frac{1}{2}$ at the $i$-th site in the original system, as another example, has $m=\frac{1}{2}$, $\mathbf{I}=\{i\}$, and $\mathbf{J}=\emptyset$. For consistency, the Pauli matrices acting on the two-level degrees of freedom (\ref{equ:superspin_notation}) share the same labels as the encoding superspin and are hence denoted by $X^{(m)}_{\mathbf{I}, \mathbf{J}}$, $Y^{(m)}_{\mathbf{I}, \mathbf{J}}$, $Z^{(m)}_{\mathbf{I}, \mathbf{J}}$, e.g., $Y^{(m)}_{\mathbf{I}, \mathbf{J}} = -i \ket{\spinUp{m}{I}{J}}\bra{\spinDown{m}{I}{J}} + h.c$. Sometimes, the context makes clear the index sets $\mathbf{I}$, $\mathbf{J}$ so we omit them for simplicity, as we have done in Sec.~\ref{sec:lr:RG:first_step}.

More conveniently, we can also view every superspin degrees of freedom $\left\{\ket{\spinUp{m}{I}{J}}, \ket{\spinDown{m}{I}{J}}\right\}$ as the two possible states of a \textit{site} indexed by $a$ in the system described by $\Heff$ at a given RG step. We can then write these states as $\left\{\ket{\Uparrow^{m}_a}, \ket{\Downarrow^{m}_a}\right\}$. We also define Pauli matrices $X^{(m)}_a$, $Y^{(m)}_a$, $Z^{(m)}_a$ for these two-level degrees of freedom, where e.g. $Y^{(m)}_a = -i\ket{\Uparrow^{m}_a}\bra{\Downarrow^{m}_a}+h.c$.

Following above notations, we write the global $\textrm{U}(1)$ and $\Z_2$ symmetries~\eqref{equ:overview:symmetries} that $\Heff$ inherits along RSRG-X in the following compact forms. We claim that, at every RG step, $\textrm{U}(1)$ and $Z_2$ symmetries are generated by, respectively,
\eq{
    M = \sum_a m_a Z^{(m_a)}_a, \;\;\; P = \prod_a X^{(m_a)}_a. \label{equ:symmetries_superspin}
}
Here, $a$ ranges over all the sites supporting $\Heff$ --- they may host superspins. One may check that $M$, $P$ in the form above~\eqref{equ:symmetries_superspin} are indeed the symmetry generators in our RG steps, to be detailed in Sec.~\ref{sec:lr:RG:subsequent_steps}.

\subsubsection{Subsequent RG Steps: A Self-Contained Description}
\label{sec:lr:RG:subsequent_steps}

Unlike those for the nearest-neighbor chains (Sec.~\ref{sec:review}), the subsequent RG steps for the interacting model~\eqref{equ:overview:model} may differ from the first. This happens when the strongest one- or two-site interaction takes a newly generated form, the $ZZ$ form for example, which may already appear in the updated $\Heff$ during the first step~\eqref{equ:LR:RG_Heff}. For an RSRG-X formalism aiming at approximating all the eigenstates to be self-contained, it is necessary to exhaust all the possible forms of the strong interaction $H_0$ and describe all the branching choices for each form.

\begin{figure}
\includegraphics[width=0.47\textwidth]{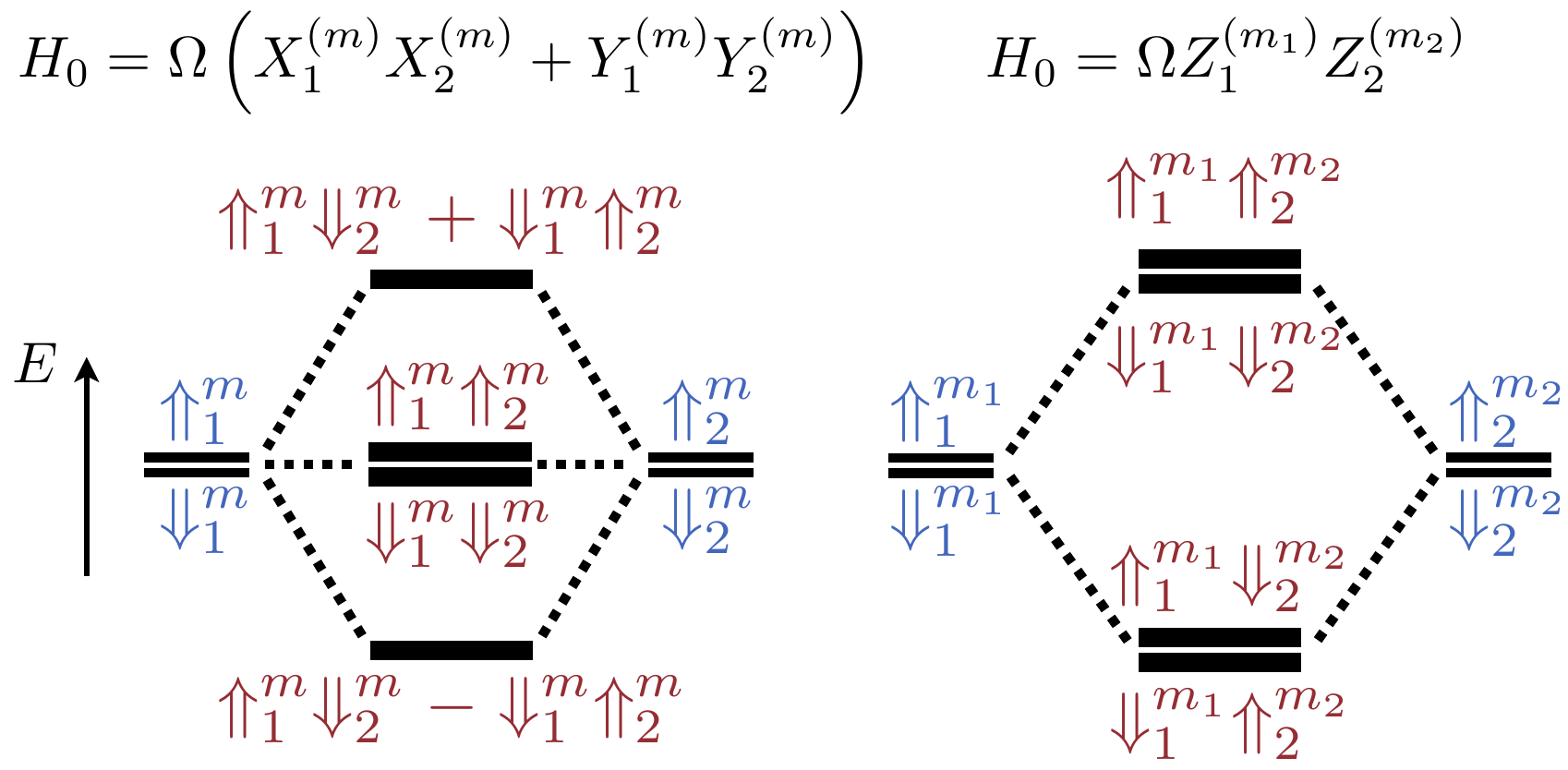}
\caption{RG steps for two-site $H_0$. The initial superspin degrees of freedom acted on by $H_{\text{eff}}$ are shown in blue, and the eigenstates of $H_0$ are shown in red. Left: $H_0 =\Omega(X_1^{(m)}X_2^{(m)}+Y_1^{(m)}Y_2^{(m)})$. In the $V_0$ subspace~\eqref{equ:LR:XY_subspaces}, with energy $E=0$ with respect to $H_0$, the two superspin-$m$'s are merged into a superspin-$2m$ with states $\ket{\Uparrow^{2m}}=\ket{\Uparrow^m_1\Uparrow^m_2}$ and $\ket{\Downarrow^{2m}}=\ket{\Downarrow^m_1\Downarrow^m_2}$. Right: $H_0 = \Omega Z^{(m_1)}_1Z^{(m_2)}_2$. In the $V'_{\pm}$ subspaces~\eqref{equ:LR:ZZ_subspaces} the superspin-$m_1$ and superspin-$m_2$ are merged into superspin-$|m_1 \pm m_2|$ degrees of freedom, respectively.}
\label{fig:hybridize}
\end{figure}

We list the following cases of $H_0$ at an RG step and outline for each case the important consequences of every branching choice, detailing precise perturbative treatments in Appendix~\ref{appendix:pseudo_codes}. Without loss of generality, we label the sites in $\Heff$ such that $H_0$ always acts within the sites $1$ and $2$, respectively hosting superspin-$m_{1, 2}$. In terms of the sites of the original system, let $1$, $2$ each refer to the index sets $(\mathbf{I}_{1, 2},\mathbf{J}_{1, 2})$. 

(i) $H_0 = \Omega\left(X_1^{(m_1)}X_2^{(m_2)}+Y_1^{(m_1)}Y_2^{(m_2)}\right)$, where the $\textrm{U}(1)$ symmetry (\ref{equ:symmetries_superspin}) mandates $m_1=m_2$, or equivalently $|\mathbf{I}_1|-|\mathbf{J}_1|=|\mathbf{I}_2|-|\mathbf{J}_2|$. The treatment of this case is similar to the first RG step (Sec.~\ref{sec:lr:RG:first_step}).  Particularly, we utilize the large gaps of order $|\Omega|$ separating the three subspaces at corresponding energies $E$,
\begin{alignat}{3}
    &V_+ &&=  \mathcal{H}_{\conj{12}} \otimes \left\{\frac{1}{\sqrt{2}} \left(\ket{\Uparrow^{m_1}_1 \Downarrow^{m_2}_2} + \ket{\Downarrow^{m_1}_1 \Uparrow^{m_2}_2} \right)\right\}, &\;\; E&= 2\Omega  \nonumber \\
    &V_0 &&=  \mathcal{H}_{\conj{12}} \otimes \mathrm{span}\left\{\ket{\Uparrow^{m_1}_1 \Uparrow^{m_2}_2}, \ket{\Downarrow^{m_1}_1 \Downarrow^{m_2}_2}\right\}, &\;\; E&= 0 \nonumber \\
    &V_- &&= \mathcal{H}_{\conj{12}} \otimes \left\{\frac{1}{\sqrt{2}} \left(\ket{\Uparrow^{m_1}_1 \Downarrow^{m_2}_2} - \ket{\Downarrow^{m_1}_1 \Uparrow^{m_2}_2} \right)\right\}, &\;\; E&= -2\Omega, \label{equ:LR:XY_subspaces}
\end{alignat}
now given in the superspin notation. As usual, $\mathcal{H}_{\conj{12}}$ denotes the Hilbert space of $\Heff$ with sites $1$, $2$ excluded. We branch into one of the subspaces (\ref{equ:LR:XY_subspaces}) of our choice to update $\Heff$. If we branch into $V_+$ or $V_-$, the sites 1 and 2 have their degrees of freedom fixed to be $\left(\ket{\uparrow^{\otimes\left(\mathbf{I}_1\sqcup \mathbf{J}_2\right)}\downarrow^{\otimes\left(\mathbf{I}_2\sqcup \mathbf{J}_1\right)}} \pm \ket{\downarrow^{\otimes\left(\mathbf{I}_1\sqcup \mathbf{J}_2\right)}\uparrow^{\otimes\left(\mathbf{I}_2\sqcup \mathbf{J}_1\right)}}\right)/\sqrt{2}$ for the two subspaces respectively and are thus cast out of the updated $\Heff$; if we branch into $V_0$, the updated $\Heff$ involves a new superspin-$(m_1+m_2)$ site in replacement of the sites $1$, $2$. Specifically, the new superspin encodes $\left\{\ket{\spinUp{n}{K}{L}}, \ket{\spinDown{n}{K}{L}}\right\}$ where $n = m_1 + m_2$, and $\mathbf{K} = \mathbf{I}_1\sqcup\mathbf{I}_2$, $\mathbf{L} = \mathbf{J}_1\sqcup\mathbf{J}_2$ are the disjoint unions of the old index sets. This kind of RG step is illustrated on the left in Fig.~\ref{fig:hybridize}.

(ii) $H_0 = \Omega Z^{(m_1)}_1 Z^{(m_2)}_2$. We utilize the large gap of order $|\Omega|$ and branch into either of the two subspaces 
\begin{alignat}{3}
    &V_+' &&= \mathcal{H}_{\conj{12}} \otimes \mathrm{span}\left\{\ket{\Uparrow^{m_1}_1 \Uparrow^{m_2}_2}, \ket{\Downarrow^{m_1}_1 \Downarrow^{m_2}_2}\right\},& \quad E &= \Omega \nonumber \\
    &V_-' &&= \mathcal{H}_{\conj{12}}\otimes  \mathrm{span}\,\left\{\ket{\Uparrow^{m_1}_1\Downarrow^{m_2}_2}, \ket{\Downarrow^{m_1}_1\Uparrow^{m_2}_2}\right\},&\quad E &= -\Omega. \label{equ:LR:ZZ_subspaces}
\end{alignat}
In both cases, a new superspin $\left\{\ket{\spinUp{n}{K}{L}}, \ket{\spinDown{n}{K}{L}}\right\}$ replaces the sites 1, 2. If we branch into $V_+'$, then $n = m_1+m_2$, $\mathbf{K} = \mathbf{I}_1\sqcup\mathbf{I}_2$, $\mathbf{L} = \mathbf{J}_1\sqcup\mathbf{J}_2$, as in (i). If we branch into $V_-'$, then (assuming without loss of generality $m_1\ge m_2$) $n=m_1-m_2$, $\mathbf{K} = \mathbf{I}_1\sqcup\mathbf{J}_2$, $\mathbf{L} = \mathbf{I}_2\sqcup\mathbf{J}_1$. These branching choices are illustrated on the right in Fig.~\ref{fig:hybridize}. Note that, if $m_1 = m_2$, then in $V'_-$ we may generate a one-site $X^{(0)}_{\mathbf{K}, \mathbf{L}}$ field term, which respects the global $\textrm{U}(1)$ and $\Z_2$ symmetries (\ref{equ:symmetries_superspin}). The situation where $H_0$ takes the form of $X^{(0)}$ is treated in (iii).

(iii) $H_0 = 2\Omega X_1^{(0)}$. Note $|\mathbf{I}_1| = |\mathbf{J}_1|$ by superspin construction (\ref{equ:superspin_notation}). This case is a special situation where a dangling $X^{(0)}$ term generated in (ii) dominates in $\Heff$; say the term resides on site 1. A large energy gap $4|\Omega|$ separates the two eigenstate subspaces
\eq{
    U_\pm = \mathcal{H}_{\overline{1}}\otimes \left\{\frac{1}{\sqrt{2}}\left(\ket{\Uparrow^{0}_1}\pm\ket{\Downarrow^{0}_1}\right)\right\}, \quad E = \pm 2\Omega,  \label{equ:LR:X_subspaces}
}
where $\mathcal{H}_{\overline{1}}$ denotes the Hilbert space on all the other sites in $\Heff$. We may branch into either one of $U_\pm$. This fixes the degrees of freedom on site $1$ to be $\left(\ket{\uparrow^{\otimes\mathbf{I}_1}\downarrow^{\otimes\mathbf{J}_1}} \pm \ket{\downarrow^{\otimes\mathbf{I}_1}\uparrow^{\otimes\mathbf{J}_1}}\right) / \sqrt{2}$ respectively for the two subspaces and thus casts the site out of the updated $\Heff$.

For all the branching choices we take in (i)--(iii), one may check that the global $\textrm{U}(1)$ and $\Z_2$ symmetries inherited by the updated $\Heff$ indeed remain as are in Equ.~\eqref{equ:symmetries_superspin}. Moreover, the updated $\Heff$ stays as a Hamiltonian for a system of superspins (Sec.~\ref{sec:LR:RG:superspin}) each encoding two-level degrees of freedom. Formally, $\Heff$ always acts on a Hilbert space of a structure $\bigotimes_{a}\left\{\ket{\Uparrow^{m_a}_a}, \ket{\Downarrow^{m_a}_a}\right\}$ and takes the generic form, under the superspin notations (Sec.~\ref{sec:LR:RG:superspin}),
\begin{widetext}
\eq{
    \Heff = \sum_{\substack{a, b \\ m_a = m_b \neq 0}} J^{\parallel}_{ab} \left(X^{(m_a)}_a X^{(m_b)}_b + Y^{(m_a)}_a Y^{(m_b)}_b\right)
    + \sum_{a, b} J^{\perp}_{ab} Z^{(m_a)}_a Z^{(m_b)}_b
    + \sum_{\substack{a \\ m_a = 0}} h_a X^{(0)}_a + \cdots, \label{equ:LR:Heff}
} 
\end{widetext}
where the ``$\cdots$'' collects the other terms. We always pick $H_0$ to be the strongest among the interactions explicitly listed~\eqref{equ:LR:Heff}. Doing so guarantees that the subsequent RG steps fall into one of the three cases (i)--(iii). We have therefore been able to describe a self-contained RSRG-X formalism for interacting random systems subject to the global $\textrm{U}(1)$ and $\Z_2$ symmetries~\eqref{equ:symmetries_superspin}.

The three explicitly listed interactions in Equ.~\eqref{equ:LR:Heff} almost exhaust all the one- and two-site interactions that respect the $\mathrm{U}(1)$ and $\Z_2$ symmetries~\eqref{equ:symmetries_superspin}. There are only two related exceptions: interactions $X^{(0)}X^{(0)}$, $Y^{(0)} Y^{(0)}$ between two superspin-0's. These interactions are but a small portion of all the kinds of interactions, since the superspin-0 sites themselves are less likely to appear; see Appendix~\ref{appendix:superspin0} for detailed reasoning. We further demonstrate in Appendix~\ref{appendix:superspin0} that these $X^{(0)}X^{(0)}$, $Y^{(0)} Y^{(0)}$ interactions are generally weak as long as the three-site interactions are weak. In Appendix~\ref{appendix:premises}, we verify numerically the rarity of strong $X^{(0)}X^{(0)}$, $Y^{(0)} Y^{(0)}$, and three-site interactions altogether for the specific models we consider~\eqref{equ:overview:model}.

\subsubsection{The RSRG-X States}
\label{sec:LR:RG:RSRGX_states}

The relative simplicity of the steps (i)--(iii) leads to RSRG-X states (Sec.~\ref{sec:review:RSRGX_states}) of a simple form

\begin{widetext}
\eq{
    \ket{\psi} = \left(\bigotimes_{\mathbf{I}, \mathbf{J}} \frac{1}{\sqrt{2}} \left(\ket{\spinups{I}\spindowns{J}} \pm \ket{\spindowns{I}\spinups{J}}  \right)\right) 
    \otimes
    \left(\ket{\spinups{K}} \otimes \ket{\spindowns{L}} \vphantom{\int}\right), \label{equ:RSRGX_state}
}
\end{widetext}
where $\mathbf{K}, \mathbf{L}$ and all the $\mathbf{I}, \mathbf{J}$ are index sets that disjointly cover the sites in the original system~\eqref{equ:overview:model}. The first tensor product component in Equ.~\eqref{equ:RSRGX_state} results from the degrees of freedom fixed along the RG steps (i)--(iii), as detailed in Sec.~\ref{sec:lr:RG:subsequent_steps}. The second component results from the final stage of RSRG-X: when the RG step terminates, the system may be left with a single superspin $\{\ket{\spinUp{m}{K}{L}}, \ket{\spinDown{m}{K}{L}}\}$ with $m\neq 0$, where the symmetries~\eqref{equ:symmetries_superspin} rule out any dynamics --- we hence fix the remaining degrees of freedom to be either $\ket{\spinups{K}} \otimes \ket{\spindowns{L}}$ or  $\ket{\spindowns{K}} \otimes \ket{\spinups{L}}$. In both cases, our formalism naturally generates RSRG-X state components with definite magnetization~\eqref{equ:overview:symmetries}.

Remarkably, these RSRG-X states~\eqref{equ:RSRGX_state} bear a formal resemblance to those of the nearest-neighbor chains~\eqref{equ:RSRGX_state_NN}. This results from our strategy of always picking $H_0$ among the interactions listed explicitly in Equ.~\eqref{equ:LR:Heff}. Following the terminologies introduced for the nearest-neighbor RSRG-X states~\eqref{equ:RSRGX_state_NN}, we still call the first group of tensor product components~\eqref{equ:RSRGX_state} the \textit{resonant parts}, and the second group the \textit{frozen part}. We will elaborate in Sec.~\ref{sec:LR:RG:picture} on how  the resonant parts contribute to many-body spin-flip dynamics that are, nevertheless, distinct from those in the nearest-neighbor chains.

\begin{figure}
\includegraphics[width=0.47\textwidth]{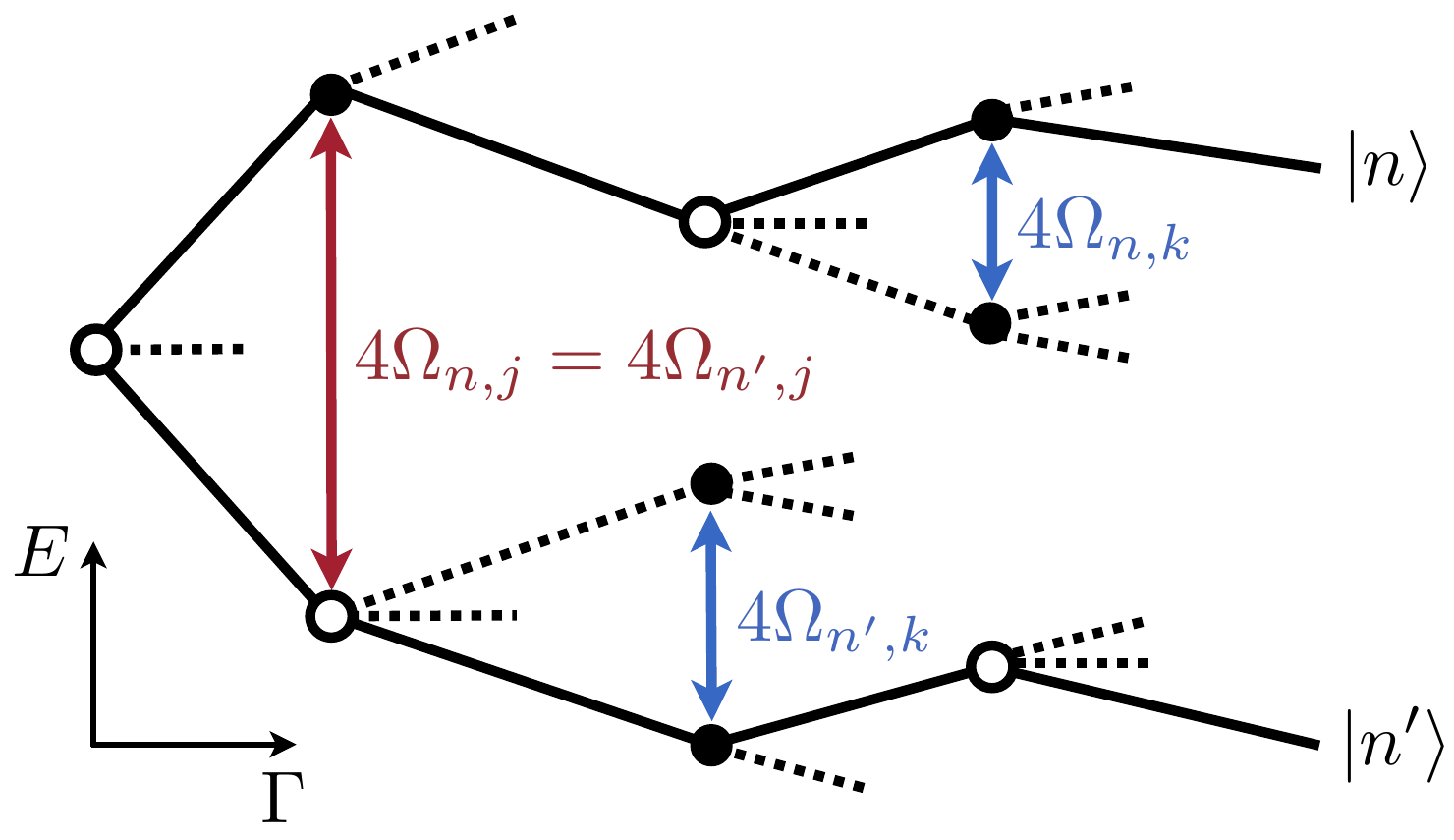}
\caption{Branching of energy levels through our RSRG-X formalism, with $E$ the energy and $\Gamma$ the RG time. The states $\ket{n}$ and $\ket{n'}$ indicated on the right are examples of RSRG-X states which we use as approximations for the actual eigenstates. Solid lines show the branching choices leading to these states. Steps where the strong interaction $H_0$ takes the form $XX+YY$ ($ZZ$) are indicated with open (solid) circles. The first branching choice, for example, involves the site $j$, where the energy scale of coherent spin-flip dynamics is $\Omega_{n,j}$. Note that the energy scales (and spatial structures) of the resonances involving another site $k$ may be different for $\ket{n}$ and for $\ket{n'}$, as illustrated.}
\label{fig:branching}
\end{figure}

Unlike the case of the nearest-neighbor chains~\eqref{equ:RSRGX_state_NN}, the grouping of sites in Equ.~\eqref{equ:RSRGX_state} is not fixed throughout all the RSRG-X states for a single realization of the randomness. Indeed, distinct branching choices (Sec.~\ref{sec:lr:RG:subsequent_steps}) may lead to, at subsequent RG steps, different types of strong interactions $H_0$ identified at different spatial locations. This consequently differentiates the superspin formation and thus the structure of the eventual RSRG-X states. In Fig.~\ref{fig:branching}, we illustrate the branching choices leading to two RSRG-X states of different structures.

\subsubsection{Dynamics: The Many-Body Spin-Flip Picture}
\label{sec:LR:RG:picture}

\begin{figure*}
\includegraphics[width=0.95\textwidth]{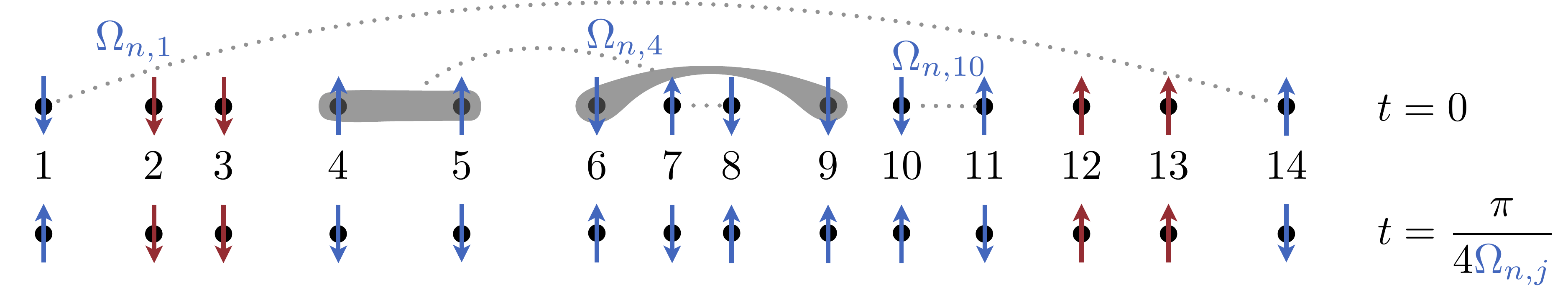}
\caption{Visualization of the contribution of an RSRG-X state $\ket{n}$, which locally approximates an actual eigenstate, to spin dynamics at sites $j$ in an interacting model~\eqref{equ:overview:model}. The red (blue) arrows represent polarizations in the frozen (resonant) part(s) of $\ket{n}$~\eqref{equ:RSRGX_state}. Each dashed line illustrates a coherent spin-flip process~\eqref{equ:Cnj_results} at frequency $4\Omega_{n, j}$ resulting from the strong interaction $H_0$ of energy $\Omega_{n, j}$ at some RG step (Sec.~\ref{sec:lr:RG:subsequent_steps}). The gray shades illustrate the formation of collectively flipping spins, due to superspins. Note that, for the illustrated configuration, $\Omega_{n, 1} = \Omega_{n, 14}$, $\Omega_{n, 4} = \Omega_{n, 5} = \Omega_{n, 6} = \Omega_{n, 9}$, $\Omega_{n, 7} = \Omega_{n, 8}$, and $\Omega_{n, 10} = \Omega_{n, 11}$.} \label{fig:manybody_picture}
\end{figure*}

The structure of RSRG-X states~\eqref{equ:RSRGX_state} for the beyond-nearest-neighbor models~\eqref{equ:overview:model} offers a valuable picture of the spin dynamics. We illustrate this picture by considering $C(t)$, the on-site infinite-temperature autocorrelation function~\eqref{equ:Czz_def} of a local spin in the original system, computed with these approximate states. The results will also eventually guide us to compute $\overline{\mathcal{S}_p}(\omega)$ numerically (Sec.~\ref{sec:LR:numerics}).

The autocorrelation function at site $j$ as predicted by RSRG-X can be viewed as an average over the RSRG-X states $\ket{n}$
\eq{
    C_j(t) = 2^{-N} \sum_n  C_{n,j}(t),   \label{equ:C_spectral_average}
}
with the contribution from every $\ket{n}$ being
\eq{
    C_{n, j}(t) = \sum_{m} \left|\braket{m|Z_j|n}\right|^2 \cos\left(\left|E_n-E_m\right|t\right), \label{equ:Cjn_sum_m}
}
where the sum is again over the RSRG-X states, and $E_{n, m}$ are the approximate energies, whose difference we will soon find. Note that our RSRG-X states~\eqref{equ:RSRGX_state} are orthonormal; indeed, we always branch into orthogonal subspaces~(\ref{equ:LR:XY_subspaces}--\ref{equ:LR:X_subspaces}) of the strong interaction $H_0$ and keep proper normalization. To compute the sum over $m$, consider the position of the probed site $Z_j$ as well as the structure~\eqref{equ:RSRGX_state} of $\ket{n}$: there are two cases, as follows. 

(a) $Z_j$ rests on the frozen part of $\ket{n}$. Then, $Z_j\ket{n} = \ket{n}$, so $\braket{m|Z_j|n} = \braket{m|n}\neq 0$ only if $m=n$, by orthogonality. Therefore, $C_{n, j}(t) = 1$.

(b) $Z_j$ rests on a resonant part of $\ket{n}$. We proceed to consider the RG step (Sec.~\ref{sec:lr:RG:subsequent_steps}) that has generated this resonance: let the energy scale $\Omega$ be $\Omega_{n, j}$ and the branching subspace be $\mathcal{V}$. Note that $\mathcal{V}$ is one between either the $V_\pm$ subspaces~\eqref{equ:LR:XY_subspaces} in case (i) or the $U_\pm$ subspaces~\eqref{equ:LR:X_subspaces} in case (iii). In either case, the energy separation between the two subspaces is $4|\Omega_{j, n}|$. Now consider the sum~\eqref{equ:Cjn_sum_m} over $m$: the matrix element $\braket{m|Z_j|n}$ between the RSRG-X states can be nonzero only if $\ket{m}$ belongs to the other subspace $Z_j \mathcal{V}$, e.g. $\ket{m} \in V_-$ for $\ket{n} \in V_+$, and we approximate $|E_n - E_m| \approx 4|\Omega_{n, j}|$ for all such $\ket{m}$. This last approximation follows essentially from the RG premise of strong randomness (Sec.~\ref{sec:review:generalities}). Therefore, when $j$ is on a resonant part of $\ket{n}$ we have $C_{n, j}(t)=\cos \left(4\Omega_{j, n} t\right)$. 

Combining (a) and (b) above, we find
\eq{
    C_{n, j}(t) = \begin{cases}
        1 & j\;\mathrm{on \; frozen\; part\; of} \ket{n} \\
        \cos(4\Omega_{n, j}t) & j\;\mathrm{on \; resonant\; part\; of} \ket{n}
    \end{cases}.\label{equ:Cnj_results}
}
Each resonant part contributes a resonance between the two spin configurations $\spinups{I}\spindowns{J}$ and $\spindowns{I}\spinups{J}$, at a frequency $4|\Omega_{n, j}|$ determined by RG histories. This can be viewed as a generalization of the nearest-neighbor dynamics where we always have $|\mathbf{I}|=|\mathbf{J}|=1$, i.e., flipping of spin-$\frac{1}{2}$ pairs (Sec.~\ref{sec:NN:analytics}). Here for the beyond-nearest-neighbor models, however, the spin-flips can now involve groups of aligned spin-$\frac{1}{2}$'s given by $\mathbf{I}, \mathbf{J}$. Note that the spins within each group may or may not be spatially separated. This generalization can be visualized as a many-body spin-flip picture, which we illustrate in Fig.~\ref{fig:manybody_picture}.

\subsection{Numerically Implementing RSRG-X}
\label{sec:LR:numerics}

The RSRG-X formalism provides a physical picture (Sec.~\ref{sec:LR:RG:picture}) that is useful in predicting the spin dynamics. Due to the increase in the complexity of RSRG-X states (Sec.~\ref{sec:lr:RG:subsequent_steps}) relative to the nearest-neighbor model, our focus will be on numerical estimates, rather than analytical calculations, of the infinite-temperature autocorrelation function $\overline{C}(t)$ (therefore $\overline{\mathcal{S}_p}(\omega)$, the spin survival probability~\eqref{equ:Spt_Czz}) in the specific models~\eqref{equ:overview:model} we consider. By comparing these estimates with calculations based on ED, we will determine the accuracy of our method. 

The contribution to $\overline{C}(t)$ probed at site $j$ from a single realization of the randomness can be viewed as a uniform sum~\eqref{equ:C_spectral_average} over $C_{n, j}(t)$, which we have computed~\eqref{equ:Cnj_results} for individual RSRG-X states $\ket{n}$. To estimate $\overline{C}(t)$, we therefore perform a simultaneous Monte Carlo average over randomness samples, RSRG-X states, and sites $j$. Notably, doing so requires us to keep track of only one RSRG-X state at a time and thus costs relatively mild computational resources. Consequently, we will be able to extend our numerical RSRG-X calculations to system sizes far beyond those accessible by ED. 

In particular, we sample the states $\ket{n}$ and find $\Omega_{n, j}$, which enters $C_{n, j} (t)$ through Equ.~\eqref{equ:Cnj_results}, by randomized branching along the RG steps. When $H_0$ takes the form in case (i) of Sec.~\ref{sec:lr:RG:subsequent_steps} we choose to branch into $V_+$, $V_0$, or $V_-$ with respective probabilities $\frac{1}{4}$, $\frac{1}{2}$, and $\frac{1}{4}$, in case (ii) we branch into $V'_+$ or $V'_-$ with equal probability, and similarly in case (iii) we branch into $U_+$ or $U_-$ with equal probability. This procedure samples $\ket{n}$ uniformly from the RSRG-X states, each being chosen with probability $2^{-N}$. Moreover, we find $\Omega_{n, j}$ along the branching path (see Fig.~\ref{fig:branching} for an illustration) leading to $\ket{n}$.

With this sampling procedure, in Sec.~\ref{sec:LR:numerics:picture} we show an abundance of many-body spin-flip processes (Sec.~\ref{sec:LR:RG:picture}) in the long-range and next-nearest-neighbor models~\eqref{equ:overview:model} of our interest, establishing the models' major difference in spin dynamics from the nearest-neighbor chain~\eqref{equ:H_NN}. Next, we move on to present RSRG-X predictions of the spin relaxation dynamics in these models. In Sec.~\ref{sec:LR:numerics:sp}, we show agreement of the disorder-averaged $\overline{\mathcal{S}_p}(\omega)$ to the ED results at low frequencies. In Sec.~\ref{sec:LR:numerics:beyond_ED}, we present $\overline{\mathcal{S}_p}(\omega)$ results obtained by RSRG-X at system sizes far beyond the reach of ED and comment on the possible phenomenologies of dynamics.

\subsubsection{Abundance of Many-Body Spin-Flips}
\label{sec:LR:numerics:picture}

\begin{figure}
    \includegraphics[width = 0.48\textwidth]{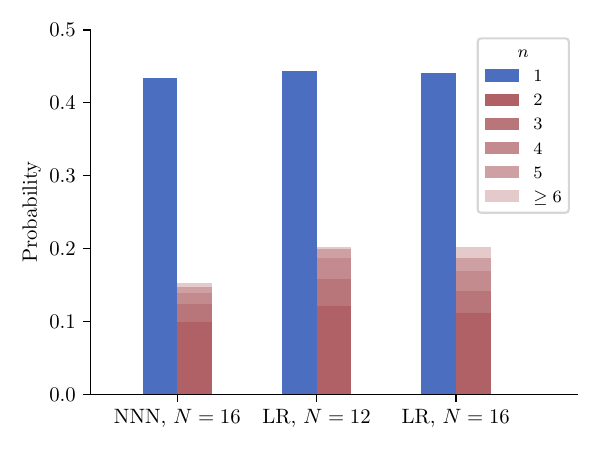}
    \caption{The probabilities that a site engages in a spin-flip process between a pair of $n$ collectively aligned spin-$\frac{1}{2}$'s, in next-nearest-neighbor (NNN) and long-range (LR) chains~\eqref{equ:overview:model} of $N$ spins. Results represent averages over $2.5\times 10^5$ RSRG-X states, each sampled from an independent realization of the randomness.}
    \label{fig:RG_superspin}
\end{figure}

Based on our RSRG-X formalism involving superspins, we have proposed in Sec.~\ref{sec:LR:RG:picture} a many-body spin-flip picture of the dynamics in the interacting models. To substantiate this picture, we provide evidence that, for the long-range and next-nearest-neighbor models we consider~\eqref{equ:overview:model}, the many-body (as opposed to two-body) spin-flip processes are indeed abundant and thus contribute significantly to the dynamics.

To be precise, we study the probabilities that a site engages in a collective spin-flip process involving a pair of $n$ aligned spin-$\frac{1}{2}$'s. For each realization of the randomness, we randomly sample an RSRG-X state with our Monte Carlo procedure (Sec~\ref{sec:LR:numerics}) and register the instances where a site belongs to a resonant part~\eqref{equ:RSRGX_state} with $|\mathbf{I}|=|\mathbf{J}| = n$. We then average over randomness until results converge.

The results (Fig.~\ref{fig:RG_superspin}) show significant contribution from $n\ge 2$ for both the next-nearest-neighbor and long-range models at various system sizes. Note that, in contrast, for the nearest-neighbor chains~\eqref{equ:H_NN} the results would feature solely $n=1$ as spin flips can only be two-body (Sec.~\ref{sec:NN:analytics}). Our RSRG-X formalism has therefore revealed a qualitative difference in dynamics between the nearest-neighbor and beyond-nearest-neighbor models.

\subsubsection{Agreement with ED: Ensemble Averaged $\overline{\mathcal{S}_p}$}
\label{sec:LR:numerics:sp}

Having illustrated the qualitative property (Sec.~\ref{sec:LR:numerics:picture}) of spin dynamics as predicted by RSRG-X, we next consider the observable $\overline{\mathcal{S}_p}(\omega)$, the disorder-averaged spin survival probability at infinite temperature~\eqref{equ:Spt_Czz}. We obtain $\overline{\mathcal{S}_p}(\omega)$ results by numerical RSRG-X with the Monte Carlo approach mentioned above in Sec.~\ref{sec:LR:numerics} and compare with ED results. See Appendix~\ref{appendix:numerics} for further details of both numerical approaches.

\begin{figure}
    \includegraphics[width = 0.48\textwidth]{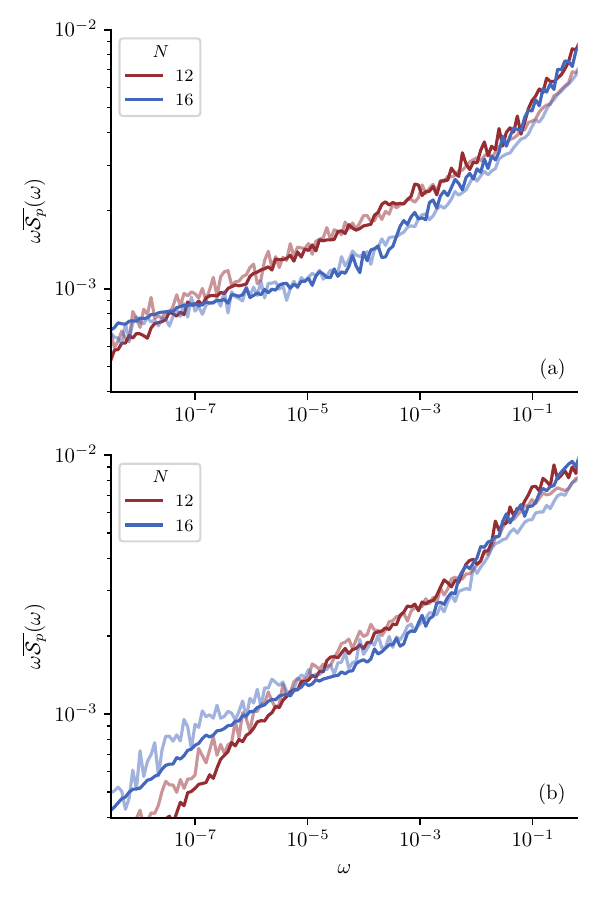}
\caption{Spin survival probability computed with ED (dark) and RSRG-X (light) for (a) the next-nearest-neighbor (b) the long-range chains~\eqref{equ:overview:model}. Each ED dataset is an  average over 1000 realizations of the randomness, while each RSRG-X dataset is sampled as in Fig.~\ref{fig:RG_superspin}.}
    \label{fig:LR_NNN_SDscale_ED_RG}
\end{figure}

We present $\overline{\mathcal{S}_p}(\omega)$ results in Fig.~\ref{fig:LR_NNN_SDscale_ED_RG} for the next-nearest-neighbor and long-range models~\eqref{equ:overview:model}. Remarkably, for the next-nearest-neighbor model we find good quantitative agreement between ED and RSRG-X results throughout the low frequency regime $\omega \lesssim 10^{-2}$ we have investigated. For the long-range model, the agreement holds for the regime $10^{-6} \lesssim \omega \lesssim 10^{-2}$, whereas moderate discrepancy appears for the lower frequencies $\omega \lesssim 10^{-6}$, a regime where we observe strong finite-size effects (Fig.~\ref{fig:LR_NNN_SDscale_ED_RG}). The agreements in the regimes above indicate that RSRG-X is capable of predicting the spin relaxation dynamics at late times at the system sizes ED has access to.

One question to ask is whether these beyond-nearest-neighbor models might share universal late-time spin relaxation dynamics~\eqref{equ:asymptote} with the nearest-neighbor model. Due to the strong finite-size effects in the low-frequency regime $\omega \lesssim 10^{-2}$ at $N=16$ for both the next-nearest-neighbor and long-range models (Fig.~\ref{fig:LR_NNN_SDscale_ED_RG}), however, it is hard to determine the phenomenology with the results presented so far. Especially, our results suggest that it is impractical to investigate such question numerically with ED. Alternatively, in Sec.~\ref{sec:LR:numerics:beyond_ED} we will address this question by scaling up our RSRG-X simulation at low frequencies, which already matches ED in the small system sizes both methods can access.

\subsubsection{Computing $\overline{\mathcal{S}_p}$ Beyond ED Capabilities}
\label{sec:LR:numerics:beyond_ED}

Now with RSRG-X, we attempt to investigate the late-time infinite-temperature spin relaxation dynamics of the next-nearest-neighbor and long-range models~\eqref{equ:overview:model} in the thermodynamic limit. Notably, our RSRG-X simulation as a Monte Carlo approach (Sec.~\ref{sec:LR:numerics}, also Appendix~\ref{appendix:numerics:RSRGX:LR}) requires only moderately increasing computer memory as the system size $N$ grows, allowing access up to at least $N=40$ for both models. The enlarging scope of $N$ clearly goes beyond the reach of ED for these essentially interacting models.

\begin{figure}
    \includegraphics[width = 0.48\textwidth]{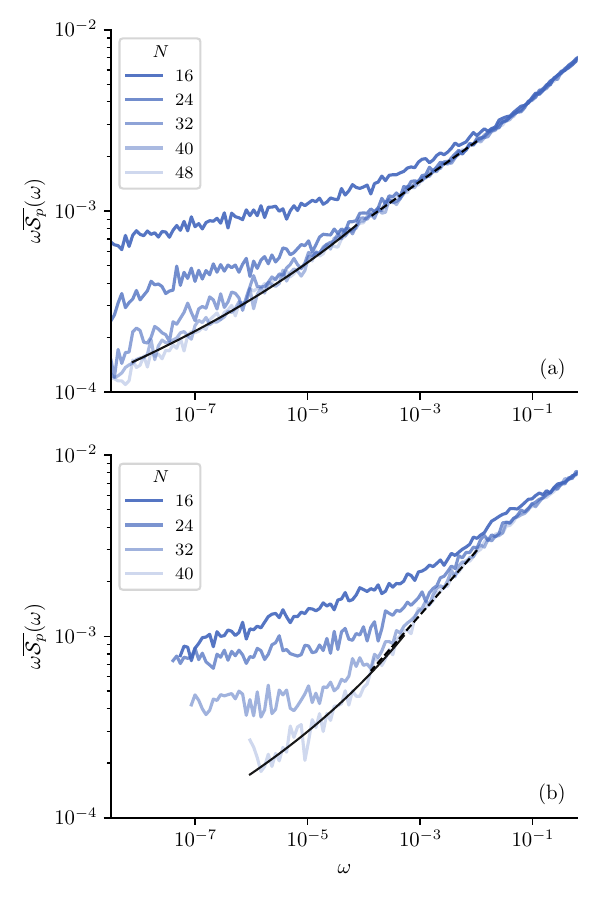}
    \caption{Numerical RSRG-X predictions of $\overline{\mathcal{S}_p}(\omega)$, the spin survival probability~\eqref{equ:Spt_Czz}, for (a) next-nearest-neighbor, (b) long-range chains of sizes $N$ with random dipolar interactions~\eqref{equ:overview:model}. Low-frequency cutoffs of data are determined so that the displayed results are insensitive to $\epsilon_{\mathrm{thres}}$, the energy threshold below which generation of interactions is neglected in the numerical RSRG-X (Appendix~\ref{appendix:numerics:RSRGX:LR}).
    Each dataset represents average over more than (a) $2 \times 10^5$, (b) $2.8\times 10^6 / N$ RSRG-X states, each sampled from an independent realization of the randomness. Solid black lines show fits of the form in Equ.~\eqref{equ:overview:Sp_freq} corresponding to the $\conj{S_p}(t) - \conj{S_p}(\infty) \sim [\log(t)]^{-2}$ asymptote (Appendix~\ref{appendix:asymptote}). Dashed black lines represent fits of the form $\omega \overline{\mathcal{S}_p}(\omega) \sim \omega^{c}$ corresponding to $\overline{S_p}(t) - \overline{S_p}(\infty)\sim t^{-c}$ in the time domain: (a) $c=0.22$, (b) $c=0.35$. Note that diffusive spin relaxation would correspond to $c=\frac{1}{2}$ and thus a straight line of steeper slope.}
    \label{fig:LR_NNN_RG_SDscale_varN}
\end{figure}

We present $\overline{\mathcal{S}_p}(\omega)$ results in Fig.~\ref{fig:LR_NNN_RG_SDscale_varN}. Surprisingly for the next-nearest-neighbor model, we already observe good convergence of the result with increasing system size for frequencies $\omega \gtrsim 10^{-7}$ at $N\gtrsim 40$. Where we observe such convergence, for $10^{-4}\lesssim \omega \lesssim 10^{-2}$, the results suggest a power-law decay $\overline{S_p}(t) - \overline{S_p}(\infty) \sim t^{-0.22}$, whereas for smaller $\omega\lesssim 10^{-4}$ there is a significant deviation from the power-law scaling. Notably, in the latter regime our results are well described by the nearest-neighbor universal scaling form~\eqref{equ:overview:Sp_freq}, $\omega \overline{\mathcal{S}_p}(\omega) = A\left[
\log\left(\omega_0/\omega\right)\right]^{-3}$ (Fig.~\ref{fig:LR_NNN_RG_SDscale_varN}\textcolor{vibrant}{a}). As we establish in Appendix~\ref{appendix:asymptote}, such a fit implies the $\conj{S_p}(t) - \conj{S_p}(\infty) \sim A / \left(\log t\right)^{2}$ asymptote~\eqref{equ:asymptote} in the late-time domain. Our results therefore suggest, despite a possible intermediate power-law decay, spin relaxation that is slower than any power law and persistent at large timescales for the next-nearest-neighbor model.

For the long-range model, our results in Fig.~\ref{fig:LR_NNN_RG_SDscale_varN}\textcolor{vibrant}{b} at accessible system sizes only show convergence upon increasing $N$ in the regime $\omega \gtrsim 10^{-4}$. In such regime, the $\overline{\mathcal{S}_p}(\omega)$ data suggest a power-law scaling $\overline{S_p}(t) - \overline{S_p}(\infty) \sim t^{-0.35}$ at late times.
At lower frequencies, we identify at large system size $N=40$ a fit of the universal scaling form~\eqref{equ:overview:Sp_freq} established for the nearest-neighbor model and also observed above for the next-nearest-neighbor model, but due to finite-size effects we are unable to confirm such scaling in the thermodynamic limit.
Nevertheless, our results indicate that the spin relaxation dynamics for the long-range model is clearly slower than being diffusive --- the latter case would be $\omega \overline{\mathcal{S}_p}(\omega) \sim \omega^{1/2}$, corresponding to a line of much steeper slope on the scales of Fig.~\ref{fig:LR_NNN_RG_SDscale_varN}\textcolor{vibrant}{b}.

In summary, our RSRG-X simulations suggest that, for the next-nearest-neighbor model, the late-time spin relaxation dynamics feature the same universal scaling form as the nearest-neighbor model (Sec.~\ref{sec:NN:analytics}) in the thermodynamic limit. For the long-range model, it remains inconclusive whether this universal form, or instead a sub-diffusive power-law decay, holds. We emphasize again that, despite possibly sharing this universality, the spin dynamics for the nearest-neighbor and beyond-nearest-neighbor models are described by qualitatively distinct physical pictures (Sec.~\ref{sec:LR:numerics:picture}).

\section{The Two-Dimensional Model}
\label{sec:2d}

\begin{figure*}
    \includegraphics[width = 0.98\textwidth]{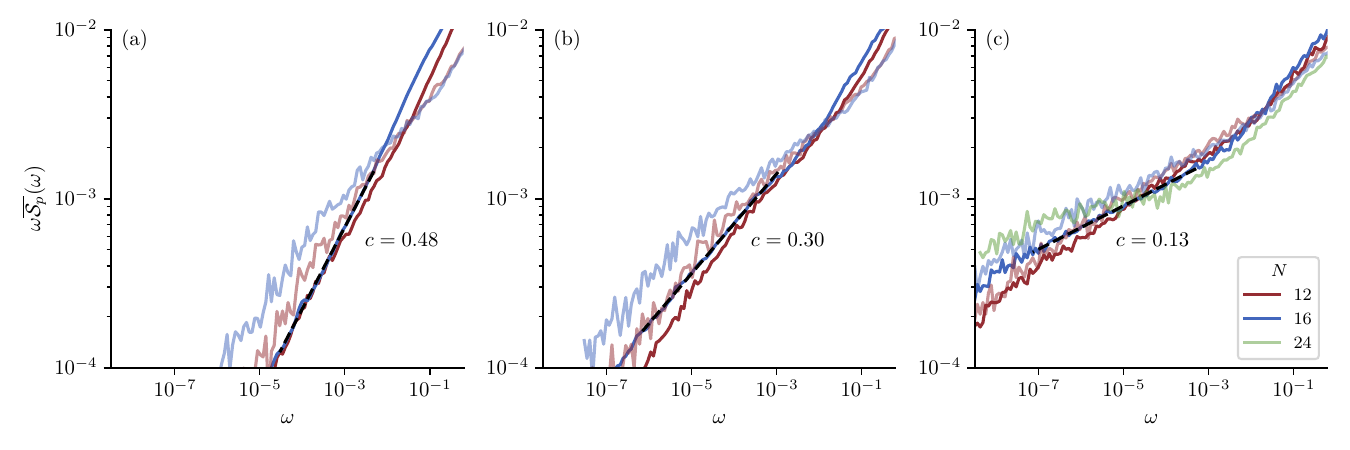}
    \caption{ED (dark) and numerical RSRG-X (light) results for $\overline{\mathcal{S}_p}(\omega)$, the spin survival probability~\eqref{equ:Spt_Czz}, for the two-dimensional model~\eqref{equ:H_2d} of randomly positioned sites with long-range interactions of strength $~r^{-\alpha}$ depending on distance $r$. Each ED dataset represents an average over more than 400 realizations of the randomness, while each RSRG-X dataset uses $2.8\times 10^6 / N$ RSRG-X states, each sampled from an independent realization of the randomness. Low-frequency cutoffs of data are determined so that the displayed results are insensitive to $\epsilon_{\mathrm{thres}}$, the energy threshold below which generation of interactions is neglected in the numerical RSRG-X (Appendix~\ref{appendix:numerics:RSRGX:LR}). Dashed black lines show fits of the form $\omega \overline{\mathcal{S}_p}(\omega) \sim \omega^c$ corresponding to $\sim t^{-c}$ decay in the time domain. Note that, in two dimensions, diffusion would lead to $c=1$.}
    \label{fig:2D_SDscale_varN_vara}
\end{figure*}

Although we have so far focused on one-dimensional systems, our RSRG-X formalism (Sec.~\ref{sec:LR:RSRGX}) can in principle be applied to systems with strongly random $XX+YY$ interactions in any geometry. It is therefore natural to ask whether the formalism could provide accurate predictions for dynamics in higher dimensions. In this section, we consider a two-dimensional model with power-law decaying interactions $\sim r^{-\alpha}$ between randomly positioned spins at distance $r$. We present preliminary numerical RSRG-X results at accessible system sizes for the infinite-temperature spin survival probability $\overline{\mathcal{S}_p}(\omega)$ at low frequencies. Our results suggest slow spin relaxation at late times. Furthermore, at large $\alpha$, RSRG-X results show good agreement to ED at low frequencies. 

The Hamiltonian of the two-dimensional model reads
\eq{
H_{2d} = \sum_{(j,k)} J_{jk} \left(X_j X_k + Y_j Y_k\right), \label{equ:H_2d}
}
where $J_{jk} = r_{jk}^{-\alpha}$, and $r_{jk}$ is the distance between sites $j$ and $k$. The position $\mathbf{x}_j = \left(x^1_j, x^2_j\right)$ of every site $j$ is randomly chosen within $[0, L)\times [0, L)$, and we define the distances under PBC, such that $r_{jk}^2 = \sum_{a=1, 2}\left(\min\left\{|x_j^a-x_k^a|, L-|x_j^a-x_k^a|\right\}\right)^2$. We still impose a hard-core constraint $r_{jk} > \delta$ and again set $\delta = 1/40$. To fix the density of sites, we scale $L = \sqrt{N}$. We focus on the cases $3 \leq \alpha \leq 6$ as are reasonable to consider under experimental setups.

The ground state of the two-dimensional $XX+YY$ model (\ref{equ:H_2d}) of hundreds of sites with generic realizations of random antiferromagnetic interactions ($J_{jk} > 0$) has been examined earlier with RSRG techniques \cite{MMHF00, Laumann12}. There, scaling up to large system sizes is possible because branching into the ground-state eigenspace $V_-$, as is in our general RSRG-X formalism (\ref{equ:LR:XY_subspaces}), does not generate superspins and thus leads to relatively simple updates of $\Heff$. A similar simplification applies to the ground-state RSRG of random Heisenberg antiferromagnets \cite{BL82}. The consensus for these prior studied models has been that the interactions do not flow to strong randomness, and hence that the true ground states are not characterized by any infinite-randomness fixed points \cite{BL82, MMHF00, Laumann12}.

In our numerical program to find generic RSRG-X states of the two-dimensional model~\eqref{equ:H_2d}, we also observe, as shown in Appendix~\ref{appendix:premises}, worse fulfillment of the RG premise of strong randomness (Sec.~\ref{sec:review:generalities}) than in one dimension. Especially, there is a higher chance that the strong interaction $H_0$ in $\Heff$ involves more than two sites, in which case our strategy of identifying $H_0$ is unjustified (Sec.~\ref{sec:lr:RG:subsequent_steps}). These observations raise potential issue to the trustworthiness of our RSRG-X simulations.

We compare the RSRG-X predictions with ED results in Fig.~\ref{fig:2D_SDscale_varN_vara} for $3\leq \alpha \leq 6$. Notably, despite the issue described above, we observe quantitative agreement in the low-frequency regime $10^{-7} \lesssim \omega \lesssim 10^{-3}$ at large $\alpha = 6$. For lower $\alpha$, qualitative agreement remains. We indeed expect better agreement at larger $\alpha$ because the positional randomness then implies a broader distribution of the initial couplings $J_{jk}$.

Additionally, both the ED and RSRG-X results in Fig.~\ref{fig:2D_SDscale_varN_vara} suggest power-law decay of $\overline{S_p}(t) \sim t^{-c}$ at late times. The extracted exponent $c$ indicates slower decay at larger $\alpha$, and for all $\alpha \ge 3$ we observe subdiffusive relaxation; note that diffusion would correspond to $c=1$ in two dimensions. We emphasize that, admittedly, the system size we have access to is too small to lead to definite conclusions in the thermodynamic limit. One possible way to lift such uncertainty is through experimental studies: a recent experiment on Rydberg atoms randomly positioned in three dimensions has studied the relaxation of the transverse (rather than conserved) component of the spin degrees of freedom and observed stretched-exponential decay of the corresponding autocorrelation function~\cite{Titus_experiment_24}. Our numerical results here in two dimensions, however, suggest qualitatively different behavior for the conserved component. To what extent the spatial dimensions might result in distinct regimes of late-time relaxation dynamics is worth further investigation.

\section{Conclusion}
\label{sec:conclusion}

We have proposed an RSRG-X formalism (Sec.~\ref{sec:LR:RSRGX}) applicable to a large family of spin-$\frac{1}{2}$ systems with random interactions featuring global $U(1)$ and $\Z_2$ symmetries. The effective Hamiltonian updated along RSRG-X typically involves superspins (Sec.~\ref{sec:LR:RG:superspin}), two-level degrees of freedom of (anti)aligned microscopic spins in the original system. Such a superspin RSRG-X resolves the dynamics in terms of contributions from coherent flips between pairs of oppositely-oriented groups of aligned spins (Sec.~\ref{sec:LR:RG:picture}). Each group may consist of several spins that are not necessarily spatially adjacent. This many-body spin-flip picture reduces to two-body in the non-interacting model of nearest-neighbor $XX+YY$, where each group only has one spin (Sec.~\ref{sec:NN:analytics}).

Our formalism generates a numerical method for calculating the disorder-averaged infinite-temperature spin survival probability, $\overline{\mathcal{S}_p}(\omega)$, for various models. Our focus has been on experimentally relevant long-range models with dipolar interactions between randomly positioned sites in one and two spatial dimensions. In one dimension, we have also studied these models with interactions truncated beyond nearest-neighbor and next-nearest-neighbor sites. For all of these one-dimensional models, our RSRG-X quantitatively predicts $\overline{\mathcal{S}_p}(\omega)$ at low frequencies, even when finite-size effects are severe. In two dimensions, we find a discrepancy between ED and RSRG-X for the dipolar model, although for more rapidly decaying power-law interactions the agreement between the two methods improves.

The numerical implementation also allows us to showcase the contributions from many-body (rather than two-body) spin-flip processes in the infinite-temperature spin dynamics (Sec.~\ref{sec:LR:numerics:picture}). Such contributions, resulting from the formation of dynamical nontrivial superspins along RSRG-X, mark a key qualitative difference between the dynamics of interacting and non-interacting random $XX+YY$ chains. These contributions have been considered unimportant at low temperatures~\cite{MDH01}, and have been neglected in previous RSRG-X studies ~\cite{MVK22, Mohdeb23}. 

That we can design a self-contained RSRG-X is a consequence of global $\mathrm{U}(1)$ and $\Z_2$ symmetries. These symmetries constrain the form of typical one- and two-site interactions along our RSRG-X to a manageable number of cases (Sec.~\ref{sec:lr:RG:subsequent_steps}). Furthermore, all such interactions, when standalone, feature largely gapped eigenspaces each containing two levels at most (see some examples in Fig.~\ref{fig:hybridize}), so branching into an eigenspace generates an effective Hamiltonian expressible in terms of two-level superspins. While the idea that symmetries get inherited and limit the permissible terms is not new to the general renormalization group approaches, our work highlights how RSRG-X can be simplified by symmetries of the system and anticipates similar works along this line. As another example, Ref.~\cite{Protopopov20} has addressed the role of $\textrm{SU}(2)$ symmetry in the RSRG-X for nearest-neighbor Heisenberg models \cite{Protopopov20}.

Although we have described individual RG steps of our RSRG-X analytically (Sec.~\ref{sec:lr:RG:subsequent_steps}, also Appendix~\ref{appendix:pseudo_codes}), it remains unclear whether an analytical analysis of RG flows, resembling that for nearest-neighbor chains (Sec.~\ref{sec:review}), is possible. A goal we find of immense theoretical interest is to identify a possibly universal description of the typical eigenstates resulting from our RSRG-X. Although the ground state of some long-range systems, for example in two dimensions, are not governed by any infinite-randomness fixed points~\cite{BL82, MMHF00, Laumann12}, for our excited-state RSRG the formation of superspins might affect the flow of the randomness. To explore possible universality, an immediate challenge is to analytically track the distribution of interactions involving three or more sites. We leave this direction for future attempts.

Our work raises general questions on the phenomenology of spin relaxation in one- and higher-dimensional systems with random dipolar interactions. In one dimension, our results for the next-nearest-neighbor model~\eqref{equ:overview:model} suggest no deviation from the late-time $\conj{S_p}(t)-\conj{S_p}(\infty)\sim \left[\log(t)\right]^{-2}$ asymptote established for the nearest-neighbor model. For the long-range model, our RSRG-X simulation suffers from strong finite-size effects, but we observe the same asymptotic behavior at the largest accessible system size $N=40$ (Fig.~\ref{fig:LR_NNN_SDscale_ED_RG}). In both cases, our results suggest slower-than-power-law decay up to surprisingly large timescale in one dimension; recall that the typical energy scale~\eqref{equ:overview:model} is fixed to unity. In two dimensions, we observe power-law, sub-diffusive decay of $\conj{S_p}(t)$ in the small system sizes to which we have access (Fig.~\ref{fig:2D_SDscale_varN_vara}). Further numerical and even analytical advances are necessary to determine the phenomenology in the thermodynamic limit. 

The presence of slow dynamics due to randomness has been explored in numerous theoretical works, including recent modifications of Lieb-Robinson bounds~\cite{Baldwin23, Baldwin24} and proofs of sub-diffusive transport~\cite{deRoeck24}, both concerning disordered spin chains with general types of local interactions. One question to ask is whether the slower-than-power-law decay of spin survival probability we have observed in a family of spin chains with $\mathrm{U}(1)$ and $\Z_2$ symmetries suggests possibly tighter bounds on transport properties in this setting. Such constraints on transport, if any, might be advanced by our RSRG-X formalism combined with a careful analysis of the subleading contributions from the perturbative treatments; indeed the general idea that strong interactions renormalize in the first step to effectively weaker ones has been applied in Ref.~\cite{Baldwin24}. Moreover, although Lieb-Robinson bounds have been derived for generic systems with long-range interactions~\cite{Lucas19, Lucas21}, our work raises questions over how these bounds are modified in systems with positional randomness.

On a practical level, the wide range of interacting models to which our RSRG-X applies include systems with long-range random interactions of the form $XX+YY$ and also even $ZZ$. Many cutting-edge experimental platforms feature such interactions and thus can probe the signatures of spin dynamics predicted by our formalism. Conversely, an important application of our method is to calibrate and benchmark emerging hardware and algorithm for quantum computation. Especially, the quantum simulation of random one-dimensional spin models will be a valuable step forward: the dynamics in these models involve interference effects of a kind not seen in translation-invariant settings. Meanwhile, random systems in one dimension are typically not as intractable as in higher dimensions; even finding the ground state of a random classical Ising model is NP-complete in two dimensions~\cite{barahona}. The rapid progress in experimental capabilities offers the promise that theory and experiment on random quantum spin models can be more strongly coupled than they have been in the past, leading to new discoveries in this rich area of physics.

\acknowledgments

Y.J.Z. and J.E.M. acknowledge support from the National Science Foundation, Challenge Institute for Quantum Computation at UC Berkeley. Y.J.Z. thanks David A. Huse and Kai Klocke for helpful discussions. S.J.G. was supported by the Gordon and Betty Moore Foundation. This research used the Lawrencium computational cluster resource provided by the IT Division at the Lawrence Berkeley National Laboratory (supported by the Director, Office of Science, Office of Basic Energy Sciences, of the U.S. Department of Energy under Contract No. DE-AC02-05CH11231).

\appendix

\section{Decay Asymptotes in Frequency and Time Domains}
\label{appendix:asymptote}
In this appendix, we show explicitly that the low-frequency part of $\overline{\mathcal{S}_p}(\omega)$ in the form of Equ.~\eqref{equ:analytics:Sp_freq} leads to the slow-decaying asymptote~\eqref{equ:asymptote} at late times and determine the parameters quantitatively. This amounts to evaluating the integral $\conj{S_p}(t) = \int d \omega\, \overline{\mathcal{S}_p}(\omega) e^{-i\omega t}$. Our goal is to show, for large $t$,
\eq{
    \conj{S_p}(t) = B + \frac{A}{\left[\log\left(2\omega_0 t/\pi\right)\right]^2}\left[1+\mathcal{O}\left(\frac{1}{\log(\omega_0 t)}\right)\right], \label{appendix:asymptote:result}
}
in alignment with Equ.~\eqref{equ:analytics:Spt} claimed in Sec.~\ref{sec:NN:analytics}.

Since the high-frequency parts of $\overline{\mathcal{S}_p}(\omega)$ do not contribute to the late-time dynamics, it suffices to perform the integral within $\omega\in \left(-\Lambda, \Lambda\right)$ for some reasonably small cutoff $\Lambda < \omega_0$. Under the context of the fixed-point derivations in Sec.~\ref{sec:NN:analytics}, for example, a convenient choice is $\Lambda = \omega_0 e^{-\Gamma_c}$. Nevertheless, we will see as expected that the result does not depend on $\Lambda$.

The dynamical part of $\conj{S_p}(t)$ is then contributed by
\al{
    I(t) &= \int_{-\Lambda}^{\Lambda}d\omega\, \left|\frac{A}{\omega\left[\log\left(\omega_0 / \omega\right)\right]^3}\right|e^{-i\omega t} \nonumber \\
    &= \int_0^\Lambda d\omega\, \frac{2A\cos(\omega t)}{\omega\left[\log\left(\omega_0/\omega\right)\right]^3}.
}
For convenience, we express the integral in terms of $\lambda = \log(\omega_0t)$, the logarithm of a dimensionless time, and perform a change of variable through $u=\left[\log(\omega_0/\omega)\right]^{-2}$,
\eq{
    I(\lambda) = A \int_{0}^{u_0}du\,\cos\left(e^{\lambda-\frac{1}{\sqrt{u}}}\right),
}
where $u_0 = \left[\log \left(\omega_0/\Lambda\right)\right]^{-2}$. We then need to show that $I(\lambda)$
decays as $1/\lambda^2$ at large $\lambda$.

To show the last proposition, separate the integral into two parts $I(l) = I_1(l) + I_2(l)$ at $u = x_0$, such that
\al{
I_1(\lambda) &= A \int_{0}^{x_0}du\,\cos\left(e^{\lambda-\frac{1}{\sqrt{u}}}\right) \\
I_2(\lambda) &= A\int_{x_0}^{u_0}du\,\cos\left(e^{\lambda-\frac{1}{\sqrt{u}}}\right).
}
We pick $x_0$ to be the smallest zero of the cosine function, $x_0 = \left[\lambda - \log(\pi/2)\right]^{-2}$. We can achieve $x_0 < u_0$ always at large enough $\lambda$ and, rigorously, when $t > \pi /(2\Lambda)$. The latter regime is naturally what one expects to be late-time, given the cutoff $\Lambda$.

To evaluate $I_1(\lambda)$, we note that the integrand is nearly constantly $1$ within the domain of integration. One way to confirm this is by rescaling
\eq{
    I_1(\lambda) = x_0 A \int_0^1dv\, \cos\left(e^{\lambda-\frac{1}{\sqrt{x_0 v}}}\right)
}
and observing that the integrand reaches $\frac{1}{2}$ at $v_{1/2} = x_0^{-1}\left[\lambda - \log(\pi/3)\right]^{-2} = 1 - \mathcal{O}(\lambda^{-1})$ approaching 1 at large $\lambda$. Hence, to the leading order we have 
\al{
    I_1(\lambda) &= x_0 A \int_{0}^{1}dv\, \left(1+ \mathcal{O}(\lambda^{-1})\right) \nonumber \\
    &= \frac{A}{\left[\lambda - \log(\pi/2)\right]^2}\left(1+\mathcal{O}(\lambda^{-1})\right), \label{appendix:asymptote:I1_result}
}
showing that $I_1(\lambda)\sim 1/\lambda^2$ at large $\lambda$.

Next, we bound $I_2(\lambda)$ by integration by part and argue that it is subleading at large $\lambda$. Indeed,
\al{
    I_2(\lambda) =&\; 2A u^{3/2}e^{\frac{1}{\sqrt{u}} - \lambda}\left.\sin\left(e^{\lambda-\frac{1}{\sqrt{u}}}\right) \right|^{u=u_0}_{u=x_0} - J(\lambda) \nonumber\\
    =&\; 2A u_0^{3/2}e^{\frac{1}{\sqrt{u_0}}- \lambda}\sin\left(e^{\lambda-\frac{1}{\sqrt{u_0}}}\right) \nonumber \\ 
    &- \frac{4A}{\pi \left[\lambda-\log(\pi/2)\right]^3} - J(\lambda), \label{equ:int_by_part}
}
where
\eq{
    J(\lambda) = A\int_{x_0}^{u_0}du\, \left(3\sqrt{u}-1\right)\sin\left(e^{\lambda-\frac{1}{\sqrt{u}}}\right)e^{\frac{1}{\sqrt{u}} - \lambda}.
}
The first two terms in the last step of Equ.~\eqref{equ:int_by_part} decays no slower than $1/\lambda^3$ at large $\lambda$, while the integral $J(\lambda)$ is bounded by
\eq{
    |J(\lambda)| \leq A\int_{x_0}^{u_0}du\, \left|3\sqrt{u} - 1\right| e^{\frac{1}{\sqrt{u}} - \lambda}.
}
As long as we have picked $\Lambda$ small enough such that $\Lambda < \omega_0 e^{-3}$, we always have $3\sqrt{u} - 1 < 0$ in the integral, and therefore
\al{
    |J(\lambda)| &\leq A \left|\int_{x_0}^{u_0}du\, \left(3\sqrt{u} - 1\right)e^{\frac{1}{\sqrt{u}} - \lambda}\right| \nonumber \\
    &= 2A \left|\left.u^{3/2} e^{\frac{1}{\sqrt{u}} - \lambda}\right|^{u = u_0}_{u=x_0}\right| \nonumber \\
    &= \left|2A u_0^{3/2} e^{\frac{1}{u_0}-l} - \frac{4A}{\pi \left[\lambda-\log(\pi/2)\right]^3}\right|,
}
decaying no slower than $1/\lambda^3$ at large $\lambda$.

In conclusion, the leading contribution to $I(\lambda)$ comes from $I_1(\lambda)$, which we have evaluated (\ref{appendix:asymptote:I1_result}). Recall that $I(t)$ contributes to the dynamical part of $\conj{S_p}(t)$, while the static part comes from the $B\delta(\omega)$ term in the spectral function (\ref{equ:analytics:Sp_freq}). Piecing the two parts together and restoring $\lambda = \log(\omega_0 t)$, we arrive at Equ.~\eqref{appendix:asymptote:result}.

\section{Interactions at two superspin-0 sites}
\label{appendix:superspin0}
In this appendix, we explain why we can expect the $X^{(0)} X^{(0)}$ and $Y^{(0)} Y^{(0)}$ types of interactions between two superspin-0 sites to be rarely the strongest in the effective Hamiltonian $\Heff$ updated along RSRG-X. Our reasoning consists of two parts. First, we point out that the superspin-0 sites themselves are rare, so the strongest interaction rarely involves two such sites. Furthermore, we argue that the $X^{(0)} X^{(0)}$ and $Y^{(0)} Y^{(0)}$ interactions are weak if the three-site interactions, as we also expect, are weak. These theoretical considerations are further supported by the numerical RSRG-X of the specific models~\eqref{equ:overview:model} we consider (Appendix~\ref{appendix:premises}).

That the superspin-0 sites themselves are less common can be seen from the branching choices (i)--(iii) we have listed (Sec.~\ref{sec:lr:RG:subsequent_steps}). Indeed, superspin-0 sites can be only generated by branching into the subspace $V_-'$ in (ii) when the strong interaction $H_0$ happens to reside on sites of identical superspins. On the other hand, a superspin-0 site gets absorbed into a new superspin-nonzero site as long as $H_0$ involves a superspin-0 site and another superspin-nonzero site, the situation falling under (ii). Since the superspin-0 sites are hard to generate but easy to diminish, we expect them to be much fewer than other sites of nonzero superspins.

To argue that the $X^{(0)}X^{(0)}$ and $Y^{(0)}Y^{(0)}$ interactions are weak as long as the three-site interactions in $\Heff$ are weak, we consider how such interactions can be generated in some prior RG step. In fact, a superspin-0 site itself can only be generated when we branch into the $V_-'$ subspace [case (ii) in Sec.~\ref{sec:lr:RG:subsequent_steps}]: let the strongest interaction $H_0$ in $\Heff$ at this step involve, without loss of generality, the sites $1$, $2$ of equal superspin-$\frac{m}{2}$ with $m>0$. Then, $\Heff^{(V_-')}$ contains in replacement of the sites 1, 2 a new superspin-0 site, which we label by $0$, as well as new interactions of the form $X^{(0)}X^{(0)}$ and $Y^{(0)}Y^{(0)}$ --- let us consider a new interaction, for example, $X_0^{(0)} X_3^{(0)}$. In light of the $\mathrm{U}(1)$ and $\Z_2$ symmetries~\eqref{equ:symmetries_superspin}, we exhaust the allowed interactions in $\Heff$ among the sites 1, 2, 3,
\al{
    H_{123} = \Omega & Z_1^{(m)} Z_2^{(m)} + \epsilon_1 Z_1^{(m)}Z_3^{(0)} + \epsilon_2 Z_2^{(m)}Z_3^{(0)}\nonumber \\
    & +\alpha(X_1^{(m)}X_2^{(m)}+Y_1^{(m)} Y_2^{(m)})  + \beta X_3^{(0)} \nonumber \\ 
    & +\omega (X_1^{(m)}X_2^{(m)} + Y_1^{(m)}Y_2^{(m)}) X_3^{(0)}.
}
Here, $\Omega$ is the strongest interaction strength, while we also expect $\omega$ as the strength of a three-site interaction to be weaker than the other one- and two-site interactions, 
\eq{
    |\omega| \ll |\epsilon_1|, |\epsilon_2|, |\alpha|, |\beta| \ll |\Omega|. \label{equ:superspin0:assumption}
} 
This is consistent with the reliability of our RSRG-X formalism, which identifies $H_0$ among the three explicitly listed one- and two-site interactions in Equ.~\eqref{equ:LR:Heff}. This sets up our argument.

Now consider branching into the $V_-'$ subspace. By perturbation theory~\eqref{equ:degen_perturb}, we write out the terms in the updated effective Hamiltonian that are of our concern,
\eq{
    H_\mathrm{eff}^{(V_-')} = (\epsilon_1 -\epsilon_2) Z_0^{(0)} Z_3^{(0)}   + 2\alpha X_0^{(0)} + \beta X_3^{(0)}  + 2 \omega X_0^{(0)} X_3^{(0)}. 
}
Note that all the interactions fall under the forms (i)--(iii) in Sec.~\ref{sec:lr:RG:subsequent_steps}, except for the interaction $X_0^{(0)} X_3^{(0)}$. This latter interaction, however, is expected to be weaker than the others, as a result of Equ.~\eqref{equ:superspin0:assumption}. 

\section{Numerical Methods}
\label{appendix:numerics}
In this section, we detail the methods adopted for both ED (Sec.~\ref{appendix:numerics:ED}) and numerical RSRG-X (Appendix~\ref{appendix:numerics:RSRGX}). Generally, the program starts with a specified realization of the randomness at finite system size. When computing ensemble-averaged quantities, we average over sufficiently many such samples. We boost the ensemble averaging by exploiting the fact that all the types of the randomness we consider are unbiased in space; for every sample, we spatially average over the reference site, i.e., the site where the $Z$ operator resides in Equ.~\eqref{equ:Spt_Czz}, before averaging over the ensemble.

\subsection{Exact Diagonalization}
\label{appendix:numerics:ED}
Throughout this work, we perform numerical ED of the system Hamiltonians~(\ref{equ:overview:model},~\ref{equ:H_2d}) at even number of sites. Crucially, we reduce the Hilbert space dimension by the non-interacting nature for the nearest-neighbor models (Appendix~\ref{appendix:numerics:ED:NN}) and by the $\mathrm{U}(1)$ and $\Z_2$ symmetries~\eqref{equ:overview:symmetries} for the beyond-nearest-neighbor and two-dimensional models (Appendix~\ref{appendix:numerics:ED:LR}). With the exactly diagonalized Hamiltonian, we compute the $\overline{\mathcal{S}_p}(\omega)$ of interest at low frequencies. For numerical convenience, in our calculations of $\overline{S_p}(t)$ in Fig.~\ref{fig:LR_ED}, we neglect contributions from pairs of eigenstates with energy separations greater than a UV cutoff $\epsilon_{\mathrm{UV}} = 1$. While this leads to errors at early times, it does not significantly affect the late-time behavior that we are interested in.

\subsubsection{Nearest-Neighbor Models}
\label{appendix:numerics:ED:NN}
For the nearest-neighbor models~\eqref{equ:H_NN} under PBC, the fact that the total $Z$ spin is conserved by the $\mathrm{U}(1)$ symmetry~\eqref{equ:overview:symmetries} allows us to decompose the Hilbert space into sectors even and odd values of $\frac{1}{2} \sum_j (1-Z_j)$. A Jordan-Wigner transformation maps the model~\eqref{equ:H_NN} to
\eq{
    H = \sum_j t_i \left(c_j^\dagger c_{j+1} + h.c.\right)
}
where in the odd sector we have
\eq{
    t_j = J_j, \quad \forall j
}
and in the even sector
\eq{
    t_j = \begin{cases}
        J_j & j = 0, ..., N-1 \\
        -J_j & j = N
    \end{cases}.
}
For either sector, nevertheless, the dynamics are governed by a single-particle Hamiltonian, so we apply the standard ED methods for numerical simulation.

\subsubsection{Beyond-Nearest-Neighbor and Two-Dimensional Models}
\label{appendix:numerics:ED:LR}
For chains with interactions beyond the nearest neighbors~\eqref{equ:overview:model} or models in two dimensions~\eqref{equ:H_2d}, we decompose the Hilbert space of the interacting model based on the $\mathrm{U}(1)$ charge~\eqref{equ:overview:symmetries}, i.e., the total $Z$ spin. As we focus on systems with even $N$ number of sites, the sectors are labeled by the quantum number $M = -N/2, ..., -1, 0, 1, ..., N/2$. Noting that the symmetry generators~\eqref{equ:overview:symmetries} satisfy $PM = -MP$, we further divide the $M=0$ sector based on the eigenvalues $\pm 1$ of $P$. Eventually, we work with the above-mentioned $N$ sectors and compute the dynamics within each of them.

\subsection{Numerical RSRG-X}
\label{appendix:numerics:RSRGX}
In principle, given a realization of the randomness we can numerically track $\Heff$ along all the possible branching choices detailed for the RG steps depending on the specific models (Secs.~\ref{sec:review:NN},~\ref{sec:LR:RSRGX}). In this way, we can obtain the complete set of RSRG-X states with the corresponding approximate eigenenergies. The information so obtained would then enable us to simulate the late-time dynamics, spin relaxation for example. However, this straightforward approach is computationally inefficient as the total number of RSRG-X states grows exponentially with system sizes. We simplify our approach to various extents respectively for the nearest-neighbor models in Appendix~\ref{appendix:numerics:RSRGX:NN} and for the general interacting models --- including specifically chains with interactions beyond the nearest neighbors~\eqref{equ:overview:model} and the two-dimensional system~\eqref{equ:H_2d} --- in Appendix~\ref{appendix:numerics:RSRGX:LR}.

\subsubsection{Nearest-Neighbor Models}
\label{appendix:numerics:RSRGX:NN}
For nearest-neighbor models~\eqref{equ:H_NN}, the results we have presented (Sec.~\ref{sec:NN:numerics}) chiefly concern the spin relaxation dynamics. To this end, we simplify the numerical RSRG-X simulation by keeping track of, rather than the $\Heff$, the magnitudes of interaction strengths. Indeed, these magnitudes determine how themselves are updated along the RG steps, and such updates are also independent of the branching choices~\eqref{equ:review:RG_step}. Moreover, these magnitudes determine the pairing $D$ of sites as well as the interaction strength $|\Omega_\alpha|$ for every pair $\alpha$ in $D$. These pieces of information are sufficient for computing $\mathcal{S}_p(\omega)$ (\ref{equ:S_p_freq_D}) for every realization of the randomness.

\subsubsection{Beyond-Nearest-Neighbor and Two-Dimensional Models}
\label{appendix:numerics:RSRGX:LR}
For the models with interactions beyond the nearest neighbors~\eqref{equ:overview:model} or in two dimensions~\eqref{equ:H_2d}, $\Heff$ contains so much more types of interactions (Sec.~\ref{sec:lr:RG:subsequent_steps}) that there is no known simpler approach than keeping track of $\Heff$ explicitly. We do so by storing $\Heff$ as a sum of numerically weighted Pauli strings in the computer program. In practice, many terms in $\Heff$ are too weak to affect the dynamics we concern. We can therefore ignore the generation of terms whose magnitude is larger than a designated threshold $\epsilon_\mathrm{thres} = 10^{-12}$. By comparing results at various $\epsilon_\mathrm{thres}$, we ensure that the data presented in this work are insensitive to our choice of such threshold.

At every RG step, we update $\Heff$ based on a branching choice. Having described in Sec.~\ref{sec:lr:RG:subsequent_steps} all the types RG steps and branching choices to perform, we realize these steps in our computer program following the perturbative treatments detailed in Appendix~\ref{appendix:pseudo_codes}. The final RG step outputs an RSRG-X state (\ref{equ:RSRGX_state}).

Within the assumption of strong randomness, we have argued that the autocorrelation function can be approximated using Eqs.~(\ref{equ:Cjn_sum_m},~\ref{equ:Cnj_results}). The approximation leads to a spectral function

\begin{widetext}
    \eq{
    \overline{\mathcal{S}_p}(\omega) = \frac{1}{2}\delta(\omega) + \left\langle \frac{1}{4N}\sum_{j=1}^N \begin{cases}
        \delta(4\Omega_{n, j}-\omega) + \delta(4\Omega_{n, j}+\omega) & j\;\mathrm{on \; resonant\; part\; of} \ket{n} \\
        2\delta(\omega) & j\;\mathrm{on \; frozen\; part\; of} \ket{n}
    \end{cases}\right\rangle\label{equ:appendix:S_p_averaging}
    }
\end{widetext}
where $\braket{\cdots}$ denotes an average over all RSRG-X states for each realization of the randomness, as well as over these realizations. As discussed in the main text, in practice we perform these averages via Monte Carlo (Sec.~\ref{sec:LR:numerics}). 

\section{Detailed Perturbative Treatments in Superspin RSRG-X}
\label{appendix:pseudo_codes}

Whereas in Sec.~\ref{sec:lr:RG:subsequent_steps} we have listed the consequent superspin formation for all the possible branching choices in our RSRG-X formalism for models with beyond-nearest-neighbor interactions~\eqref{equ:overview:model}, in this section we describe the technical perturbative treatments we adopt to update the effective Hamiltonian $\Heff$ for every choice. With such description, a computer program capable of computing multiplication of Pauli strings can carry out the RSRG-X numerically; we remark on further implementation details in Appendix~\ref{appendix:numerics:RSRGX:LR}. In the following Appendix~\ref{sec:pseudo_codes:general}, we lay out a general framework with which we describe the perturbative treatments, and in Appendix~\ref{sec:pseudo_codes:detailed} we supply detailed description for every branching choice.

\subsection{General Perturbative Treatment}
\label{sec:pseudo_codes:general}

We outline our general strategy before detailing the perturbative treatment for every choice of branching subspace $\mathcal{V}$. Our goal is to describe the updated Hamiltonian $\Heff^{(\mathcal{V})}$ due to branching into $\mathcal{V}$, given the $\Heff$ an RG step begins with: as in Equ.~\eqref{equ:H_separation}, we separate $\Heff$ into the strong interaction $H_0$, the interactions $H_1$ (other than $H_0$) that overlap with $H_0$ at some site, and the interactions $H_\perp$ that do not involve any sites supporting $H_0$. For convenience, let $H_0$ be supported on superspin site(s) indexed by an index set $\mathbf{S}$ [e.g., $\mathbf{S} = \{1, 2\}$ in cases (i), (ii), and $\mathbf{S} = \{1\}$ in case (iii) in Sec.~\ref{sec:lr:RG:subsequent_steps}]. It follows that $H_\perp$ is supported on $\overline{\mathbf{S}}$, by which we denote all the other superspin sites. Eventually, we express $\Heff^{(\mathcal{V})}$ in separate terms as are contributed by the perturbation theory~\eqref{equ:degen_perturb},
\eq{
    \Heff^{(\mathcal{V})} = E_0 + H_\mathrm{I} + H_\mathrm{II} + H_\perp. \label{equ:pseudo_codes:general:results}
}
Here, $H_\perp$ directly enters $\Heff^{(\mathcal{V})}$ as the sites $\overline{\mathbf{S}}$ are unaffected by the perturbative treatment, $E_0$ is the zeroth-order energy i.e. the energy of $H_0$ in $\mathcal{V}$, and $H_\mathrm{I}$ ($H_\mathrm{II}$) is the first- (second-) order contribution resulting from the perturbative interactions $H_1$.

Note that, in our formalism, the eigenspaces of $H_0$, when standalone, are either one- or two-fold degenerate (Sec.~\ref{sec:lr:RG:subsequent_steps}). We denote the standalone eigenstates of $H_0$ generally by $\ket{\chi_\alpha}$ with respective zeroth-order energies $E_\alpha = \braket{\chi_\alpha | H_0 | \chi_\alpha}$. The branching subspace $\mathcal{V}$ then corresponds to either one or two states among $\ket{\chi_\alpha}$: we discuss each situation as follows.

\subsubsection{No Degeneracy}
\label{sec:pseudo_codes:general:1}
Without loss of generality, now we denote the branching subspace $\mathcal{V}$ by
\eq{
    \mathcal{V} = \ket{\chi_1} \bra{\chi_1} \otimes \mathcal{H}_{\overline{\mathbf{S}}}.
}
Here, $\mathcal{H}_{\overline{\mathbf{S}}}$ is the Hilbert space on sites $\overline{\mathbf{S}}$, which are unaffected by the strong interaction $H_0$. The perturbation theory~\eqref{equ:degen_perturb} for deriving $\Heff^{(\mathcal{V})}$ involves projection into the subspace $\mathcal{V}$, facilitated by a projector $P = \ket{\chi_1}\bra{\chi_1} \otimes I_{\overline{\mathbf{S}}}$, where we have included the identity operator $I_{\overline{\mathbf{S}}}$ on $\mathcal{H}_{\overline{\mathbf{S}}}$.

The first- and second-order terms, $H_\mathrm{I}$ and $H_\mathrm{II}$, in our construction~\eqref{equ:pseudo_codes:general:results}
then follow in the form of partial traces over the old site(s) $\mathbf{S}$, which is then decimated,
\eq{
    H_\mathrm{I} = \tr_{\mathbf{S}} \left(P H_1\right), \quad
    H_\mathrm{II} = \tr_{\mathbf{S}} \left[P H_1 \left(K \otimes I_{\overline{\mathbf{S}}}\right) H_1\right].
}
In the second expression, $K$ accounts for virtual processes through all the other eigenstates, which in the present case are all gapped from $\mathcal{V}$,
\eq{
    K = \sum_{\alpha \neq 1} \frac{1}{E_0 - E_\alpha} \ket{\chi_\alpha} \bra{\chi_\alpha}.
}

\subsubsection{Two-fold Degeneracy}
\label{sec:pseudo_codes:general:2}
In this case, we have without loss of generality
\eq{
    \mathcal{V} = \mathcal{H}_\mathrm{new} \otimes \mathcal{H}_{\overline{\mathbf{S}}}, \quad \mathcal{H}_\mathrm{new} = \mathrm{span}\left\{\ket{\chi_{\alpha}}\right\}_{\alpha=1, 2}, \label{equ:pseudo_codes:general:V_structure_2}
}
where $\mathcal{H}_{\overline{\mathbf{S}}}$ denotes as above the Hilbert space on sites $\overline{\mathbf{S}}$ untouched by $H_0$, and $\mathcal{H}_\mathrm{new}$
is the superspin degrees of freedom on the new site (Sec.~\ref{sec:LR:RG:superspin}), on which the Pauli operators $\tilde{X}$, $\tilde{Y}$, $\tilde{Z}$ and the identity $\tilde{I}$ are defined, e.g., $\tilde{Y} = -i \ket{\chi_1}\bra{\chi_2} + h.c$. To express the updated Hamiltonian~\eqref{equ:pseudo_codes:general:results} in terms of the Pauli strings acting on the branching subspace $\mathcal{V}$ following its tensor-product structure~\eqref{equ:pseudo_codes:general:V_structure_2}, for each contribution $\mu\in\left\{\mathrm{I}, \mathrm{II}\right\}$ we decompose
\eq{
H_\mu = \tilde{X} \otimes H_\mu^{X} + \tilde{Y} \otimes H_\mu^{Y} + \tilde{Z} \otimes H_\mu^{Z} + \tilde{I} \otimes H_\mu^{I}.\label{equ:pseudo_codes:general:decomposition}
}
Generally, we can find $H_\mu^\sigma$ for $\sigma\in\left\{X, Y, Z, I\right\}$ in this decomposition~\eqref{equ:pseudo_codes:general:decomposition} by partial trace $H_\mu^\sigma = \frac{1}{2}\tr_{\mathbf{S}}\left(P_\sigma H_\mu\right)$ over the old site(s) $\mathbf{S}$, which is then replaced by the new site. Here, $P_\sigma = \tilde{\sigma} \otimes I_{\overline{\mathbf{S}}}$ is the corresponding projector, and we have included $I_{\overline{\mathbf{S}}}$, the identity acting on the unaffected Hilbert space $\mathcal{H}_{\overline{\mathbf{S}}}$.

Next, we identify $H_\mathrm{I}$ and $H_\mathrm{II}$  respectively in the first and second orders of the  perturbation theory~\eqref{equ:degen_perturb},
\eq{
    H_\mathrm{I} = \sum_{\sigma\in\left\{X,Y,Z,I\right\}} \tilde{\sigma}\otimes \frac{1}{2} \tr_\mathbf{S} \left(P_\sigma H_1\right),
}
and the second-order results follow similarly,
\eq{
    H_\mathrm{II} = \sum_{\sigma\in\left\{X,Y,Z,I\right\}} \tilde{\sigma}\otimes \frac{1}{2} \tr_\mathbf{S} \left[P_\sigma H_1 \left(K\otimes I_{\overline{\mathbf{S}}}\right) H_1\right].
}
Here, $K$ accounts for the virtual processes through the $H_0$ eigenstates $\left\{\ket{\chi_\alpha}\right\}_{\alpha\neq 1, 2}$ gapped from $\mathcal{V}$,
\eq{
    K = \sum_{\alpha\neq 1, 2} \frac{1}{E_0 - E_\alpha} \ket{\chi_\alpha} \bra{\chi_\alpha}.
}

\subsection{Detailed Treatments}
\label{sec:pseudo_codes:detailed}
With the general strategy (Appendix~\ref{sec:pseudo_codes:general}), the derivation of detailed perturbative treatments for every branching choice is straightforward. We elaborate on the results here, organized according to the three cases of strong interaction $H_0$ listed in Sec.~\ref{sec:lr:RG:subsequent_steps}.

For case (i), $H_0 = \Omega\left(X_1^{(m_1)}X_2^{(m_2)}+Y_1^{(m_1)}Y_2^{(m_2)}\right)$. Denote $m = m_1 = m_2$ from now on. The local eigenstates of $H_0$ with respective energies $E_0$ are, organized by branching subspaces,

\begin{alignat}{4}
    &V_+ &:&\quad \ket{\chi_+} &&= \frac{1}{\sqrt{2}}\left(\ket{\Uparrow_1^{m} \Downarrow_2^{m}} + \ket{\Downarrow_1^{m} \Uparrow_2^{m}}\right);&&\quad E_0 = 2\Omega \nonumber \\
    &V_0 &:&\quad \ket{\chi_{\uparrow\uparrow}} &&= \ket{\Uparrow_1^{m}\Uparrow_2^{m}},
    \;\ket{\chi_{\downarrow\downarrow}} = \ket{\Downarrow_1^{m}\Downarrow_2^{m}};&&\quad E_0 = 0 \nonumber \\
    &V_- &:&\quad \ket{\chi_-} &&= \frac{1}{\sqrt{2}}\left(\ket{\Uparrow_1^{m} \Downarrow_2^{m}} + \ket{\Downarrow_1^{m} \Uparrow_2^{m}}\right);&&\quad E_0 = -2\Omega.
\end{alignat}
Branching into either of $V_\pm$ corresponding to a non-degenerate local eigenstate follows the approach outlined in Sec.~\ref{sec:pseudo_codes:general:1}, with specifically $\ket{\chi_1}$ set to $\ket{\chi_\pm}$ and
\newcommand{\chidd}{\chi_{\downarrow\downarrow}}
\newcommand{\chiuu}{\chi_{\uparrow\uparrow}}
\newcommand{\chiud}{\chi_{\uparrow\downarrow}}
\newcommand{\chidu}{\chi_{\downarrow\uparrow}}
\eq{
    K = \pm \frac{1}{2\Omega} \left(\ket{\chiuu}\bra{\chiuu} + \ket{\chidd}\bra{\chidd}\right) \pm \frac{1}{4\Omega} \ket{\chi_\mp}\bra{\chi_\mp}.
}
For the branching choice of $V_0$, we follow Sec.~\ref{sec:pseudo_codes:general:2} where we set $\ket{\chi_1} = \ket{\chiuu}$, $\ket{\chi_2} = \ket{\chidd}$, and
\eq{
    K = \frac{1}{2\Omega} \left(\ket{\chi_-}\bra{\chi_-}-\ket{\chi_+}\bra{\chi_+}\right).
}

For case (ii), $H_0 = \Omega Z^{(m_1)}_1 Z^{(m_2)}_2$, assuming without loss of generality $m_1 \ge m_2$. The local spectrum of $H_0$ is, organized corresponding to the branching subspaces,
\begin{alignat}{5}
    & V'_+ &:\quad& \ket{\chiuu} &&= \ket{\Uparrow_1^{m_1}\Uparrow_2^{m_2}}&,\;& 
    \ket{\chidd} = \ket{\Downarrow_1^{m_1}\Downarrow_2^{m_2}} &;\quad& E_0 = \Omega \nonumber \\
    & V'_- &:\quad& \ket{\chiud} &&= \ket{\Uparrow_1^{m_1}\Downarrow_2^{m_2}}&,\;& 
    \ket{\chidu} = \ket{\Downarrow_1^{m_1}\Uparrow_2^{m_2}} &;\quad& E_0 = -\Omega. \nonumber
\end{alignat}
One may branch into $V_+'$ following Sec.~\ref{sec:pseudo_codes:general:2} with specifically $\ket{\chi_1} = \ket{\chiuu}$, $\ket{\chi_2} = \ket{\chidd}$, and 
\eq{
    K = \frac{1}{2\Omega}\left(\ket{\chiud}\bra{\chiud} + \ket{\chidu}\bra{\chidu} \right).
}
As for another branching choice $V_-'$, one may set $\ket{\chi_1} = \ket{\chiud}, \ket{\chi_2} = \ket{\chidu}$, and
\eq{
    K = -\frac{1}{2\Omega}\left(\ket{\chiuu}\bra{\chiuu} + \ket{\chidd}\bra{\chidd} \right).
}

For case (iii), $H_0 = 2\Omega X_1^{(0)}$ has a non-degenerate local eigenstate for every branching subspace,
\eq{
U_\pm:\quad \ket{\chi_\pm'} = \frac{1}{\sqrt{2}} \left(\ket{\Uparrow^0_1} \pm \ket{\Downarrow_1^0}\right);\quad E_0 = \pm 2\Omega.
}
Branching into each of $U_\pm$ then amounts to following the procedure detailed in Sec.~\ref{sec:pseudo_codes:general:1}, with $\ket{\chi_1} = \ket{\chi_\pm'}$ and 
\eq{
    K = \pm \frac{1}{4\Omega} \ket{\chi_\mp'}\bra{\chi_\mp'}.
}

\section{Testing the RG Premises}
\label{appendix:premises}
Our RSRG-X formalism for the randomly interacting $XX+YY$ models relies on the perturbative treatments in the RG steps to update $\Heff$ (Sec.~\ref{sec:lr:RG:subsequent_steps}). Such treatments are valid if the strong interaction $H_0$ we pick is much larger than $H_1$, the overlapping interactions~\eqref{equ:H_separation}, which nontrivially enter the perturbation theory~\eqref{equ:degen_perturb}. In this appendix, we present numerical evidence that this premise is better satisfied at later RG steps by all the models~(\ref{equ:overview:model}, \ref{equ:H_2d}) we have studied in one and two dimensions.

To quantify the extent to which this premise is satisfied, we consider what we define to be the \textit{RG quality} at a given RG step
\eq{
    Q = \log\left(\frac{\left\Vert H_0\right\Vert}{\left\Vert H_1\right\Vert}\right), \label{equ:RG_quality}
}
where $\left\Vert \cdots \right\Vert$ denotes the operator (infinity) norm, as usual. The larger is $Q$, the better controlled is the perturbation theory~\eqref{equ:degen_perturb} carried out at this step. We then investigate, during the Monte Carlo sampling of RSRG-X states (Sec.~\ref{sec:LR:numerics}), what the probability is to encounter a step of quality $Q$. 

\begin{table}
    \begin{ruledtabular}
        \begin{tabular}{crd}
        Model & $N$& \mbox{Rate}\\
        \hline
            & 16 & .000320(10) \\
         LR & 24 & .000243(8) \\
            & 32 & .000193(6) \\
        \hline
            & 16 & .000436(16) \\
        NNN & 24 & .000336(17) \\
            & 32 & .000344(12) \\
        \hline 
        2D, $\alpha=3$ & 16 & .01085(11) \\
        2D, $\alpha=4$ & 16 & .00901(11) \\
        2D, $\alpha=6$ & 16 & .00808(12) \\
        2D, $\alpha=6$ & 24 & .00753(10) \\
        \end{tabular}
    \end{ruledtabular}
    \caption{Observed rates at which, at an RG step, the $H_0$ we have identified and treated as the leading term in the perturbative analysis overlaps with an actually stronger interaction. The uncertainties in parentheses are estimated by the square root of counts divided by the total steps.}
    \label{tab:multi_site_H0}
\end{table}

\begin{figure*}
    \includegraphics[width=0.32\textwidth]{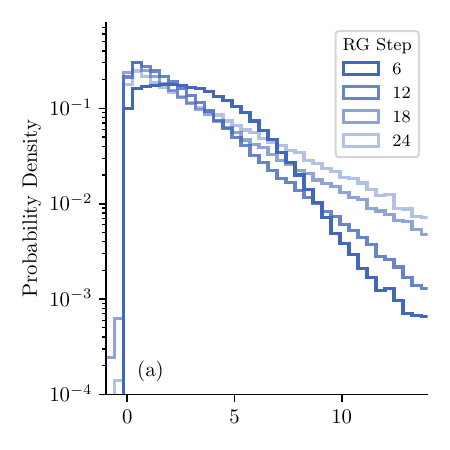}
    \includegraphics[width=0.32\textwidth]{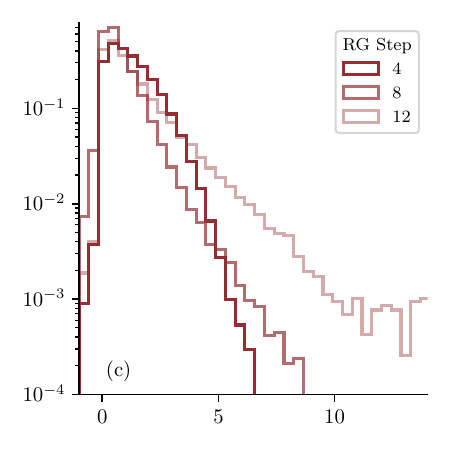}
    \includegraphics[width=0.32\textwidth]{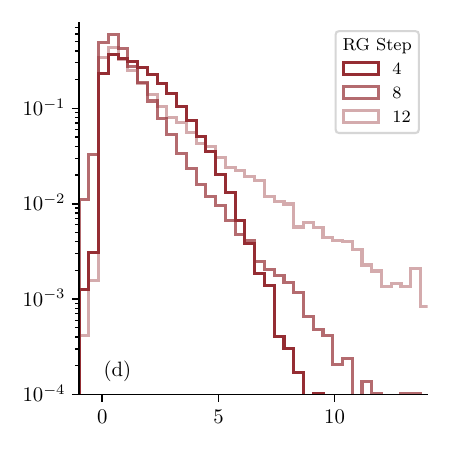} \\
    \includegraphics[width=0.32\textwidth]{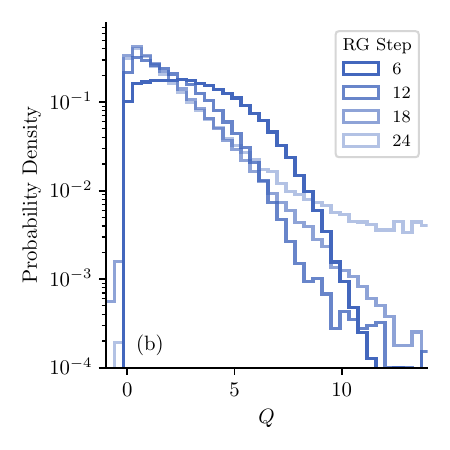}
    \includegraphics[width=0.32\textwidth]{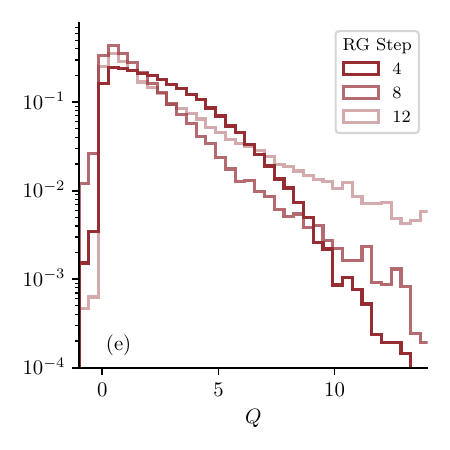}
    \includegraphics[width=0.32\textwidth]{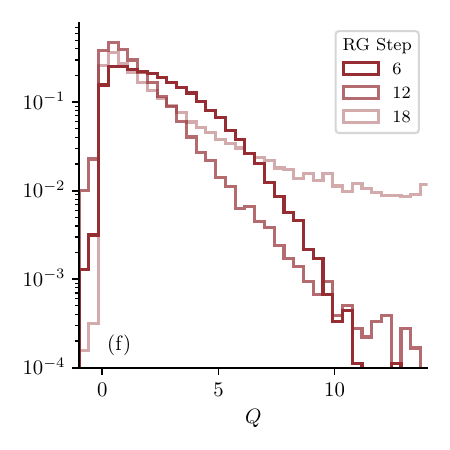}

    \caption{Probability distribution, across over $8\times 10^4$ realizations of the randomness, of the RG quality $Q$~\eqref{equ:RG_quality} at various RG steps for the random systems we have studied: the one-dimensional chains~\eqref{equ:overview:model} at $N=32$ with (a) the next-nearest-neighbor and (b) the long-range interactions, and the two-dimensional systems~\eqref{equ:H_2d} with interactions $\sim r^{-\alpha}$ at $N=16$ with (c) $\alpha = 3$, (d) $\alpha=4$, (e) $\alpha=6$, and (f) at $N=24$ with $\alpha=6$.} \label{fig:RG_qual}
\end{figure*}

We illustrate the probability distribution of $Q$ at enlarging RG steps in Fig.~\ref{fig:RG_qual} for the several models we have considered. Remarkably, we observe heavier tails of large $Q$ especially at later RG steps $\gtrsim N/2$. This observation suggests that the perturbative treatments are indeed better controlled as RSRG-X proceeds at late stages. Since the energy scale of $\Heff$ meanwhile decreases at later RG steps, this observation supports our general approach of applying RSRG-X to the prediction of low-frequency, i.e., late-time, dynamics.

Fig.~\ref{fig:RG_qual} also shows that it is possible to have $Q < 0$, or equivalently $\left\Vert H_0 \right\Vert < \left\Vert H_1 \right\Vert$. This feature results from our design (Sec.~\ref{sec:lr:RG:subsequent_steps}) of identifying $H_0$ only among the three kinds of interactions listed explicitly in Equ.~\eqref{equ:LR:RG_Heff}. Admittedly, such design may potentially lead to untrustworthy perturbative treatments if, among the overlapping interactions $H_1$, there is some interaction that is both stronger than $H_0$ and of a form other than these three kinds, resulting in $Q<0$. Such an interaction, for concreteness, may take a form of $X^{(0)} X^{(0)}$ and $Y^{(0)} Y^{(0)}$ between superspin-0 sites, or of interactions involving three or more sites (Sec.~\ref{sec:lr:RG:subsequent_steps}). Based on the data we have gathered, however, the probability for an RG step to encounter $Q<0$ is smaller than $0.0005$ for the one-dimensional models~\eqref{equ:overview:model} and than $0.011$ in two dimensions~\eqref{equ:H_2d}; see Table~\ref{tab:multi_site_H0}. Additionally, such a probability generally decreases at enlarging system sizes to which we have access (Table~\ref{tab:multi_site_H0}). These observations support our design of the RSRG-X formalism (Sec.~\ref{sec:lr:RG:subsequent_steps}).

We note that, still, for the two-dimensional systems the probability to encounter $Q<0$ at an RG step is much larger  than for the one-dimensional chains (Table~\ref{tab:multi_site_H0}). Among the two-dimensional systems themselves, this probability is larger at lower $\alpha$, where we expect weaker initial randomness. Indeed, we also find less spread to high RG qualities at lower $\alpha$ in Fig.~\ref{fig:RG_qual}. These observations suggest worse fulfillment of the RG premise (Sec.~\ref{sec:review:generalities}) for perturbative treatments in two dimensions, especially at smaller $\alpha\lesssim 4$. Such worse fulfillment aligns with the weaker initial randomness and also with the worse agreement between the ED and numerical RSRG-X results (Sec.~\ref{sec:2d}).

% End of appendices.

% \nocite{*}

% \bibliography{ref}% Produces the bibliography via BibTeX.

\begin{thebibliography}{61}%
    \makeatletter
    \providecommand \@ifxundefined [1]{%
     \@ifx{#1\undefined}
    }%
    \providecommand \@ifnum [1]{%
     \ifnum #1\expandafter \@firstoftwo
     \else \expandafter \@secondoftwo
     \fi
    }%
    \providecommand \@ifx [1]{%
     \ifx #1\expandafter \@firstoftwo
     \else \expandafter \@secondoftwo
     \fi
    }%
    \providecommand \natexlab [1]{#1}%
    \providecommand \enquote  [1]{``#1''}%
    \providecommand \bibnamefont  [1]{#1}%
    \providecommand \bibfnamefont [1]{#1}%
    \providecommand \citenamefont [1]{#1}%
    \providecommand \href@noop [0]{\@secondoftwo}%
    \providecommand \href [0]{\begingroup \@sanitize@url \@href}%
    \providecommand \@href[1]{\@@startlink{#1}\@@href}%
    \providecommand \@@href[1]{\endgroup#1\@@endlink}%
    \providecommand \@sanitize@url [0]{\catcode `\\12\catcode `\$12\catcode `\&12\catcode `\#12\catcode `\^12\catcode `\_12\catcode `\%12\relax}%
    \providecommand \@@startlink[1]{}%
    \providecommand \@@endlink[0]{}%
    \providecommand \url  [0]{\begingroup\@sanitize@url \@url }%
    \providecommand \@url [1]{\endgroup\@href {#1}{\urlprefix }}%
    \providecommand \urlprefix  [0]{URL }%
    \providecommand \Eprint [0]{\href }%
    \providecommand \doibase [0]{https://doi.org/}%
    \providecommand \selectlanguage [0]{\@gobble}%
    \providecommand \bibinfo  [0]{\@secondoftwo}%
    \providecommand \bibfield  [0]{\@secondoftwo}%
    \providecommand \translation [1]{[#1]}%
    \providecommand \BibitemOpen [0]{}%
    \providecommand \bibitemStop [0]{}%
    \providecommand \bibitemNoStop [0]{.\EOS\space}%
    \providecommand \EOS [0]{\spacefactor3000\relax}%
    \providecommand \BibitemShut  [1]{\csname bibitem#1\endcsname}%
    \let\auto@bib@innerbib\@empty
    %</preamble>
    \bibitem [{\citenamefont {Choi}\ \emph {et~al.}(2017)\citenamefont {Choi}, \citenamefont {Choi}, \citenamefont {Kucsko}, \citenamefont {Maurer}, \citenamefont {Shields}, \citenamefont {Sumiya}, \citenamefont {Onoda}, \citenamefont {Isoya}, \citenamefont {Demler}, \citenamefont {Jelezko}, \citenamefont {Yao},\ and\ \citenamefont {Lukin}}]{Choi17}%
      \BibitemOpen
      \bibfield  {author} {\bibinfo {author} {\bibfnamefont {J.}~\bibnamefont {Choi}}, \bibinfo {author} {\bibfnamefont {S.}~\bibnamefont {Choi}}, \bibinfo {author} {\bibfnamefont {G.}~\bibnamefont {Kucsko}}, \bibinfo {author} {\bibfnamefont {P.~C.}\ \bibnamefont {Maurer}}, \bibinfo {author} {\bibfnamefont {B.~J.}\ \bibnamefont {Shields}}, \bibinfo {author} {\bibfnamefont {H.}~\bibnamefont {Sumiya}}, \bibinfo {author} {\bibfnamefont {S.}~\bibnamefont {Onoda}}, \bibinfo {author} {\bibfnamefont {J.}~\bibnamefont {Isoya}}, \bibinfo {author} {\bibfnamefont {E.}~\bibnamefont {Demler}}, \bibinfo {author} {\bibfnamefont {F.}~\bibnamefont {Jelezko}}, \bibinfo {author} {\bibfnamefont {N.~Y.}\ \bibnamefont {Yao}},\ and\ \bibinfo {author} {\bibfnamefont {M.~D.}\ \bibnamefont {Lukin}},\ }\bibfield  {title} {\bibinfo {title} {Depolarization dynamics in a strongly interacting solid-state spin ensemble},\ }\href {https://doi.org/10.1103/PhysRevLett.118.093601} {\bibfield  {journal} {\bibinfo  {journal} {Phys. Rev. Lett.}\ }\textbf {\bibinfo {volume} {118}},\ \bibinfo {pages} {093601} (\bibinfo {year} {2017})}\BibitemShut {NoStop}%
    \bibitem [{\citenamefont {Kucsko}\ \emph {et~al.}(2018)\citenamefont {Kucsko}, \citenamefont {Choi}, \citenamefont {Choi}, \citenamefont {Maurer}, \citenamefont {Zhou}, \citenamefont {Landig}, \citenamefont {Sumiya}, \citenamefont {Onoda}, \citenamefont {Isoya}, \citenamefont {Jelezko}, \citenamefont {Demler}, \citenamefont {Yao},\ and\ \citenamefont {Lukin}}]{Kucsko18}%
      \BibitemOpen
      \bibfield  {author} {\bibinfo {author} {\bibfnamefont {G.}~\bibnamefont {Kucsko}}, \bibinfo {author} {\bibfnamefont {S.}~\bibnamefont {Choi}}, \bibinfo {author} {\bibfnamefont {J.}~\bibnamefont {Choi}}, \bibinfo {author} {\bibfnamefont {P.~C.}\ \bibnamefont {Maurer}}, \bibinfo {author} {\bibfnamefont {H.}~\bibnamefont {Zhou}}, \bibinfo {author} {\bibfnamefont {R.}~\bibnamefont {Landig}}, \bibinfo {author} {\bibfnamefont {H.}~\bibnamefont {Sumiya}}, \bibinfo {author} {\bibfnamefont {S.}~\bibnamefont {Onoda}}, \bibinfo {author} {\bibfnamefont {J.}~\bibnamefont {Isoya}}, \bibinfo {author} {\bibfnamefont {F.}~\bibnamefont {Jelezko}}, \bibinfo {author} {\bibfnamefont {E.}~\bibnamefont {Demler}}, \bibinfo {author} {\bibfnamefont {N.~Y.}\ \bibnamefont {Yao}},\ and\ \bibinfo {author} {\bibfnamefont {M.~D.}\ \bibnamefont {Lukin}},\ }\bibfield  {title} {\bibinfo {title} {Critical thermalization of a disordered dipolar spin system in diamond},\ }\href {https://doi.org/10.1103/PhysRevLett.121.023601} {\bibfield  {journal} {\bibinfo  {journal} {Phys. Rev. Lett.}\ }\textbf {\bibinfo {volume} {121}},\ \bibinfo {pages} {023601} (\bibinfo {year} {2018})}\BibitemShut {NoStop}%
    \bibitem [{\citenamefont {Zu}\ \emph {et~al.}(2021)\citenamefont {Zu}, \citenamefont {Machado}, \citenamefont {Ye}, \citenamefont {Choi}, \citenamefont {Kobrin}, \citenamefont {Mittiga}, \citenamefont {Hsieh}, \citenamefont {Bhattacharyya}, \citenamefont {Markham}, \citenamefont {Twitchen}, \citenamefont {Jarmola}, \citenamefont {Budker}, \citenamefont {Laumann}, \citenamefont {Moore},\ and\ \citenamefont {Yao}}]{Zu21}%
      \BibitemOpen
      \bibfield  {author} {\bibinfo {author} {\bibfnamefont {C.}~\bibnamefont {Zu}}, \bibinfo {author} {\bibfnamefont {F.}~\bibnamefont {Machado}}, \bibinfo {author} {\bibfnamefont {B.}~\bibnamefont {Ye}}, \bibinfo {author} {\bibfnamefont {S.}~\bibnamefont {Choi}}, \bibinfo {author} {\bibfnamefont {B.}~\bibnamefont {Kobrin}}, \bibinfo {author} {\bibfnamefont {T.}~\bibnamefont {Mittiga}}, \bibinfo {author} {\bibfnamefont {S.}~\bibnamefont {Hsieh}}, \bibinfo {author} {\bibfnamefont {P.}~\bibnamefont {Bhattacharyya}}, \bibinfo {author} {\bibfnamefont {M.}~\bibnamefont {Markham}}, \bibinfo {author} {\bibfnamefont {D.}~\bibnamefont {Twitchen}}, \bibinfo {author} {\bibfnamefont {A.}~\bibnamefont {Jarmola}}, \bibinfo {author} {\bibfnamefont {D.}~\bibnamefont {Budker}}, \bibinfo {author} {\bibfnamefont {C.~R.}\ \bibnamefont {Laumann}}, \bibinfo {author} {\bibfnamefont {J.~E.}\ \bibnamefont {Moore}},\ and\ \bibinfo {author} {\bibfnamefont {N.~Y.}\ \bibnamefont {Yao}},\ }\bibfield  {title} {{\selectlanguage {english}\bibinfo {title} {Emergent hydrodynamics in a strongly interacting dipolar spin ensemble}},\ }\href {https://doi.org/10.1038/s41586-021-03763-1} {\bibfield  {journal} {\bibinfo  {journal} {Nature}\ }\textbf {\bibinfo {volume} {597}},\ \bibinfo {pages} {45} (\bibinfo {year} {2021})}\BibitemShut {NoStop}%
    \bibitem [{\citenamefont {Davis}\ \emph {et~al.}(2023)\citenamefont {Davis}, \citenamefont {Ye}, \citenamefont {Machado}, \citenamefont {Meynell}, \citenamefont {Wu}, \citenamefont {Mittiga}, \citenamefont {Schenken}, \citenamefont {Joos}, \citenamefont {Kobrin}, \citenamefont {Lyu}, \citenamefont {Wang}, \citenamefont {Bluvstein}, \citenamefont {Choi}, \citenamefont {Zu}, \citenamefont {Jayich},\ and\ \citenamefont {Yao}}]{Davis23}%
      \BibitemOpen
      \bibfield  {author} {\bibinfo {author} {\bibfnamefont {E.~J.}\ \bibnamefont {Davis}}, \bibinfo {author} {\bibfnamefont {B.}~\bibnamefont {Ye}}, \bibinfo {author} {\bibfnamefont {F.}~\bibnamefont {Machado}}, \bibinfo {author} {\bibfnamefont {S.~A.}\ \bibnamefont {Meynell}}, \bibinfo {author} {\bibfnamefont {W.}~\bibnamefont {Wu}}, \bibinfo {author} {\bibfnamefont {T.}~\bibnamefont {Mittiga}}, \bibinfo {author} {\bibfnamefont {W.}~\bibnamefont {Schenken}}, \bibinfo {author} {\bibfnamefont {M.}~\bibnamefont {Joos}}, \bibinfo {author} {\bibfnamefont {B.}~\bibnamefont {Kobrin}}, \bibinfo {author} {\bibfnamefont {Y.}~\bibnamefont {Lyu}}, \bibinfo {author} {\bibfnamefont {Z.}~\bibnamefont {Wang}}, \bibinfo {author} {\bibfnamefont {D.}~\bibnamefont {Bluvstein}}, \bibinfo {author} {\bibfnamefont {S.}~\bibnamefont {Choi}}, \bibinfo {author} {\bibfnamefont {C.}~\bibnamefont {Zu}}, \bibinfo {author} {\bibfnamefont {A.~C.~B.}\ \bibnamefont {Jayich}},\ and\ \bibinfo {author} {\bibfnamefont {N.~Y.}\ \bibnamefont {Yao}},\ }\bibfield  {title} {{\selectlanguage {english}\bibinfo {title} {Probing many-body dynamics in a two-dimensional dipolar spin ensemble}},\ }\href {https://doi.org/10.1038/s41567-023-01944-5} {\bibfield  {journal} {\bibinfo  {journal} {Nat. Phys.}\ }\textbf {\bibinfo {volume} {19}},\ \bibinfo {pages} {836} (\bibinfo {year} {2023})}\BibitemShut {NoStop}%
    \bibitem [{\citenamefont {Browaeys}\ and\ \citenamefont {Lahaye}(2020)}]{Browaeys_review}%
      \BibitemOpen
      \bibfield  {author} {\bibinfo {author} {\bibfnamefont {A.}~\bibnamefont {Browaeys}}\ and\ \bibinfo {author} {\bibfnamefont {T.}~\bibnamefont {Lahaye}},\ }\bibfield  {title} {\bibinfo {title} {Many-body physics with individually controlled {Rydberg} atoms},\ }\href {https://doi.org/10.1038/s41567-019-0733-z} {\bibfield  {journal} {\bibinfo  {journal} {Nat. Phys.}\ }\textbf {\bibinfo {volume} {16}},\ \bibinfo {pages} {132} (\bibinfo {year} {2020})}\BibitemShut {NoStop}%
    \bibitem [{\citenamefont {Su}\ \emph {et~al.}(2023)\citenamefont {Su}, \citenamefont {Douglas}, \citenamefont {Szurek}, \citenamefont {Groth}, \citenamefont {Ozturk}, \citenamefont {Krahn}, \citenamefont {H{\'e}bert}, \citenamefont {Phelps}, \citenamefont {Ebadi}, \citenamefont {Dickerson}, \citenamefont {Ferlaino}, \citenamefont {Markovi{\'c}},\ and\ \citenamefont {Greiner}}]{Su23}%
      \BibitemOpen
      \bibfield  {author} {\bibinfo {author} {\bibfnamefont {L.}~\bibnamefont {Su}}, \bibinfo {author} {\bibfnamefont {A.}~\bibnamefont {Douglas}}, \bibinfo {author} {\bibfnamefont {M.}~\bibnamefont {Szurek}}, \bibinfo {author} {\bibfnamefont {R.}~\bibnamefont {Groth}}, \bibinfo {author} {\bibfnamefont {S.~F.}\ \bibnamefont {Ozturk}}, \bibinfo {author} {\bibfnamefont {A.}~\bibnamefont {Krahn}}, \bibinfo {author} {\bibfnamefont {A.~H.}\ \bibnamefont {H{\'e}bert}}, \bibinfo {author} {\bibfnamefont {G.~A.}\ \bibnamefont {Phelps}}, \bibinfo {author} {\bibfnamefont {S.}~\bibnamefont {Ebadi}}, \bibinfo {author} {\bibfnamefont {S.}~\bibnamefont {Dickerson}}, \bibinfo {author} {\bibfnamefont {F.}~\bibnamefont {Ferlaino}}, \bibinfo {author} {\bibfnamefont {O.}~\bibnamefont {Markovi{\'c}}},\ and\ \bibinfo {author} {\bibfnamefont {M.}~\bibnamefont {Greiner}},\ }\bibfield  {title} {{\selectlanguage {english}\bibinfo {title} {Dipolar quantum solids emerging in a {Hubbard} quantum simulator}},\ }\href {https://doi.org/10.1038/s41586-023-06614-3} {\bibfield  {journal} {\bibinfo  {journal} {Nature}\ }\textbf {\bibinfo {volume} {622}},\ \bibinfo {pages} {724} (\bibinfo {year} {2023})}\BibitemShut {NoStop}%
    \bibitem [{\citenamefont {Bloch}\ \emph {et~al.}(2023)\citenamefont {Bloch}, \citenamefont {Hofer}, \citenamefont {Cohen}, \citenamefont {Browaeys},\ and\ \citenamefont {Ferrier-Barbut}}]{bloch2023trapping}%
      \BibitemOpen
      \bibfield  {author} {\bibinfo {author} {\bibfnamefont {D.}~\bibnamefont {Bloch}}, \bibinfo {author} {\bibfnamefont {B.}~\bibnamefont {Hofer}}, \bibinfo {author} {\bibfnamefont {S.~R.}\ \bibnamefont {Cohen}}, \bibinfo {author} {\bibfnamefont {A.}~\bibnamefont {Browaeys}},\ and\ \bibinfo {author} {\bibfnamefont {I.}~\bibnamefont {Ferrier-Barbut}},\ }\bibfield  {title} {\bibinfo {title} {Trapping and imaging single {D}ysprosium atoms in optical tweezer arrays},\ }\href {https://doi.org/10.1103/PhysRevLett.131.203401} {\bibfield  {journal} {\bibinfo  {journal} {Phys. Rev. Lett.}\ }\textbf {\bibinfo {volume} {131}},\ \bibinfo {pages} {203401} (\bibinfo {year} {2023})}\BibitemShut {NoStop}%
    \bibitem [{\citenamefont {Gr\"un}\ \emph {et~al.}(2024)\citenamefont {Gr\"un}, \citenamefont {White}, \citenamefont {Ortu}, \citenamefont {Di~Carli}, \citenamefont {Edri}, \citenamefont {Lepers}, \citenamefont {Mark},\ and\ \citenamefont {Ferlaino}}]{gruen2024optical}%
      \BibitemOpen
      \bibfield  {author} {\bibinfo {author} {\bibfnamefont {D.~S.}\ \bibnamefont {Gr\"un}}, \bibinfo {author} {\bibfnamefont {S.~J.~M.}\ \bibnamefont {White}}, \bibinfo {author} {\bibfnamefont {A.}~\bibnamefont {Ortu}}, \bibinfo {author} {\bibfnamefont {A.}~\bibnamefont {Di~Carli}}, \bibinfo {author} {\bibfnamefont {H.}~\bibnamefont {Edri}}, \bibinfo {author} {\bibfnamefont {M.}~\bibnamefont {Lepers}}, \bibinfo {author} {\bibfnamefont {M.~J.}\ \bibnamefont {Mark}},\ and\ \bibinfo {author} {\bibfnamefont {F.}~\bibnamefont {Ferlaino}},\ }\bibfield  {title} {\bibinfo {title} {Optical tweezer arrays of {E}rbium atoms},\ }\href {https://doi.org/10.1103/PhysRevLett.133.223402} {\bibfield  {journal} {\bibinfo  {journal} {Phys. Rev. Lett.}\ }\textbf {\bibinfo {volume} {133}},\ \bibinfo {pages} {223402} (\bibinfo {year} {2024})}\BibitemShut {NoStop}%
    \bibitem [{\citenamefont {Cornish}\ \emph {et~al.}(2024)\citenamefont {Cornish}, \citenamefont {Tarbutt},\ and\ \citenamefont {Hazzard}}]{cornish2024quantum}%
      \BibitemOpen
      \bibfield  {author} {\bibinfo {author} {\bibfnamefont {S.~L.}\ \bibnamefont {Cornish}}, \bibinfo {author} {\bibfnamefont {M.~R.}\ \bibnamefont {Tarbutt}},\ and\ \bibinfo {author} {\bibfnamefont {K.~R.}\ \bibnamefont {Hazzard}},\ }\bibfield  {title} {\bibinfo {title} {Quantum computation and quantum simulation with ultracold molecules},\ }\href {https://doi.org/10.1038/s41567-024-02453-9} {\bibfield  {journal} {\bibinfo  {journal} {Nat. Phys.}\ }\textbf {\bibinfo {volume} {20}},\ \bibinfo {pages} {730} (\bibinfo {year} {2024})}\BibitemShut {NoStop}%
    \bibitem [{\citenamefont {Peng}\ \emph {et~al.}(2023)\citenamefont {Peng}, \citenamefont {Ye}, \citenamefont {Yao},\ and\ \citenamefont {Cappellaro}}]{Peng23}%
      \BibitemOpen
      \bibfield  {author} {\bibinfo {author} {\bibfnamefont {P.}~\bibnamefont {Peng}}, \bibinfo {author} {\bibfnamefont {B.}~\bibnamefont {Ye}}, \bibinfo {author} {\bibfnamefont {N.~Y.}\ \bibnamefont {Yao}},\ and\ \bibinfo {author} {\bibfnamefont {P.}~\bibnamefont {Cappellaro}},\ }\bibfield  {title} {{\selectlanguage {english}\bibinfo {title} {Exploiting disorder to probe spin and energy hydrodynamics}},\ }\href {https://doi.org/10.1038/s41567-023-02024-4} {\bibfield  {journal} {\bibinfo  {journal} {Nat. Phys.}\ }\textbf {\bibinfo {volume} {19}},\ \bibinfo {pages} {1027} (\bibinfo {year} {2023})}\BibitemShut {NoStop}%
    \bibitem [{\citenamefont {Franz}\ \emph {et~al.}(2024)\citenamefont {Franz}, \citenamefont {Geier}, \citenamefont {Hainaut}, \citenamefont {Braemer}, \citenamefont {Thaicharoen}, \citenamefont {Hornung}, \citenamefont {Braun}, \citenamefont {G\"arttner}, \citenamefont {Z\"urn},\ and\ \citenamefont {Weidem\"uller}}]{Titus_experiment_24}%
      \BibitemOpen
      \bibfield  {author} {\bibinfo {author} {\bibfnamefont {T.}~\bibnamefont {Franz}}, \bibinfo {author} {\bibfnamefont {S.}~\bibnamefont {Geier}}, \bibinfo {author} {\bibfnamefont {C.}~\bibnamefont {Hainaut}}, \bibinfo {author} {\bibfnamefont {A.}~\bibnamefont {Braemer}}, \bibinfo {author} {\bibfnamefont {N.}~\bibnamefont {Thaicharoen}}, \bibinfo {author} {\bibfnamefont {M.}~\bibnamefont {Hornung}}, \bibinfo {author} {\bibfnamefont {E.}~\bibnamefont {Braun}}, \bibinfo {author} {\bibfnamefont {M.}~\bibnamefont {G\"arttner}}, \bibinfo {author} {\bibfnamefont {G.}~\bibnamefont {Z\"urn}},\ and\ \bibinfo {author} {\bibfnamefont {M.}~\bibnamefont {Weidem\"uller}},\ }\bibfield  {title} {\bibinfo {title} {Observation of anisotropy-independent magnetization dynamics in spatially disordered {Heisenberg} spin systems},\ }\href {https://doi.org/10.1103/PhysRevResearch.6.033131} {\bibfield  {journal} {\bibinfo  {journal} {Phys. Rev. Res.}\ }\textbf {\bibinfo {volume} {6}},\ \bibinfo {pages} {033131} (\bibinfo {year} {2024})}\BibitemShut {NoStop}%
    \bibitem [{\citenamefont {Barredo}\ \emph {et~al.}(2015)\citenamefont {Barredo}, \citenamefont {Labuhn}, \citenamefont {Ravets}, \citenamefont {Lahaye}, \citenamefont {Browaeys},\ and\ \citenamefont {Adams}}]{Browaeys_chain}%
      \BibitemOpen
      \bibfield  {author} {\bibinfo {author} {\bibfnamefont {D.}~\bibnamefont {Barredo}}, \bibinfo {author} {\bibfnamefont {H.}~\bibnamefont {Labuhn}}, \bibinfo {author} {\bibfnamefont {S.}~\bibnamefont {Ravets}}, \bibinfo {author} {\bibfnamefont {T.}~\bibnamefont {Lahaye}}, \bibinfo {author} {\bibfnamefont {A.}~\bibnamefont {Browaeys}},\ and\ \bibinfo {author} {\bibfnamefont {C.~S.}\ \bibnamefont {Adams}},\ }\bibfield  {title} {\bibinfo {title} {Coherent excitation transfer in a spin chain of three {Rydberg} atoms},\ }\href {https://doi.org/10.1103/PhysRevLett.114.113002} {\bibfield  {journal} {\bibinfo  {journal} {Phys. Rev. Lett.}\ }\textbf {\bibinfo {volume} {114}},\ \bibinfo {pages} {113002} (\bibinfo {year} {2015})}\BibitemShut {NoStop}%
    \bibitem [{\citenamefont {Igl{\'o}i}\ and\ \citenamefont {Monthus}(2005)}]{IM05}%
      \BibitemOpen
      \bibfield  {author} {\bibinfo {author} {\bibfnamefont {F.}~\bibnamefont {Igl{\'o}i}}\ and\ \bibinfo {author} {\bibfnamefont {C.}~\bibnamefont {Monthus}},\ }\bibfield  {title} {\bibinfo {title} {Strong disorder {RG} approach of random systems},\ }\href {https://doi.org/https://doi.org/10.1016/j.physrep.2005.02.006} {\bibfield  {journal} {\bibinfo  {journal} {Phys. Rep.}\ }\textbf {\bibinfo {volume} {412}},\ \bibinfo {pages} {277} (\bibinfo {year} {2005})}\BibitemShut {NoStop}%
    \bibitem [{\citenamefont {Refael}\ and\ \citenamefont {Altman}(2013)}]{RA13}%
      \BibitemOpen
      \bibfield  {author} {\bibinfo {author} {\bibfnamefont {G.}~\bibnamefont {Refael}}\ and\ \bibinfo {author} {\bibfnamefont {E.}~\bibnamefont {Altman}},\ }\bibfield  {title} {{\selectlanguage {english}\bibinfo {title} {Strong disorder renormalization group primer and the superfluid{\textendash}insulator transition}},\ }\href {https://doi.org/10.1016/j.crhy.2013.09.005} {\bibfield  {journal} {\bibinfo  {journal} {C. R. Phys.}\ }\textbf {\bibinfo {volume} {14}},\ \bibinfo {pages} {725} (\bibinfo {year} {2013})}\BibitemShut {NoStop}%
    \bibitem [{\citenamefont {Igl{\'o}i}\ and\ \citenamefont {Monthus}(2018)}]{IM18}%
      \BibitemOpen
      \bibfield  {author} {\bibinfo {author} {\bibfnamefont {F.}~\bibnamefont {Igl{\'o}i}}\ and\ \bibinfo {author} {\bibfnamefont {C.}~\bibnamefont {Monthus}},\ }\bibfield  {title} {\bibinfo {title} {Strong disorder {RG} approach -- a short review of recent developments},\ }\href {https://doi.org/10.1140/epjb/e2018-90434-8} {\bibfield  {journal} {\bibinfo  {journal} {Eur. Phys. J. B}\ }\textbf {\bibinfo {volume} {91}} (\bibinfo {year} {2018})}\BibitemShut {NoStop}%
    \bibitem [{\citenamefont {Huse}(2023)}]{Huse23}%
      \BibitemOpen
      \bibfield  {author} {\bibinfo {author} {\bibfnamefont {D.~A.}\ \bibnamefont {Huse}},\ }\href {https://arxiv.org/abs/2304.08572} {\bibinfo {title} {Strong-randomness renormalization groups}} (\bibinfo {year} {2023}),\ \Eprint {https://arxiv.org/abs/2304.08572} {arXiv:2304.08572 [cond-mat.stat-mech]} \BibitemShut {NoStop}%
    \bibitem [{\citenamefont {Ma}\ \emph {et~al.}(1979)\citenamefont {Ma}, \citenamefont {Dasgupta},\ and\ \citenamefont {Hu}}]{MDH79}%
      \BibitemOpen
      \bibfield  {author} {\bibinfo {author} {\bibfnamefont {S.-k.}\ \bibnamefont {Ma}}, \bibinfo {author} {\bibfnamefont {C.}~\bibnamefont {Dasgupta}},\ and\ \bibinfo {author} {\bibfnamefont {C.-k.}\ \bibnamefont {Hu}},\ }\bibfield  {title} {\bibinfo {title} {Random antiferromagnetic chain},\ }\href {https://doi.org/10.1103/PhysRevLett.43.1434} {\bibfield  {journal} {\bibinfo  {journal} {Phys. Rev. Lett.}\ }\textbf {\bibinfo {volume} {43}},\ \bibinfo {pages} {1434} (\bibinfo {year} {1979})}\BibitemShut {NoStop}%
    \bibitem [{\citenamefont {Dasgupta}\ and\ \citenamefont {Ma}(1980)}]{DM80}%
      \BibitemOpen
      \bibfield  {author} {\bibinfo {author} {\bibfnamefont {C.}~\bibnamefont {Dasgupta}}\ and\ \bibinfo {author} {\bibfnamefont {S.-k.}\ \bibnamefont {Ma}},\ }\bibfield  {title} {\bibinfo {title} {Low-temperature properties of the random {Heisenberg} antiferromagnetic chain},\ }\href {https://doi.org/10.1103/PhysRevB.22.1305} {\bibfield  {journal} {\bibinfo  {journal} {Phys. Rev. B}\ }\textbf {\bibinfo {volume} {22}},\ \bibinfo {pages} {1305} (\bibinfo {year} {1980})}\BibitemShut {NoStop}%
    \bibitem [{\citenamefont {Bhatt}\ and\ \citenamefont {Lee}(1982)}]{BL82}%
      \BibitemOpen
      \bibfield  {author} {\bibinfo {author} {\bibfnamefont {R.~N.}\ \bibnamefont {Bhatt}}\ and\ \bibinfo {author} {\bibfnamefont {P.~A.}\ \bibnamefont {Lee}},\ }\bibfield  {title} {\bibinfo {title} {Scaling studies of highly disordered spin-\textonehalf{} antiferromagnetic systems},\ }\href {https://doi.org/10.1103/PhysRevLett.48.344} {\bibfield  {journal} {\bibinfo  {journal} {Phys. Rev. Lett.}\ }\textbf {\bibinfo {volume} {48}},\ \bibinfo {pages} {344} (\bibinfo {year} {1982})}\BibitemShut {NoStop}%
    \bibitem [{\citenamefont {Fisher}(1992)}]{Fisher92}%
      \BibitemOpen
      \bibfield  {author} {\bibinfo {author} {\bibfnamefont {D.~S.}\ \bibnamefont {Fisher}},\ }\bibfield  {title} {\bibinfo {title} {Random transverse field {Ising} spin chains},\ }\href {https://doi.org/10.1103/PhysRevLett.69.534} {\bibfield  {journal} {\bibinfo  {journal} {Phys. Rev. Lett.}\ }\textbf {\bibinfo {volume} {69}},\ \bibinfo {pages} {534} (\bibinfo {year} {1992})}\BibitemShut {NoStop}%
    \bibitem [{\citenamefont {Fisher}(1994)}]{Fisher94}%
      \BibitemOpen
      \bibfield  {author} {\bibinfo {author} {\bibfnamefont {D.~S.}\ \bibnamefont {Fisher}},\ }\bibfield  {title} {{\selectlanguage {english}\bibinfo {title} {Random antiferromagnetic quantum spin chains}},\ }\href {https://doi.org/10.1103/PhysRevB.50.3799} {\bibfield  {journal} {\bibinfo  {journal} {Phys. Rev. B}\ }\textbf {\bibinfo {volume} {50}},\ \bibinfo {pages} {3799} (\bibinfo {year} {1994})}\BibitemShut {NoStop}%
    \bibitem [{\citenamefont {Westerberg}\ \emph {et~al.}(1995)\citenamefont {Westerberg}, \citenamefont {Furusaki}, \citenamefont {Sigrist},\ and\ \citenamefont {Lee}}]{Westerberg95}%
      \BibitemOpen
      \bibfield  {author} {\bibinfo {author} {\bibfnamefont {E.}~\bibnamefont {Westerberg}}, \bibinfo {author} {\bibfnamefont {A.}~\bibnamefont {Furusaki}}, \bibinfo {author} {\bibfnamefont {M.}~\bibnamefont {Sigrist}},\ and\ \bibinfo {author} {\bibfnamefont {P.~A.}\ \bibnamefont {Lee}},\ }\bibfield  {title} {\bibinfo {title} {Random quantum spin chains: A real-space renormalization group study},\ }\href {https://doi.org/10.1103/PhysRevLett.75.4302} {\bibfield  {journal} {\bibinfo  {journal} {Phys. Rev. Lett.}\ }\textbf {\bibinfo {volume} {75}},\ \bibinfo {pages} {4302} (\bibinfo {year} {1995})}\BibitemShut {NoStop}%
    \bibitem [{\citenamefont {Fisher}(1995)}]{Fisher95}%
      \BibitemOpen
      \bibfield  {author} {\bibinfo {author} {\bibfnamefont {D.~S.}\ \bibnamefont {Fisher}},\ }\bibfield  {title} {\bibinfo {title} {Critical behavior of random transverse-field {Ising} spin chains},\ }\href {https://doi.org/10.1103/PhysRevB.51.6411} {\bibfield  {journal} {\bibinfo  {journal} {Phys. Rev. B}\ }\textbf {\bibinfo {volume} {51}},\ \bibinfo {pages} {6411} (\bibinfo {year} {1995})}\BibitemShut {NoStop}%
    \bibitem [{\citenamefont {Westerberg}\ \emph {et~al.}(1997)\citenamefont {Westerberg}, \citenamefont {Furusaki}, \citenamefont {Sigrist},\ and\ \citenamefont {Lee}}]{Westerberg97}%
      \BibitemOpen
      \bibfield  {author} {\bibinfo {author} {\bibfnamefont {E.}~\bibnamefont {Westerberg}}, \bibinfo {author} {\bibfnamefont {A.}~\bibnamefont {Furusaki}}, \bibinfo {author} {\bibfnamefont {M.}~\bibnamefont {Sigrist}},\ and\ \bibinfo {author} {\bibfnamefont {P.~A.}\ \bibnamefont {Lee}},\ }\bibfield  {title} {\bibinfo {title} {Low-energy fixed points of random quantum spin chains},\ }\href {https://doi.org/10.1103/PhysRevB.55.12578} {\bibfield  {journal} {\bibinfo  {journal} {Phys. Rev. B}\ }\textbf {\bibinfo {volume} {55}},\ \bibinfo {pages} {12578} (\bibinfo {year} {1997})}\BibitemShut {NoStop}%
    \bibitem [{\citenamefont {Damle}\ \emph {et~al.}(2000)\citenamefont {Damle}, \citenamefont {Motrunich},\ and\ \citenamefont {Huse}}]{DMH00}%
      \BibitemOpen
      \bibfield  {author} {\bibinfo {author} {\bibfnamefont {K.}~\bibnamefont {Damle}}, \bibinfo {author} {\bibfnamefont {O.}~\bibnamefont {Motrunich}},\ and\ \bibinfo {author} {\bibfnamefont {D.~A.}\ \bibnamefont {Huse}},\ }\bibfield  {title} {{\selectlanguage {english}\bibinfo {title} {Dynamics and transport in random antiferromagnetic spin chains}},\ }\href {https://doi.org/10.1103/PhysRevLett.84.3434} {\bibfield  {journal} {\bibinfo  {journal} {Phys. Rev. Lett.}\ }\textbf {\bibinfo {volume} {84}},\ \bibinfo {pages} {3434} (\bibinfo {year} {2000})}\BibitemShut {NoStop}%
    \bibitem [{\citenamefont {Motrunich}\ \emph {et~al.}(2001)\citenamefont {Motrunich}, \citenamefont {Damle},\ and\ \citenamefont {Huse}}]{MDH01}%
      \BibitemOpen
      \bibfield  {author} {\bibinfo {author} {\bibfnamefont {O.}~\bibnamefont {Motrunich}}, \bibinfo {author} {\bibfnamefont {K.}~\bibnamefont {Damle}},\ and\ \bibinfo {author} {\bibfnamefont {D.~A.}\ \bibnamefont {Huse}},\ }\bibfield  {title} {{\selectlanguage {english}\bibinfo {title} {Dynamics and transport in random quantum systems governed by strong-randomness fixed points}},\ }\href {https://doi.org/10.1103/PhysRevB.63.134424} {\bibfield  {journal} {\bibinfo  {journal} {Phys. Rev. B}\ }\textbf {\bibinfo {volume} {63}},\ \bibinfo {pages} {134424} (\bibinfo {year} {2001})}\BibitemShut {NoStop}%
    \bibitem [{\citenamefont {Refael}\ \emph {et~al.}(2002)\citenamefont {Refael}, \citenamefont {Kehrein},\ and\ \citenamefont {Fisher}}]{RKF02}%
      \BibitemOpen
      \bibfield  {author} {\bibinfo {author} {\bibfnamefont {G.}~\bibnamefont {Refael}}, \bibinfo {author} {\bibfnamefont {S.}~\bibnamefont {Kehrein}},\ and\ \bibinfo {author} {\bibfnamefont {D.~S.}\ \bibnamefont {Fisher}},\ }\bibfield  {title} {\bibinfo {title} {Spin reduction transition in spin-$\frac{3}{2}$ random {Heisenberg} chains},\ }\href {https://doi.org/10.1103/PhysRevB.66.060402} {\bibfield  {journal} {\bibinfo  {journal} {Phys. Rev. B}\ }\textbf {\bibinfo {volume} {66}},\ \bibinfo {pages} {060402} (\bibinfo {year} {2002})}\BibitemShut {NoStop}%
    \bibitem [{\citenamefont {Damle}\ and\ \citenamefont {Huse}(2002)}]{DH02}%
      \BibitemOpen
      \bibfield  {author} {\bibinfo {author} {\bibfnamefont {K.}~\bibnamefont {Damle}}\ and\ \bibinfo {author} {\bibfnamefont {D.~A.}\ \bibnamefont {Huse}},\ }\bibfield  {title} {\bibinfo {title} {Permutation-symmetric multicritical points in random antiferromagnetic spin chains},\ }\href {https://doi.org/10.1103/PhysRevLett.89.277203} {\bibfield  {journal} {\bibinfo  {journal} {Phys. Rev. Lett.}\ }\textbf {\bibinfo {volume} {89}},\ \bibinfo {pages} {277203} (\bibinfo {year} {2002})}\BibitemShut {NoStop}%
    \bibitem [{\citenamefont {Refael}\ and\ \citenamefont {Moore}(2004)}]{RM04}%
      \BibitemOpen
      \bibfield  {author} {\bibinfo {author} {\bibfnamefont {G.}~\bibnamefont {Refael}}\ and\ \bibinfo {author} {\bibfnamefont {J.~E.}\ \bibnamefont {Moore}},\ }\bibfield  {title} {\bibinfo {title} {Entanglement entropy of random quantum critical points in one dimension},\ }\href {https://doi.org/10.1103/PhysRevLett.93.260602} {\bibfield  {journal} {\bibinfo  {journal} {Phys. Rev. Lett.}\ }\textbf {\bibinfo {volume} {93}},\ \bibinfo {pages} {260602} (\bibinfo {year} {2004})}\BibitemShut {NoStop}%
    \bibitem [{\citenamefont {Mohdeb}\ \emph {et~al.}(2020)\citenamefont {Mohdeb}, \citenamefont {Vahedi}, \citenamefont {Moure}, \citenamefont {Roshani}, \citenamefont {Lee}, \citenamefont {Bhatt}, \citenamefont {Kettemann},\ and\ \citenamefont {Haas}}]{Mohdeb20}%
      \BibitemOpen
      \bibfield  {author} {\bibinfo {author} {\bibfnamefont {Y.}~\bibnamefont {Mohdeb}}, \bibinfo {author} {\bibfnamefont {J.}~\bibnamefont {Vahedi}}, \bibinfo {author} {\bibfnamefont {N.}~\bibnamefont {Moure}}, \bibinfo {author} {\bibfnamefont {A.}~\bibnamefont {Roshani}}, \bibinfo {author} {\bibfnamefont {H.-Y.}\ \bibnamefont {Lee}}, \bibinfo {author} {\bibfnamefont {R.~N.}\ \bibnamefont {Bhatt}}, \bibinfo {author} {\bibfnamefont {S.}~\bibnamefont {Kettemann}},\ and\ \bibinfo {author} {\bibfnamefont {S.}~\bibnamefont {Haas}},\ }\bibfield  {title} {\bibinfo {title} {Entanglement properties of disordered quantum spin chains with long-range antiferromagnetic interactions},\ }\href {https://doi.org/10.1103/PhysRevB.102.214201} {\bibfield  {journal} {\bibinfo  {journal} {Phys. Rev. B}\ }\textbf {\bibinfo {volume} {102}},\ \bibinfo {pages} {214201} (\bibinfo {year} {2020})}\BibitemShut {NoStop}%
    \bibitem [{\citenamefont {Roberts}\ and\ \citenamefont {Motrunich}(2021)}]{Roberts21}%
      \BibitemOpen
      \bibfield  {author} {\bibinfo {author} {\bibfnamefont {B.}~\bibnamefont {Roberts}}\ and\ \bibinfo {author} {\bibfnamefont {O.~I.}\ \bibnamefont {Motrunich}},\ }\bibfield  {title} {\bibinfo {title} {Infinite randomness with continuously varying critical exponents in the random {XYZ} spin chain},\ }\href {https://doi.org/10.1103/PhysRevB.104.214208} {\bibfield  {journal} {\bibinfo  {journal} {Phys. Rev. B}\ }\textbf {\bibinfo {volume} {104}},\ \bibinfo {pages} {214208} (\bibinfo {year} {2021})}\BibitemShut {NoStop}%
    \bibitem [{\citenamefont {Vosk}\ and\ \citenamefont {Altman}(2013)}]{VA13}%
      \BibitemOpen
      \bibfield  {author} {\bibinfo {author} {\bibfnamefont {R.}~\bibnamefont {Vosk}}\ and\ \bibinfo {author} {\bibfnamefont {E.}~\bibnamefont {Altman}},\ }\bibfield  {title} {{\selectlanguage {english}\bibinfo {title} {Many-body localization in one dimension as a dynamical renormalization group fixed point}},\ }\href {https://doi.org/10.1103/PhysRevLett.110.067204} {\bibfield  {journal} {\bibinfo  {journal} {Phys. Rev. Lett.}\ }\textbf {\bibinfo {volume} {110}},\ \bibinfo {pages} {067204} (\bibinfo {year} {2013})}\BibitemShut {NoStop}%
    \bibitem [{\citenamefont {Pekker}\ \emph {et~al.}(2014)\citenamefont {Pekker}, \citenamefont {Refael}, \citenamefont {Altman}, \citenamefont {Demler},\ and\ \citenamefont {Oganesyan}}]{Pek14}%
      \BibitemOpen
      \bibfield  {author} {\bibinfo {author} {\bibfnamefont {D.}~\bibnamefont {Pekker}}, \bibinfo {author} {\bibfnamefont {G.}~\bibnamefont {Refael}}, \bibinfo {author} {\bibfnamefont {E.}~\bibnamefont {Altman}}, \bibinfo {author} {\bibfnamefont {E.}~\bibnamefont {Demler}},\ and\ \bibinfo {author} {\bibfnamefont {V.}~\bibnamefont {Oganesyan}},\ }\bibfield  {title} {{\selectlanguage {english}\bibinfo {title} {Hilbert-glass transition: New universality of temperature-tuned many-body dynamical quantum criticality}},\ }\href {https://doi.org/10.1103/PhysRevX.4.011052} {\bibfield  {journal} {\bibinfo  {journal} {Phys. Rev. X}\ }\textbf {\bibinfo {volume} {4}},\ \bibinfo {pages} {011052} (\bibinfo {year} {2014})}\BibitemShut {NoStop}%
    \bibitem [{\citenamefont {Vosk}\ and\ \citenamefont {Altman}(2014)}]{VA14}%
      \BibitemOpen
      \bibfield  {author} {\bibinfo {author} {\bibfnamefont {R.}~\bibnamefont {Vosk}}\ and\ \bibinfo {author} {\bibfnamefont {E.}~\bibnamefont {Altman}},\ }\bibfield  {title} {\bibinfo {title} {Dynamical quantum phase transitions in random spin chains},\ }\href {https://doi.org/10.1103/PhysRevLett.112.217204} {\bibfield  {journal} {\bibinfo  {journal} {Phys. Rev. Lett.}\ }\textbf {\bibinfo {volume} {112}},\ \bibinfo {pages} {217204} (\bibinfo {year} {2014})}\BibitemShut {NoStop}%
    \bibitem [{\citenamefont {Huang}\ and\ \citenamefont {Moore}(2014)}]{HM14}%
      \BibitemOpen
      \bibfield  {author} {\bibinfo {author} {\bibfnamefont {Y.}~\bibnamefont {Huang}}\ and\ \bibinfo {author} {\bibfnamefont {J.~E.}\ \bibnamefont {Moore}},\ }\bibfield  {title} {{\selectlanguage {english}\bibinfo {title} {Excited-state entanglement and thermal mutual information in random spin chains}},\ }\href {https://doi.org/10.1103/PhysRevB.90.220202} {\bibfield  {journal} {\bibinfo  {journal} {Phys. Rev. B}\ }\textbf {\bibinfo {volume} {90}},\ \bibinfo {pages} {220202} (\bibinfo {year} {2014})}\BibitemShut {NoStop}%
    \bibitem [{\citenamefont {Vasseur}\ \emph {et~al.}(2015)\citenamefont {Vasseur}, \citenamefont {Potter},\ and\ \citenamefont {Parameswaran}}]{VPP15}%
      \BibitemOpen
      \bibfield  {author} {\bibinfo {author} {\bibfnamefont {R.}~\bibnamefont {Vasseur}}, \bibinfo {author} {\bibfnamefont {A.~C.}\ \bibnamefont {Potter}},\ and\ \bibinfo {author} {\bibfnamefont {S.~A.}\ \bibnamefont {Parameswaran}},\ }\bibfield  {title} {{\selectlanguage {english}\bibinfo {title} {Quantum criticality of hot random spin chains}},\ }\href {https://doi.org/10.1103/PhysRevLett.114.217201} {\bibfield  {journal} {\bibinfo  {journal} {Phys. Rev. Lett.}\ }\textbf {\bibinfo {volume} {114}},\ \bibinfo {pages} {217201} (\bibinfo {year} {2015})}\BibitemShut {NoStop}%
    \bibitem [{\citenamefont {You}\ \emph {et~al.}(2016)\citenamefont {You}, \citenamefont {Qi},\ and\ \citenamefont {Xu}}]{YQX16}%
      \BibitemOpen
      \bibfield  {author} {\bibinfo {author} {\bibfnamefont {Y.-Z.}\ \bibnamefont {You}}, \bibinfo {author} {\bibfnamefont {X.-L.}\ \bibnamefont {Qi}},\ and\ \bibinfo {author} {\bibfnamefont {C.}~\bibnamefont {Xu}},\ }\bibfield  {title} {{\selectlanguage {english}\bibinfo {title} {Entanglement holographic mapping of many-body localized system by spectrum bifurcation renormalization group}},\ }\href {https://doi.org/10.1103/PhysRevB.93.104205} {\bibfield  {journal} {\bibinfo  {journal} {Phys. Rev. B}\ }\textbf {\bibinfo {volume} {93}},\ \bibinfo {pages} {104205} (\bibinfo {year} {2016})}\BibitemShut {NoStop}%
    \bibitem [{\citenamefont {Vasseur}\ \emph {et~al.}(2016)\citenamefont {Vasseur}, \citenamefont {Friedman}, \citenamefont {Parameswaran},\ and\ \citenamefont {Potter}}]{VFPP16}%
      \BibitemOpen
      \bibfield  {author} {\bibinfo {author} {\bibfnamefont {R.}~\bibnamefont {Vasseur}}, \bibinfo {author} {\bibfnamefont {A.~J.}\ \bibnamefont {Friedman}}, \bibinfo {author} {\bibfnamefont {S.~A.}\ \bibnamefont {Parameswaran}},\ and\ \bibinfo {author} {\bibfnamefont {A.~C.}\ \bibnamefont {Potter}},\ }\bibfield  {title} {{\selectlanguage {english}\bibinfo {title} {Particle-hole symmetry, many-body localization, and topological edge modes}},\ }\href {https://doi.org/10.1103/PhysRevB.93.134207} {\bibfield  {journal} {\bibinfo  {journal} {Phys. Rev. B}\ }\textbf {\bibinfo {volume} {93}},\ \bibinfo {pages} {134207} (\bibinfo {year} {2016})}\BibitemShut {NoStop}%
    \bibitem [{\citenamefont {Monthus}(2018)}]{Monthus18}%
      \BibitemOpen
      \bibfield  {author} {\bibinfo {author} {\bibfnamefont {C.}~\bibnamefont {Monthus}},\ }\bibfield  {title} {\bibinfo {title} {Strong disorder real-space renormalization for the many-body-localized phase of random {Majorana} models},\ }\href {https://doi.org/10.1088/1751-8121/aaad14} {\bibfield  {journal} {\bibinfo  {journal} {J. Phys. A: Math. Theor.}\ }\textbf {\bibinfo {volume} {51}},\ \bibinfo {pages} {115304} (\bibinfo {year} {2018})}\BibitemShut {NoStop}%
    \bibitem [{\citenamefont {Protopopov}\ \emph {et~al.}(2020)\citenamefont {Protopopov}, \citenamefont {Panda}, \citenamefont {Parolini}, \citenamefont {Scardicchio}, \citenamefont {Demler},\ and\ \citenamefont {Abanin}}]{Protopopov20}%
      \BibitemOpen
      \bibfield  {author} {\bibinfo {author} {\bibfnamefont {I.~V.}\ \bibnamefont {Protopopov}}, \bibinfo {author} {\bibfnamefont {R.~K.}\ \bibnamefont {Panda}}, \bibinfo {author} {\bibfnamefont {T.}~\bibnamefont {Parolini}}, \bibinfo {author} {\bibfnamefont {A.}~\bibnamefont {Scardicchio}}, \bibinfo {author} {\bibfnamefont {E.}~\bibnamefont {Demler}},\ and\ \bibinfo {author} {\bibfnamefont {D.~A.}\ \bibnamefont {Abanin}},\ }\bibfield  {title} {\bibinfo {title} {Non-{Abelian} symmetries and disorder: A broad nonergodic regime and anomalous thermalization},\ }\href {https://doi.org/10.1103/PhysRevX.10.011025} {\bibfield  {journal} {\bibinfo  {journal} {Phys. Rev. X}\ }\textbf {\bibinfo {volume} {10}},\ \bibinfo {pages} {011025} (\bibinfo {year} {2020})}\BibitemShut {NoStop}%
    \bibitem [{\citenamefont {Mohdeb}\ \emph {et~al.}(2022)\citenamefont {Mohdeb}, \citenamefont {Vahedi},\ and\ \citenamefont {Kettemann}}]{MVK22}%
      \BibitemOpen
      \bibfield  {author} {\bibinfo {author} {\bibfnamefont {Y.}~\bibnamefont {Mohdeb}}, \bibinfo {author} {\bibfnamefont {J.}~\bibnamefont {Vahedi}},\ and\ \bibinfo {author} {\bibfnamefont {S.}~\bibnamefont {Kettemann}},\ }\bibfield  {title} {{\selectlanguage {english}\bibinfo {title} {Excited-eigenstate entanglement properties of {XX} spin chains with random long-range interactions}},\ }\href {https://doi.org/10.1103/PhysRevB.106.104201} {\bibfield  {journal} {\bibinfo  {journal} {Phys. Rev. B}\ }\textbf {\bibinfo {volume} {106}},\ \bibinfo {pages} {104201} (\bibinfo {year} {2022})}\BibitemShut {NoStop}%
    \bibitem [{\citenamefont {Braemer}\ \emph {et~al.}(2022)\citenamefont {Braemer}, \citenamefont {Franz}, \citenamefont {Weidem\"uller},\ and\ \citenamefont {G\"arttner}}]{Titus_theory_22}%
      \BibitemOpen
      \bibfield  {author} {\bibinfo {author} {\bibfnamefont {A.}~\bibnamefont {Braemer}}, \bibinfo {author} {\bibfnamefont {T.}~\bibnamefont {Franz}}, \bibinfo {author} {\bibfnamefont {M.}~\bibnamefont {Weidem\"uller}},\ and\ \bibinfo {author} {\bibfnamefont {M.}~\bibnamefont {G\"arttner}},\ }\bibfield  {title} {\bibinfo {title} {Pair localization in dipolar systems with tunable positional disorder},\ }\href {https://doi.org/10.1103/PhysRevB.106.134212} {\bibfield  {journal} {\bibinfo  {journal} {Phys. Rev. B}\ }\textbf {\bibinfo {volume} {106}},\ \bibinfo {pages} {134212} (\bibinfo {year} {2022})}\BibitemShut {NoStop}%
    \bibitem [{\citenamefont {Mohdeb}\ \emph {et~al.}(2023)\citenamefont {Mohdeb}, \citenamefont {Vahedi}, \citenamefont {Bhatt}, \citenamefont {Haas},\ and\ \citenamefont {Kettemann}}]{Mohdeb23}%
      \BibitemOpen
      \bibfield  {author} {\bibinfo {author} {\bibfnamefont {Y.}~\bibnamefont {Mohdeb}}, \bibinfo {author} {\bibfnamefont {J.}~\bibnamefont {Vahedi}}, \bibinfo {author} {\bibfnamefont {R.~N.}\ \bibnamefont {Bhatt}}, \bibinfo {author} {\bibfnamefont {S.}~\bibnamefont {Haas}},\ and\ \bibinfo {author} {\bibfnamefont {S.}~\bibnamefont {Kettemann}},\ }\bibfield  {title} {\bibinfo {title} {Global quench dynamics and the growth of entanglement entropy in disordered spin chains with tunable range interactions},\ }\href {https://doi.org/10.1103/PhysRevB.108.L140203} {\bibfield  {journal} {\bibinfo  {journal} {Phys. Rev. B}\ }\textbf {\bibinfo {volume} {108}},\ \bibinfo {pages} {L140203} (\bibinfo {year} {2023})}\BibitemShut {NoStop}%
    \bibitem [{\citenamefont {Braemer}\ \emph {et~al.}(2024)\citenamefont {Braemer}, \citenamefont {Vahedi},\ and\ \citenamefont {G\"arttner}}]{BVG24}%
      \BibitemOpen
      \bibfield  {author} {\bibinfo {author} {\bibfnamefont {A.}~\bibnamefont {Braemer}}, \bibinfo {author} {\bibfnamefont {J.}~\bibnamefont {Vahedi}},\ and\ \bibinfo {author} {\bibfnamefont {M.}~\bibnamefont {G\"arttner}},\ }\bibfield  {title} {\bibinfo {title} {Cluster truncated {Wigner} approximation for bond-disordered {Heisenberg} spin models},\ }\href {https://doi.org/10.1103/PhysRevB.110.054204} {\bibfield  {journal} {\bibinfo  {journal} {Phys. Rev. B}\ }\textbf {\bibinfo {volume} {110}},\ \bibinfo {pages} {054204} (\bibinfo {year} {2024})}\BibitemShut {NoStop}%
    \bibitem [{\citenamefont {Aramthottil}\ \emph {et~al.}(2024)\citenamefont {Aramthottil}, \citenamefont {Sierant}, \citenamefont {Lewenstein},\ and\ \citenamefont {Zakrzewski}}]{Aramthottil24}%
      \BibitemOpen
      \bibfield  {author} {\bibinfo {author} {\bibfnamefont {A.~S.}\ \bibnamefont {Aramthottil}}, \bibinfo {author} {\bibfnamefont {P.}~\bibnamefont {Sierant}}, \bibinfo {author} {\bibfnamefont {M.}~\bibnamefont {Lewenstein}},\ and\ \bibinfo {author} {\bibfnamefont {J.}~\bibnamefont {Zakrzewski}},\ }\bibfield  {title} {\bibinfo {title} {Phenomenology of many-body localization in bond-disordered spin chains},\ }\href {https://doi.org/10.1103/PhysRevLett.133.196302} {\bibfield  {journal} {\bibinfo  {journal} {Phys. Rev. Lett.}\ }\textbf {\bibinfo {volume} {133}},\ \bibinfo {pages} {196302} (\bibinfo {year} {2024})}\BibitemShut {NoStop}%
    \bibitem [{\citenamefont {Gopalakrishnan}\ \emph {et~al.}(2015)\citenamefont {Gopalakrishnan}, \citenamefont {M\"uller}, \citenamefont {Khemani}, \citenamefont {Knap}, \citenamefont {Demler},\ and\ \citenamefont {Huse}}]{Gopalakrishnan2015}%
      \BibitemOpen
      \bibfield  {author} {\bibinfo {author} {\bibfnamefont {S.}~\bibnamefont {Gopalakrishnan}}, \bibinfo {author} {\bibfnamefont {M.}~\bibnamefont {M\"uller}}, \bibinfo {author} {\bibfnamefont {V.}~\bibnamefont {Khemani}}, \bibinfo {author} {\bibfnamefont {M.}~\bibnamefont {Knap}}, \bibinfo {author} {\bibfnamefont {E.}~\bibnamefont {Demler}},\ and\ \bibinfo {author} {\bibfnamefont {D.~A.}\ \bibnamefont {Huse}},\ }\bibfield  {title} {\bibinfo {title} {Low-frequency conductivity in many-body localized systems},\ }\href {https://doi.org/10.1103/PhysRevB.92.104202} {\bibfield  {journal} {\bibinfo  {journal} {Phys. Rev. B}\ }\textbf {\bibinfo {volume} {92}},\ \bibinfo {pages} {104202} (\bibinfo {year} {2015})}\BibitemShut {NoStop}%
    \bibitem [{\citenamefont {Crowley}\ and\ \citenamefont {Chandran}(2022)}]{Crowley2022}%
      \BibitemOpen
      \bibfield  {author} {\bibinfo {author} {\bibfnamefont {P.~J.~D.}\ \bibnamefont {Crowley}}\ and\ \bibinfo {author} {\bibfnamefont {A.}~\bibnamefont {Chandran}},\ }\bibfield  {title} {\bibinfo {title} {{A constructive theory of the numerically accessible many-body localized to thermal crossover}},\ }\href {https://doi.org/10.21468/SciPostPhys.12.6.201} {\bibfield  {journal} {\bibinfo  {journal} {SciPost Phys.}\ }\textbf {\bibinfo {volume} {12}},\ \bibinfo {pages} {201} (\bibinfo {year} {2022})}\BibitemShut {NoStop}%
    \bibitem [{\citenamefont {Garratt}\ \emph {et~al.}(2021)\citenamefont {Garratt}, \citenamefont {Roy},\ and\ \citenamefont {Chalker}}]{Garratt2021}%
      \BibitemOpen
      \bibfield  {author} {\bibinfo {author} {\bibfnamefont {S.~J.}\ \bibnamefont {Garratt}}, \bibinfo {author} {\bibfnamefont {S.}~\bibnamefont {Roy}},\ and\ \bibinfo {author} {\bibfnamefont {J.~T.}\ \bibnamefont {Chalker}},\ }\bibfield  {title} {\bibinfo {title} {Local resonances and parametric level dynamics in the many-body localized phase},\ }\href {https://doi.org/10.1103/PhysRevB.104.184203} {\bibfield  {journal} {\bibinfo  {journal} {Phys. Rev. B}\ }\textbf {\bibinfo {volume} {104}},\ \bibinfo {pages} {184203} (\bibinfo {year} {2021})}\BibitemShut {NoStop}%
    \bibitem [{\citenamefont {Garratt}\ and\ \citenamefont {Roy}(2022)}]{Garratt2022}%
      \BibitemOpen
      \bibfield  {author} {\bibinfo {author} {\bibfnamefont {S.~J.}\ \bibnamefont {Garratt}}\ and\ \bibinfo {author} {\bibfnamefont {S.}~\bibnamefont {Roy}},\ }\bibfield  {title} {\bibinfo {title} {Resonant energy scales and local observables in the many-body localized phase},\ }\href {https://doi.org/10.1103/PhysRevB.106.054309} {\bibfield  {journal} {\bibinfo  {journal} {Phys. Rev. B}\ }\textbf {\bibinfo {volume} {106}},\ \bibinfo {pages} {054309} (\bibinfo {year} {2022})}\BibitemShut {NoStop}%
    \bibitem [{\citenamefont {Long}\ \emph {et~al.}(2023)\citenamefont {Long}, \citenamefont {Crowley}, \citenamefont {Khemani},\ and\ \citenamefont {Chandran}}]{Long2023}%
      \BibitemOpen
      \bibfield  {author} {\bibinfo {author} {\bibfnamefont {D.~M.}\ \bibnamefont {Long}}, \bibinfo {author} {\bibfnamefont {P.~J.~D.}\ \bibnamefont {Crowley}}, \bibinfo {author} {\bibfnamefont {V.}~\bibnamefont {Khemani}},\ and\ \bibinfo {author} {\bibfnamefont {A.}~\bibnamefont {Chandran}},\ }\bibfield  {title} {\bibinfo {title} {Phenomenology of the prethermal many-body localized regime},\ }\href {https://doi.org/10.1103/PhysRevLett.131.106301} {\bibfield  {journal} {\bibinfo  {journal} {Phys. Rev. Lett.}\ }\textbf {\bibinfo {volume} {131}},\ \bibinfo {pages} {106301} (\bibinfo {year} {2023})}\BibitemShut {NoStop}%
    \bibitem [{\citenamefont {Igl{\'o}i}\ \emph {et~al.}(2000)\citenamefont {Igl{\'o}i}, \citenamefont {Juh{\'a}sz},\ and\ \citenamefont {Rieger}}]{IJR00}%
      \BibitemOpen
      \bibfield  {author} {\bibinfo {author} {\bibfnamefont {F.}~\bibnamefont {Igl{\'o}i}}, \bibinfo {author} {\bibfnamefont {R.}~\bibnamefont {Juh{\'a}sz}},\ and\ \bibinfo {author} {\bibfnamefont {H.}~\bibnamefont {Rieger}},\ }\bibfield  {title} {{\selectlanguage {english}\bibinfo {title} {Random antiferromagnetic quantum spin chains: {Exact} results from scaling of rare regions}},\ }\href {https://doi.org/10.1103/PhysRevB.61.11552} {\bibfield  {journal} {\bibinfo  {journal} {Phys. Rev. B}\ }\textbf {\bibinfo {volume} {61}},\ \bibinfo {pages} {11552} (\bibinfo {year} {2000})}\BibitemShut {NoStop}%
    \bibitem [{\citenamefont {Sachdev}(2011)}]{Sachdev11}%
      \BibitemOpen
      \bibfield  {author} {\bibinfo {author} {\bibfnamefont {S.}~\bibnamefont {Sachdev}},\ }\href@noop {} {\emph {\bibinfo {title} {Quantum Phase Transitions}}},\ \bibinfo {edition} {2nd}\ ed.\ (\bibinfo  {publisher} {Cambridge University Press},\ \bibinfo {year} {2011})\BibitemShut {NoStop}%
    \bibitem [{Note1()}]{Note1}%
      \BibitemOpen
      \bibinfo {note} {One can also formulate the RSRG-X and derive the same update rule by directly working with the single particle Hamiltonian in the fermion language.}\BibitemShut {Stop}%
    \bibitem [{\citenamefont {Motrunich}\ \emph {et~al.}(2000)\citenamefont {Motrunich}, \citenamefont {Mau}, \citenamefont {Huse},\ and\ \citenamefont {Fisher}}]{MMHF00}%
      \BibitemOpen
      \bibfield  {author} {\bibinfo {author} {\bibfnamefont {O.}~\bibnamefont {Motrunich}}, \bibinfo {author} {\bibfnamefont {S.-C.}\ \bibnamefont {Mau}}, \bibinfo {author} {\bibfnamefont {D.~A.}\ \bibnamefont {Huse}},\ and\ \bibinfo {author} {\bibfnamefont {D.~S.}\ \bibnamefont {Fisher}},\ }\bibfield  {title} {\bibinfo {title} {Infinite-randomness quantum {Ising} critical fixed points},\ }\href {https://doi.org/10.1103/PhysRevB.61.1160} {\bibfield  {journal} {\bibinfo  {journal} {Phys. Rev. B}\ }\textbf {\bibinfo {volume} {61}},\ \bibinfo {pages} {1160} (\bibinfo {year} {2000})}\BibitemShut {NoStop}%
    \bibitem [{\citenamefont {Laumann}\ \emph {et~al.}(2012)\citenamefont {Laumann}, \citenamefont {Huse}, \citenamefont {Ludwig}, \citenamefont {Refael}, \citenamefont {Trebst},\ and\ \citenamefont {Troyer}}]{Laumann12}%
      \BibitemOpen
      \bibfield  {author} {\bibinfo {author} {\bibfnamefont {C.~R.}\ \bibnamefont {Laumann}}, \bibinfo {author} {\bibfnamefont {D.~A.}\ \bibnamefont {Huse}}, \bibinfo {author} {\bibfnamefont {A.~W.~W.}\ \bibnamefont {Ludwig}}, \bibinfo {author} {\bibfnamefont {G.}~\bibnamefont {Refael}}, \bibinfo {author} {\bibfnamefont {S.}~\bibnamefont {Trebst}},\ and\ \bibinfo {author} {\bibfnamefont {M.}~\bibnamefont {Troyer}},\ }\bibfield  {title} {\bibinfo {title} {Strong-disorder renormalization for interacting non-{Abelian} anyon systems in two dimensions},\ }\href {https://doi.org/10.1103/PhysRevB.85.224201} {\bibfield  {journal} {\bibinfo  {journal} {Phys. Rev. B}\ }\textbf {\bibinfo {volume} {85}},\ \bibinfo {pages} {224201} (\bibinfo {year} {2012})}\BibitemShut {NoStop}%
    \bibitem [{\citenamefont {Baldwin}\ \emph {et~al.}(2023)\citenamefont {Baldwin}, \citenamefont {Ehrenberg}, \citenamefont {Guo},\ and\ \citenamefont {Gorshkov}}]{Baldwin23}%
      \BibitemOpen
      \bibfield  {author} {\bibinfo {author} {\bibfnamefont {C.~L.}\ \bibnamefont {Baldwin}}, \bibinfo {author} {\bibfnamefont {A.}~\bibnamefont {Ehrenberg}}, \bibinfo {author} {\bibfnamefont {A.~Y.}\ \bibnamefont {Guo}},\ and\ \bibinfo {author} {\bibfnamefont {A.~V.}\ \bibnamefont {Gorshkov}},\ }\bibfield  {title} {\bibinfo {title} {Disordered {Lieb-Robinson} bounds in one dimension},\ }\href {https://doi.org/10.1103/PRXQuantum.4.020349} {\bibfield  {journal} {\bibinfo  {journal} {PRX Quantum}\ }\textbf {\bibinfo {volume} {4}},\ \bibinfo {pages} {020349} (\bibinfo {year} {2023})}\BibitemShut {NoStop}%
    \bibitem [{\citenamefont {Baldwin}(2024)}]{Baldwin24}%
      \BibitemOpen
      \bibfield  {author} {\bibinfo {author} {\bibfnamefont {C.~L.}\ \bibnamefont {Baldwin}},\ }\href {https://arxiv.org/abs/2409.17242} {\bibinfo {title} {Sub-ballistic operator growth in spin chains with heavy-tailed random fields}} (\bibinfo {year} {2024}),\ \Eprint {https://arxiv.org/abs/2409.17242} {arXiv:2409.17242 [cond-mat.dis-nn]} \BibitemShut {NoStop}%
    \bibitem [{\citenamefont {Roeck}\ \emph {et~al.}(2024)\citenamefont {Roeck}, \citenamefont {Giacomin}, \citenamefont {Huveneers},\ and\ \citenamefont {Prosniak}}]{deRoeck24}%
      \BibitemOpen
      \bibfield  {author} {\bibinfo {author} {\bibfnamefont {W.~D.}\ \bibnamefont {Roeck}}, \bibinfo {author} {\bibfnamefont {L.}~\bibnamefont {Giacomin}}, \bibinfo {author} {\bibfnamefont {F.}~\bibnamefont {Huveneers}},\ and\ \bibinfo {author} {\bibfnamefont {O.}~\bibnamefont {Prosniak}},\ }\href {https://arxiv.org/abs/2408.04338} {\bibinfo {title} {Absence of normal heat conduction in strongly disordered interacting quantum chains}} (\bibinfo {year} {2024}),\ \Eprint {https://arxiv.org/abs/2408.04338} {arXiv:2408.04338 [math-ph]} \BibitemShut {NoStop}%
    \bibitem [{\citenamefont {Chen}\ and\ \citenamefont {Lucas}(2019)}]{Lucas19}%
      \BibitemOpen
      \bibfield  {author} {\bibinfo {author} {\bibfnamefont {C.-F.}\ \bibnamefont {Chen}}\ and\ \bibinfo {author} {\bibfnamefont {A.}~\bibnamefont {Lucas}},\ }\bibfield  {title} {\bibinfo {title} {Finite speed of quantum scrambling with long range interactions},\ }\href {https://doi.org/10.1103/PhysRevLett.123.250605} {\bibfield  {journal} {\bibinfo  {journal} {Phys. Rev. Lett.}\ }\textbf {\bibinfo {volume} {123}},\ \bibinfo {pages} {250605} (\bibinfo {year} {2019})}\BibitemShut {NoStop}%
    \bibitem [{\citenamefont {Tran}\ \emph {et~al.}(2021)\citenamefont {Tran}, \citenamefont {Guo}, \citenamefont {Baldwin}, \citenamefont {Ehrenberg}, \citenamefont {Gorshkov},\ and\ \citenamefont {Lucas}}]{Lucas21}%
      \BibitemOpen
      \bibfield  {author} {\bibinfo {author} {\bibfnamefont {M.~C.}\ \bibnamefont {Tran}}, \bibinfo {author} {\bibfnamefont {A.~Y.}\ \bibnamefont {Guo}}, \bibinfo {author} {\bibfnamefont {C.~L.}\ \bibnamefont {Baldwin}}, \bibinfo {author} {\bibfnamefont {A.}~\bibnamefont {Ehrenberg}}, \bibinfo {author} {\bibfnamefont {A.~V.}\ \bibnamefont {Gorshkov}},\ and\ \bibinfo {author} {\bibfnamefont {A.}~\bibnamefont {Lucas}},\ }\bibfield  {title} {\bibinfo {title} {{Lieb-Robinson} light cone for power-law interactions},\ }\href {https://doi.org/10.1103/PhysRevLett.127.160401} {\bibfield  {journal} {\bibinfo  {journal} {Phys. Rev. Lett.}\ }\textbf {\bibinfo {volume} {127}},\ \bibinfo {pages} {160401} (\bibinfo {year} {2021})}\BibitemShut {NoStop}%
    \bibitem [{\citenamefont {Barahona}(1982)}]{barahona}%
      \BibitemOpen
      \bibfield  {author} {\bibinfo {author} {\bibfnamefont {F.}~\bibnamefont {Barahona}},\ }\bibfield  {title} {\bibinfo {title} {On the computational complexity of {Ising} spin glass models},\ }\href {https://doi.org/10.1088/0305-4470/15/10/028} {\bibfield  {journal} {\bibinfo  {journal} {J. Phys. A: Math. Gen.}\ }\textbf {\bibinfo {volume} {15}},\ \bibinfo {pages} {3241} (\bibinfo {year} {1982})}\BibitemShut {NoStop}%
    \end{thebibliography}


\end{document}